\documentclass{article}

\usepackage{latexsym}
\usepackage{amsmath}
\usepackage{amssymb}
\usepackage{amsthm}
\usepackage{amscd}

\usepackage{graphicx}
\usepackage{subfigure}
\usepackage{psfrag}

\usepackage{bm}
\usepackage{cancel}

\usepackage[mathscr]{eucal}

\newcommand{\textfrac}[2]{{\textstyle \frac{#1}{#2}}}

\newcommand{\Aplus}{{\fontfamily{phv}\fontseries{b}\selectfont A}${}_{\boldsymbol{+}}$}
\newcommand{\Bplus}{{\fontfamily{phv}\fontseries{b}\selectfont B}${}_{\boldsymbol{+}}$}
\newcommand{\Cplus}{{\fontfamily{phv}\fontseries{b}\selectfont C}${}_{\boldsymbol{+}}$}
\newcommand{\Dplus}{{\fontfamily{phv}\fontseries{b}\selectfont D}${}_{\boldsymbol{+}}$}
\newcommand{\Aminus}{{\fontfamily{phv}\fontseries{b}\selectfont A}${}_{\boldsymbol{-}}$}
\newcommand{\Bminus}{{\fontfamily{phv}\fontseries{b}\selectfont B}${}_{\boldsymbol{-}}$}
\newcommand{\Cminus}{{\fontfamily{phv}\fontseries{b}\selectfont C}${}_{\boldsymbol{-}}$}
\newcommand{\Dminus}{{\fontfamily{phv}\fontseries{b}\selectfont D}${}_{\boldsymbol{-}}$}
\newcommand{\A}{{\fontfamily{phv}\fontseries{b}\selectfont A}}
\newcommand{\B}{{\fontfamily{phv}\fontseries{b}\selectfont B}}
\newcommand{\C}{{\fontfamily{phv}\fontseries{b}\selectfont C}}
\newcommand{\D}{{\fontfamily{phv}\fontseries{b}\selectfont D}}

\newcommand{\Azero}{{\fontfamily{phv}\fontseries{b}\selectfont A}${}_{\hspace{0.1em}\boldsymbol{0}\,}$}

\newcommand{\Azeroplus}{{\fontfamily{phv}\fontseries{b}\selectfont A}${}_{\hspace{0.1em}\boldsymbol{0+}}$}

\newcommand{\Azerominus}{{\fontfamily{phv}\fontseries{b}\selectfont A}${}_{\hspace{0.1em}\boldsymbol{0-}}$}

\newcommand{\Aplusminus}{{\fontfamily{phv}\fontseries{b}\selectfont A}${}_{\boldsymbol{\pm}}$}

\newcommand{\weg}{\:\,}

\DeclareMathOperator{\tr}{tr}
\DeclareMathOperator{\sgn}{sgn}

\theoremstyle{plain}
\newtheorem{Theorem}{Theorem}

\newtheorem{Definition}{Definition}
\newtheorem{Lemma}{Lemma}

\theoremstyle{remark}
\newtheorem*{Remark}{Remark}
\newtheorem{Assumption}{Assumption}
\newtheorem*{Example}{Example}

\title{Dynamics of Bianchi type~I solutions of the Einstein equations with anisotropic matter}
\author{Simone Calogero\footnote{E-Mail: calogero@ugr.es}\\[0.2cm]
Departamento de Matem\'atica Aplicada\\
Facultad de Ciencias, Universidad de Granada\\
18071 Granada, Spain\\[0.5cm]
J.~Mark Heinzle\thanks{E-Mail: Mark.Heinzle@univie.ac.at}\\[0.2cm]
Gravitational Physics\\ Faculty of Physics, University of Vienna\\
1090 Vienna, Austria}

\date { }

\begin{document}

\maketitle

\begin{abstract}
We analyze the global dynamics of Bianchi type I solutions of the Einstein equations
with anisotropic matter. The matter model is not specified explicitly 
but only through a set of mild and physically motivated assumptions; thereby
our analysis covers matter models as different from each other as, e.g., 
collisionless matter, elastic matter and magnetic fields.
The main result we prove 
is the existence of an `anisotropy classification' for the
asymptotic behaviour of Bianchi type~I cosmologies.
The type of asymptotic behaviour of generic solutions 
is determined by one single parameter
that describes certain properties 
of the anisotropic matter model under extreme conditions.
The anisotropy classification comprises the following types.
The convergent type \Aplus: Each solution converges to a Kasner
solution as the singularity is approached and each Kasner
solution is a possible past asymptotic state.
The convergent types \Bplus\ and \Cplus: Each solution converges to a Kasner
solution as the singularity is approached; however, the set of Kasner
solutions that are possible past asymptotic states is restricted.
The oscillatory type \Dplus: Each solution oscillates between
different Kasner solutions as the singularity is approached.
Furthermore, we investigate non-generic asymptotic behaviour and 
the future asymptotic behaviour of solutions.
\end{abstract}

\section{Introduction}
\label{intro}

A pivotal feature in the study of spatially homogeneous
cosmologies is the fact that the (asymptotic) dynamics of
cosmological solutions of the Einstein equations is largely determined by
the spatial geometry of the initial spacelike hypersurface.
This is reflected in the systematics of the 
Bianchi classification of spatially homogeneous cosmologies.
In contrast, a systematic analysis of the influence of the matter model 
on the (asymptotic) dynamics of solutions is not available at present.
It is known that the dynamics of solutions (and the 
asymptotics towards the initial singularity, in particular) strongly depends 
on the matter source that is considered; the relatively simple behaviour
of the vacuum and the perfect fluid case, see, e.g.,~\cite{WE}, 
is replaced by a considerably more intricate behaviour in the case of 
anisotropic matter models~\cite{CH, HU, L, RU, S}.
The purpose of this paper is to analyze systematically and 
in detail in which way anisotropies of the matter source determine 
the dynamics of cosmological solutions.

In our analysis we consider spatially homogeneous solutions of Bianchi type~I.
The reason 
for this choice 
is that Bianchi type~I is the `foundation' of the Bianchi classification.
Generally, the dynamics of the `higher' types in the Bianchi classification
is based on the dynamics of the `lower' types; for instance,
the asymptotics of solutions of 
Bianchi types~VIII and~IX can only be understood in terms of 
solutions of Bianchi type~I and~II, see~\cite{HU2} for an up-to-date discussion.
Likewise, our understanding of cosmological models in general is conjectured
to be built on the dynamics of spatially homogeneous models 
and thus on the dynamics of Bianchi type~I models in particular,
see~\cite{HUR} and references therein.

The dynamics of Bianchi type~I cosmological models where the matter is a perfect fluid 
is well-known~\cite{WE}. In this context, it is customary to assume that the 
the perfect fluid is represented by an energy density $\rho$ and pressure $p$ that obey
a linear equation of state $p = w \rho$ with $w = \mathrm{const} \in (-1, 1)$.
Each Bianchi type~I perfect fluid solution isotropizes for late times; this means that each
solution approaches a (flat) Friedmann-Robertson-Walker (FRW) solution for late times.
Towards the initial singularity, each solution approaches a Kasner solution (Bianchi type~I vacuum
solution), i.e., to leading order in the limit $t\rightarrow 0$ the metric is described by
\begin{equation*}
d s^2 = -d t^2 + a_1 \,t^{2 p_1} d x^1\otimes dx^1 + a_2\, t^{2 p_2} d x^2\otimes d x^2 + a_3 \,t^{2 p_3} d x^3\otimes dx^3\:,
\end{equation*}
where $a_1$, $a_2$, $a_3$ are positive constants, and $p_1$, $p_2$, $p_3$ are
the Kasner exponents, which are constants that satisfy $p_1 + p_2 + p_3 = p_1^2 +p_2^2 +p_3^2 = 1$.
Conversely (and importantly), for each Kasner solution (including the Taub solution, for which 
$(p_1,p_2,p_3) = (1,0,0)$ and permutations) 
there exists a Bianchi type~I perfect fluid model that converges to this solution
as $t\rightarrow 0$; in other words, each Kasner solution is a possible past asymptotic state.

\begin{figure}[Ht]
\begin{center}
\subfigure[\Cplus\ ($1<\beta<2$)]{
\psfrag{y1}[cc][cc][0.8][0]{$\Sigma_1$}
\psfrag{y2}[cc][cc][0.8][0]{$\Sigma_2$}
\psfrag{y3}[cc][cc][0.8][0]{$\Sigma_3$}
\psfrag{1-1}[cc][cc][0.7][90]{$\mathbf{\Sigma_1=2/\beta}$}
\psfrag{2-1}[cc][cc][0.7][30]{$\mathbf{\Sigma_2=2/\beta}$}
\psfrag{3-1}[cc][cc][0.7][-30]{$\mathbf{\Sigma_3=2/\beta}$}
\psfrag{D1}[cc][cc][0.8][90]{$\partial\mathcal{D}_1$}
\psfrag{D2}[cc][cc][0.8][30]{$\partial\mathcal{D}_2$}
\psfrag{D3}[cc][cc][0.8][-30]{$\partial\mathcal{D}_3$}
\psfrag{d1}[cc][cc][0.6][0]{$\mathrm{D}_1$}
\psfrag{d2}[cc][cc][0.6][0]{$\mathrm{D}_2$}
\psfrag{d3}[cc][cc][0.6][0]{$\mathrm{D}_3$}
\includegraphics[width=0.3\textwidth]{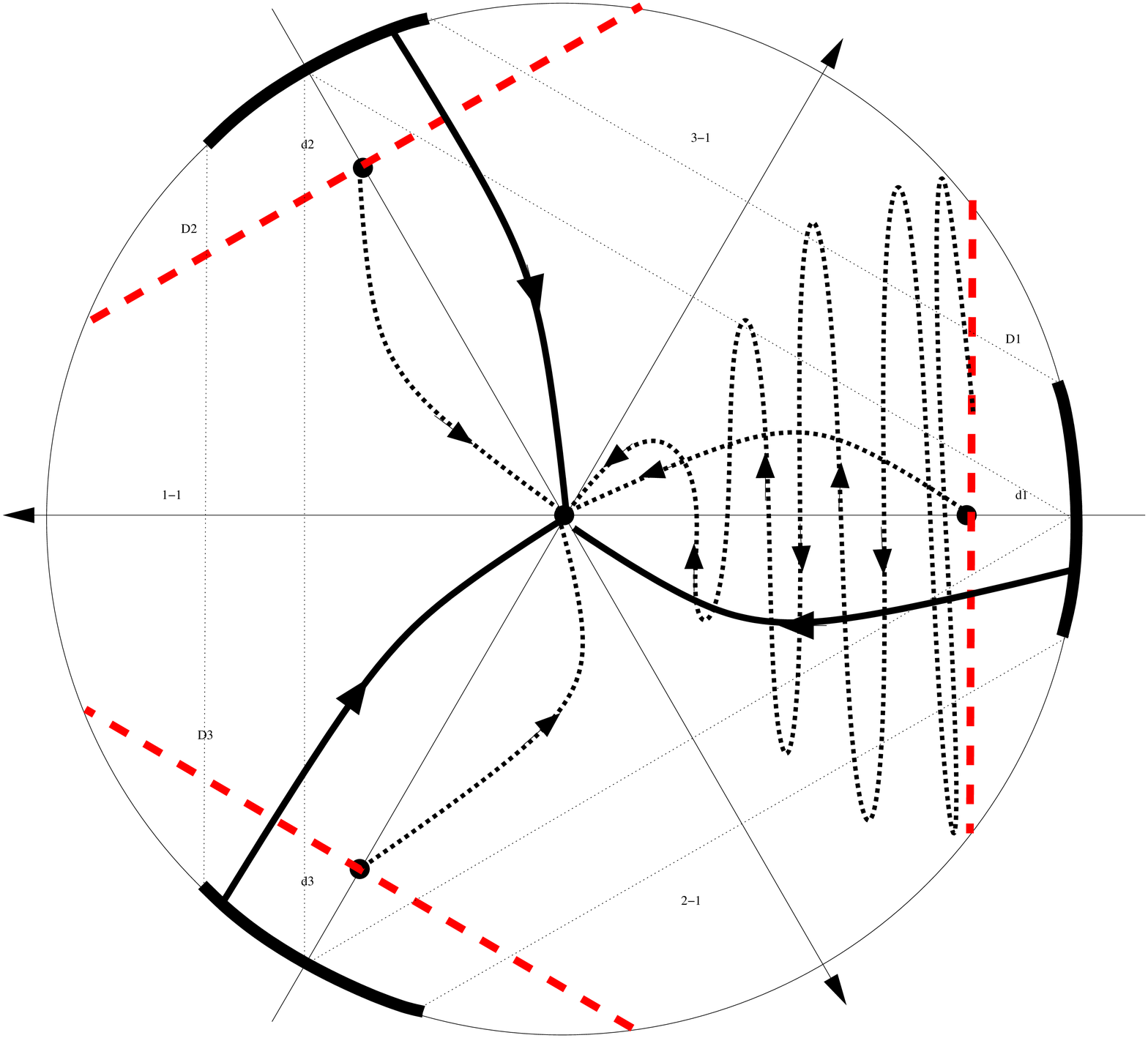}}
\quad
\subfigure[\Bplus\ ($\beta=1$)]{
\psfrag{y1}[cc][cc][0.8][0]{$\Sigma_1$}
\psfrag{y2}[cc][cc][0.8][0]{$\Sigma_2$}
\psfrag{y3}[cc][cc][0.8][0]{$\Sigma_3$}
\psfrag{D1}[cc][cc][0.8][90]{$\partial\mathcal{D}_1$}
\psfrag{D2}[cc][cc][0.8][30]{$\partial\mathcal{D}_2$}
\psfrag{D3}[cc][cc][0.8][-30]{$\partial\mathcal{D}_3$}
\psfrag{d1}[cc][cc][0.6][0]{$\mathrm{D}_1$}
\psfrag{d2}[cc][cc][0.6][0]{$\mathrm{D}_2$}
\psfrag{d3}[cc][cc][0.6][0]{$\mathrm{D}_3$}
\psfrag{T1}[cc][cc][0.6][0]{$\mathrm{T}_1$}
\psfrag{T2}[cc][cc][0.6][0]{$\mathrm{T}_2$}
\psfrag{T3}[cc][cc][0.6][0]{$\mathrm{T}_3$}
\includegraphics[width=0.3\textwidth]{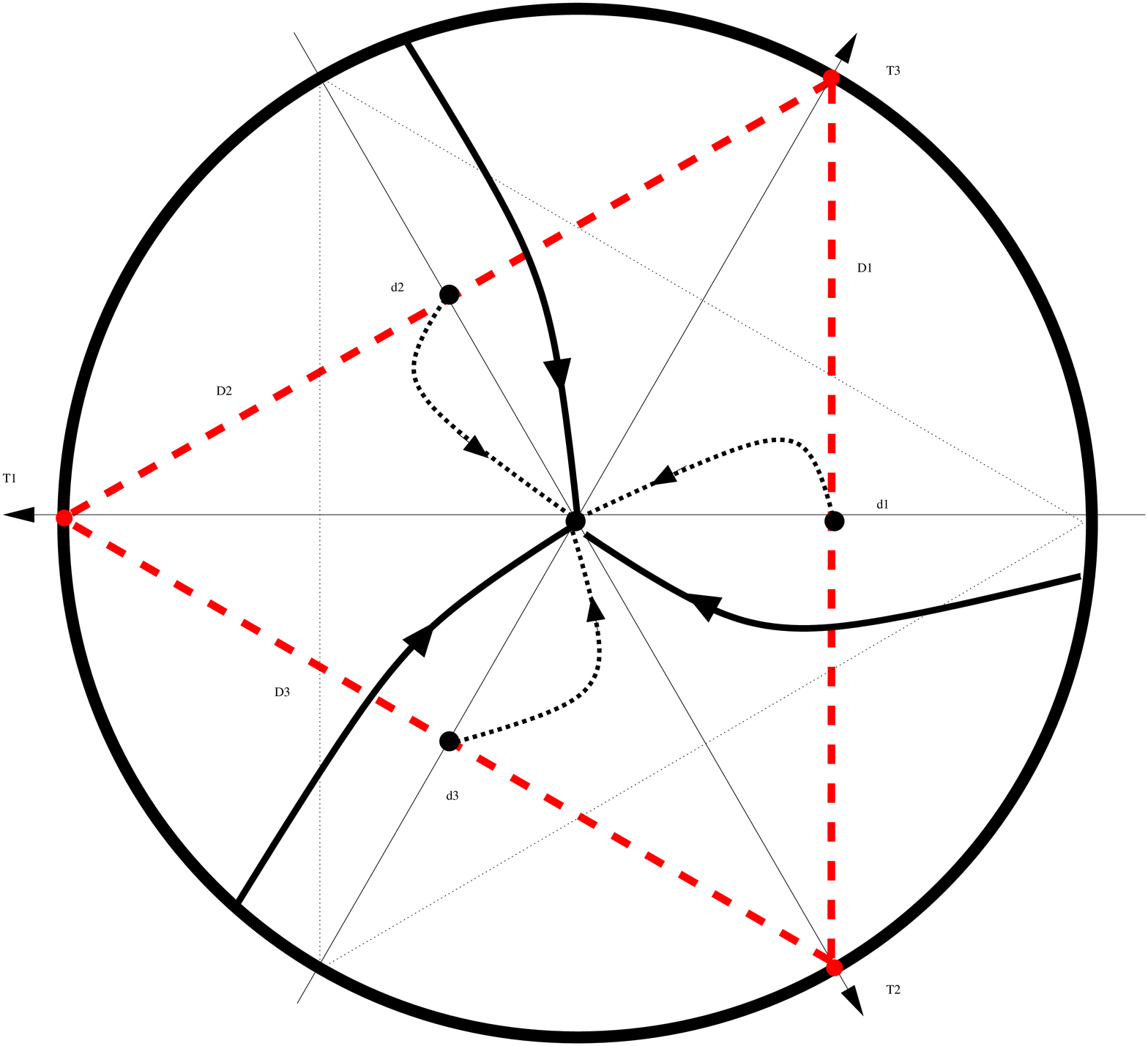}}\quad
\subfigure[\Aplus\ ($0<\beta<1$)]{
\psfrag{y1}[cc][cc][1][0]{$\Sigma_1$}
\psfrag{y2}[cc][cc][1][0]{$\Sigma_2$}
\psfrag{y3}[cc][cc][1][0]{$\Sigma_3$}
\psfrag{d1}[cc][cc][0.6][0]{$\mathrm{D}_1$}
\psfrag{d2}[cc][cc][0.6][0]{$\mathrm{D}_2$}
\psfrag{d3}[cc][cc][0.6][0]{$\mathrm{D}_3$}
\psfrag{r1}[cc][cc][0.6][0]{$\mathrm{R}_1$}
\psfrag{r2}[cc][cc][0.6][0]{$\mathrm{R}_2$}
\psfrag{r3}[cc][cc][0.6][0]{$\mathrm{R}_3$}
\includegraphics[width=0.3\textwidth]{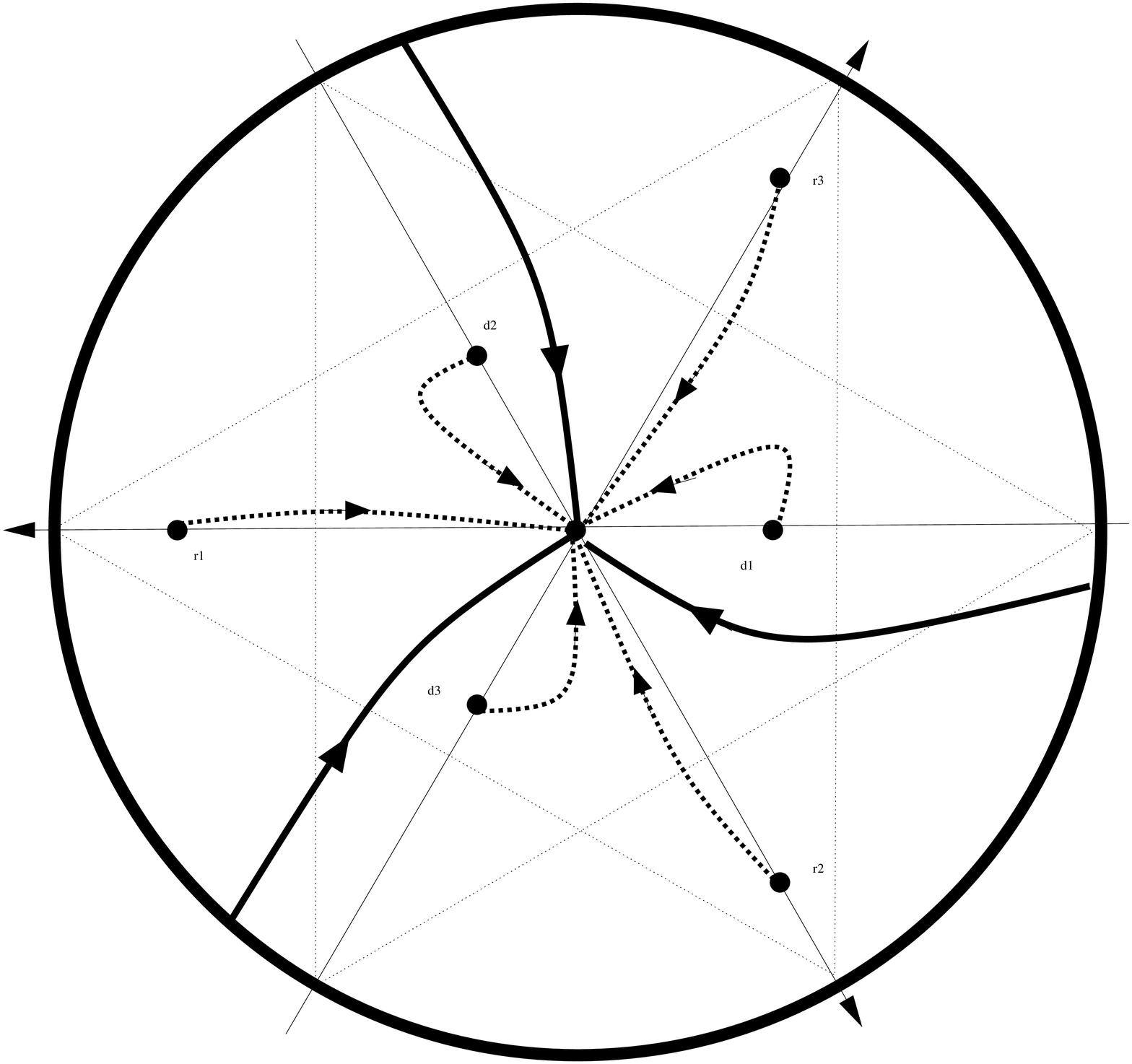}}\\
\subfigure[\Aminus\ ($-1<\beta <0$)]{
\psfrag{y1}[cc][cc][1][0]{$\Sigma_1$}
\psfrag{y2}[cc][cc][1][0]{$\Sigma_2$}
\psfrag{y3}[cc][cc][1][0]{$\Sigma_3$}
\psfrag{d1}[cc][cc][0.6][0]{$\mathrm{D}_1$}
\psfrag{d2}[cc][cc][0.6][0]{$\mathrm{D}_2$}
\psfrag{d3}[cc][cc][0.6][0]{$\mathrm{D}_3$}
\psfrag{r1}[cc][cc][0.6][0]{$\mathrm{R}_1$}
\psfrag{r2}[cc][cc][0.6][0]{$\mathrm{R}_2$}
\psfrag{r3}[cc][cc][0.6][0]{$\mathrm{R}_3$}
\includegraphics[width=0.3\textwidth]{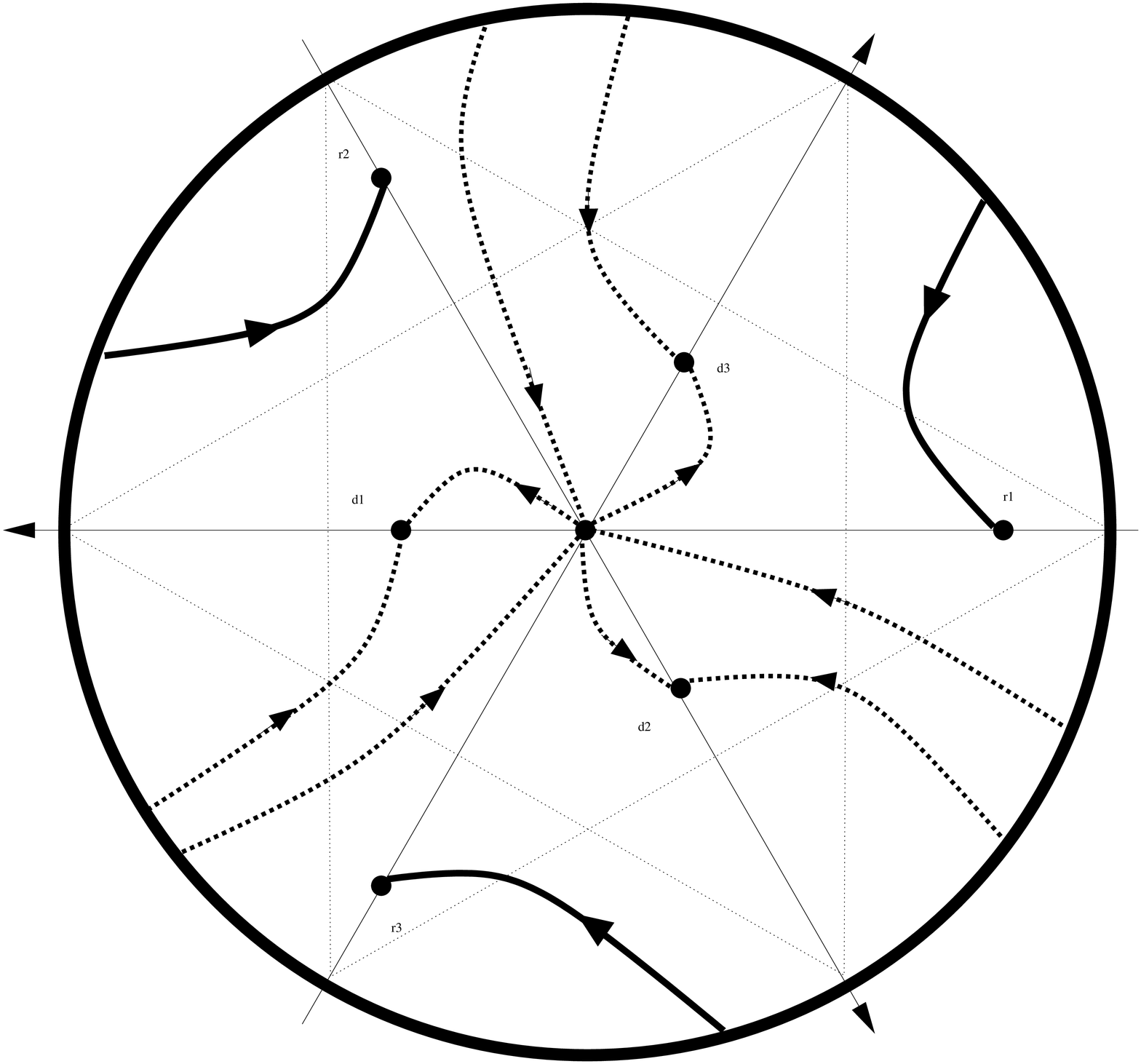}}\quad
\subfigure[\Bminus\ ($\beta=-1$)]{
\psfrag{y1}[cc][cc][1][0]{$\Sigma_1$}
\psfrag{y2}[cc][cc][1][0]{$\Sigma_2$}
\psfrag{y3}[cc][cc][1][0]{$\Sigma_3$}
\psfrag{d1}[cc][cc][0.6][0]{$\mathrm{D}_1$}
\psfrag{d2}[cc][cc][0.6][0]{$\mathrm{D}_2$}
\psfrag{d3}[cc][cc][0.6][0]{$\mathrm{D}_3$}
\psfrag{r1}[cc][cc][0.6][0]{$\mathrm{Q}_1$}
\psfrag{r2}[cc][cc][0.6][0]{$\mathrm{Q}_2$}
\psfrag{r3}[cc][cc][0.6][0]{$\mathrm{Q}_3$}
\includegraphics[width=0.3\textwidth]{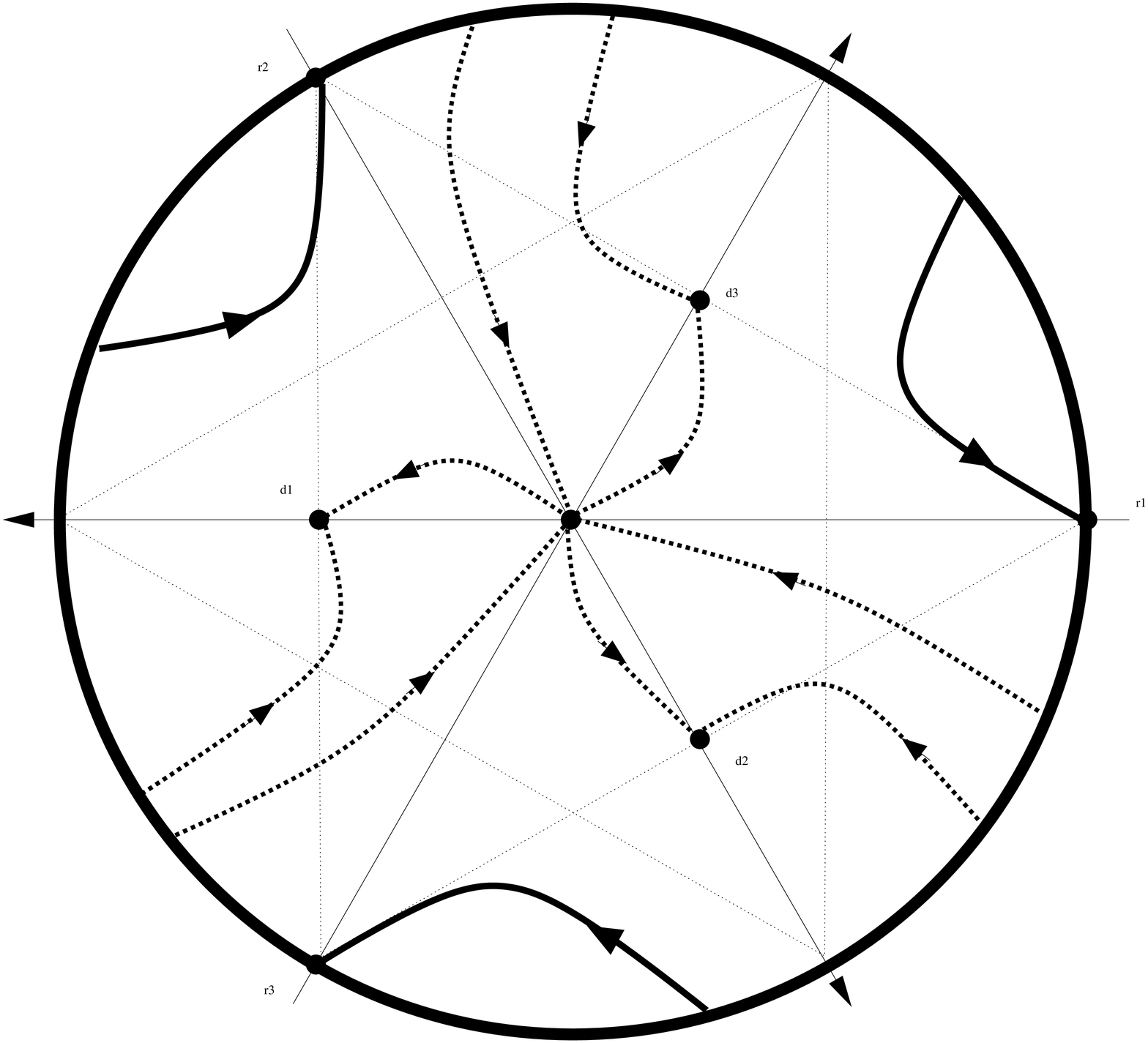}}\quad
\subfigure[\Cminus\ ($-2<\beta<-1$)]{
\psfrag{y1}[cc][cc][1][0]{$\Sigma_1$}
\psfrag{y2}[cc][cc][1][0]{$\Sigma_2$}
\psfrag{y3}[cc][cc][1][0]{$\Sigma_3$}
\psfrag{d1}[cc][cc][0.6][0]{$\mathrm{D}_1$}
\psfrag{d2}[cc][cc][0.6][0]{$\mathrm{D}_2$}
\psfrag{d3}[cc][cc][0.6][0]{$\mathrm{D}_3$}
\psfrag{r1}[cc][cc][0.6][0]{$\mathrm{Q}_1$}
\psfrag{r2}[cc][cc][0.6][0]{$\mathrm{Q}_2$}
\psfrag{r3}[cc][cc][0.6][0]{$\mathrm{Q}_3$}
\includegraphics[width=0.3\textwidth]{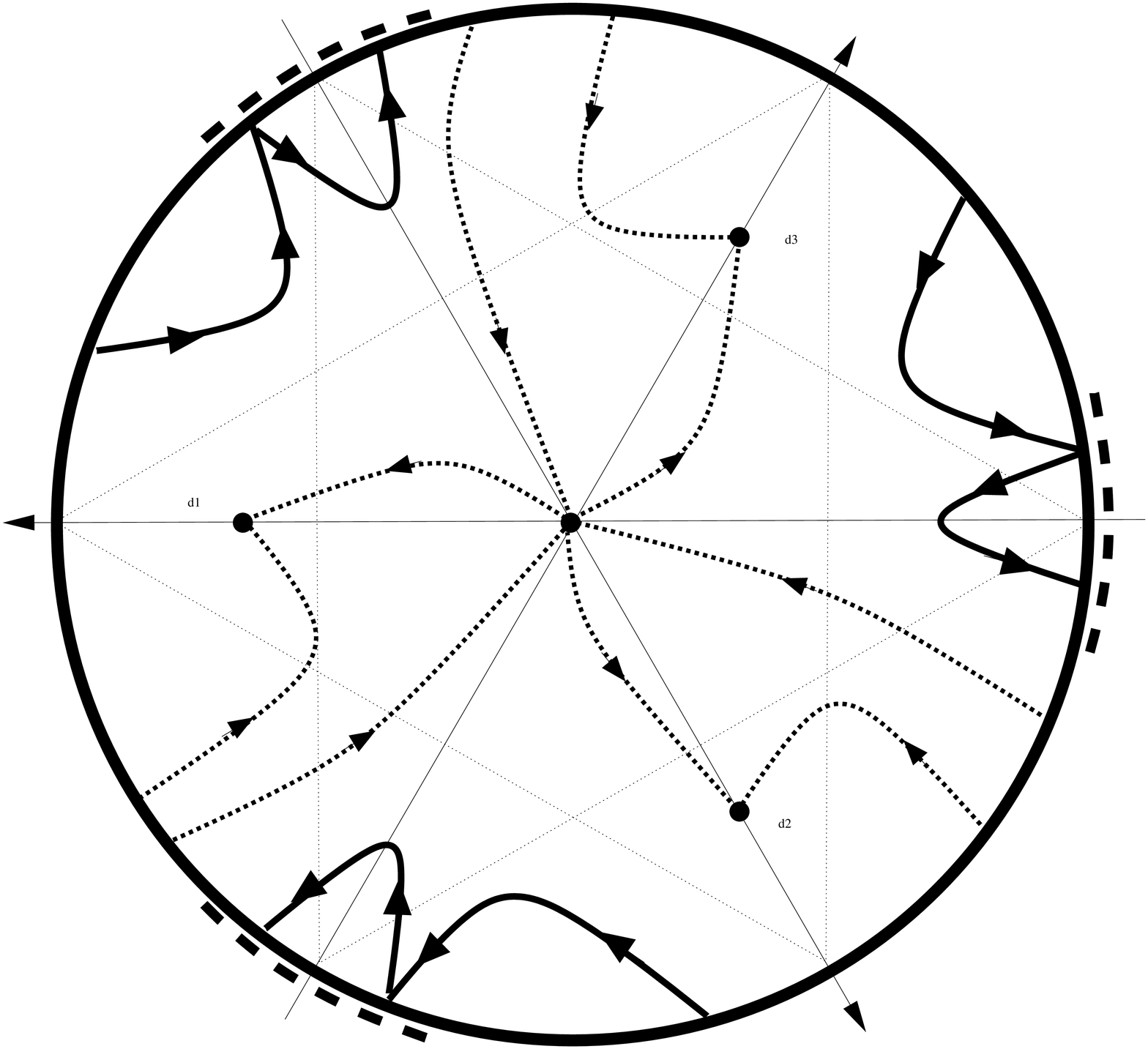}}
\caption{A schematic depiction of the projection onto the Kasner disc of interior orbits. 
By abuse of notation, we use the same letter to denote a fixed point and its projection 
onto the Kasner disc. The center of the disc is the projection of the fixed point $\mathrm{F}$ 
(which represents the isotropic solution). Generic orbits are in bold. In the cases \Bplus/\Cplus, the dashed 
lines represent the heteroclinic cycles/network. In the case \Bplus, the heteroclinic 
network could also attract orbits as $\tau\rightarrow -\infty$ (not depicted). In the case \Aminus, 
there are Kasner states which act both as $\alpha$-\ and $\omega$-limit of interior orbits.}
\label{Final}
\end{center}
\end{figure}

In this paper we are concerned with Bianchi type~I cosmological models with \textit{anisotropic matter}.
We do not specify the matter model explicitly; on the contrary, the assumptions that we impose 
on the matter source are so mild that we can treat several matter models at the same time, 
examples being matter models as different from each other 
as collisionless matter, elastic matter (for a wide variety of constitutive equations) and magnetic fields
(aligned along one of the axes).
The main results we derive are presented as Theorems~\ref{futurethm}--\ref{pastthm2} in Section~\ref{global} 
and summarized in Figure~\ref{Final}, but let us give a rough non-technical overview: 

First consider the asymptotic behaviour of models towards the future. 
For `conventional' matter sources---where `conventional' means that the isotropic state of these anisotropic matter models
is energetically favorable---we find that each associated Bianchi type~I solution 
isotropizes towards the future, i.e., each solution approaches a FRW solution for late times. 
However, there also exist matter models whose isotropic state
is unstable. In this case, the future asymptotic behaviour is
completely different; isotropization occurs but it is non-generic;
generically, solutions approach different self-similar solutions,
e.g., there exist solutions that approach Kasner solutions as $t\rightarrow \infty$.
It should be emphasized that these models are often 
compatible with the energy conditions.

Second consider the asymptotic behaviour towards the singularity. 
We observe a rather diverse asymptotic behaviour, but 
the past dynamics of \textit{generic} Bianchi type~I solutions
is again intimately connected with the Kasner states.
Interestingly enough, the details of the past asymptotic dynamics of solutions
are governed by one particular parameter, $\beta$, that describes
certain `asymptotic properties' of the matter (the properties of the matter
under extreme conditions---extreme stress, etc.---that are found close to the spacetime singularity); 
this parameter can also be regarded as a measure for
the degree of (dominant) energy condition violation under extreme conditions.
(Clearly, we always assume that the dominant energy condition is satisfied 
under `normal conditions' of the matter.)
If the matter satisfies the dominant energy condition also under extreme conditions,
then each 
Bianchi type~I solution approaches a Kasner solution,
and conversely, each Kasner solution is a possible past asymptotic state;
the Kasner solutions are on an equal footing in this respect.
(The case where the energy condition is satisfied only marginally
is slightly different; the Taub solutions play a special role in that case.)
If the dominant energy condition is violated under extreme conditions, 
this result breaks down;
however, we must distinguish between `under-critical' violations
of the energy condition (corresponding to a small value of $\beta$)
and `over-critical' violations (corresponding to a large value of $\beta$). 
For an under-critical violation of the energy condition, 
each Bianchi type~I solution approaches a Kasner state,
but the converse is false, i.e., there exist Kasner solutions that are excluded
as possible past asymptotic states. Only Kasner solutions that are 
`sufficiently different' from the Taub solution, i.e., only Kasner solutions
with $(p_1,p_2, p_3)$ sufficiently different from $(1,0,0)$ and permutations,
are attractors for Bianchi type~I solutions.
The set of potential asymptotic states is smaller if the degree
of the energy condition violation is larger.
Finally, for an over-critical violation of the energy condition,
Bianchi type~I solutions with anisotropic matter do not converge
to a Kasner state as $t\rightarrow 0$, but they follow
a sequence of Kasner states (`epochs'), in each of
which their behaviour is approximately described by a Kasner solution.
This is a type of behaviour that resembles the Mixmaster oscillatory
behaviour that is expected for Bianchi type~VIII and~IX (vacuum) cosmologies
(and generic cosmological models); note, however, that both the details
and the origin of these oscillations are very different---in the present case, 
the oscillatory behaviour is not related to the Mixmaster map; it is a consequence
of (over-critical) energy condition violation; in the Mixmaster case,
the reason is the geometry of the problem.
Independently of the energy conditions, there also exist a variety of Bianchi type~I 
solutions with anisotropic matter that exhibit 
a past asymptotic behaviour that is not connected with Kasner asymptotics;
however, these are non-generic solutions; we prove that 
there exist essentially three different types of non-generic asymptotics.

The outline of the paper is as follows. In Section~\ref{setup} we discuss the Einstein equations 
in Bianchi type~I symmetry and introduce our main assumption on the anisotropic matter model:
The stress-energy tensor is assumed to be represented by a function of the metric and the initial
data of the matter source. We shall restrict ourselves to the study of diagonal models, 
i.e., solutions for which the metric 
is diagonal in the standard coordinates of Bianchi type~I symmetric spacetimes. In 
Section~\ref{dynamicalsection} we rewrite the Einstein equations in terms of 
new variables and show that the essential dynamics is described by a system of 
autonomous differential equations on a four dimensional compact state 
space---the \textit{reduced dynamical system}. 
Section~\ref{boundaries} is by far the longest and most technical part of the paper. 
It contains a detailed analysis of the flow induced by the reduced dynamical system 
on the boundary of the state space. 
In Section~\ref{local} we study the local 
stability properties of the fixed points of the reduced dynamical system. 
Finally, in
Section~\ref{global} we present the main results: 
Theorems~\ref{futurethm}--\ref{pastthm2} and Figure~\ref{Final}.
The proofs are based on the results of Sections~\ref{boundaries}
and~\ref{local} and make use of
techniques from the theory of dynamical systems. 
The purpose of Section~\ref{mattermodels} is to bring to life the
general assumptions on the matter source that are made in Sections~\ref{setup}--\ref{local}.
We give three important examples of matter models to which our analysis applies straightforwardly: 
collisionless (Vlasov) matter, elastic matter and magnetic fields. 
Finally, Section~\ref{conclusions} 
contains some concluding remarks 
together with an outlook of possible developments of the present work. 
      
\section{The Einstein equations in Bianchi type~I}
\label{setup}

We consider a spatially homogeneous spacetime of Bianchi type~I.
The spacetime metric can be written as
\begin{equation}\label{metric}
d s^2 = -d t^2 + g_{i j}(t) d x^i d x^j \qquad (i,j=1,2,3)\:,
\end{equation}
where $g_{i j}$ is the induced Riemannian metric on the spatially
homogeneous surfaces $t=\mathrm{const}$. 
By $k_{i j}$ we denote the second fundamental form of the
surfaces $t=\mathrm{const}$.
The energy-momentum tensor $T_{\mu\nu}$ ($\mu,\nu=0,1,2,3$) 
contains the energy density $\rho = T_{00}$ and the momentum density 
$j_k= T_{0k}$. 
Einstein's equations, in units $c=1=8 \pi G$,
decompose into evolution equations and constraints.
The evolution equations are
\begin{subequations}\label{einstein}
\begin{equation}\label{evolution}
\partial_t g_{i j} = -2 g_{i l} k^l_{\weg j} \:,\quad
\partial_t k^i_{\weg j} = \tr k\: k^i_{\weg j} - T^i_{\weg j} +
\frac{1}{2} \delta^i_{\weg j} ( \tr T -\rho) \:,
\end{equation}
the Hamiltonian constraint reads 
\begin{equation}\label{constraints} (\tr k)^2
- k^i_{\weg j} k^j_{\weg i} - 2 \rho = 0\:,
\end{equation}
\end{subequations}
and the momentum constraint is $j_k = 0$ (which is due to the fact that Bianchi I spacetimes are spatially flat).
The latter does not impose any restriction on the gravitational degrees of freedom,
which are represented by $g_{ij}$ and $k^i_{\ j}$,
but only on the matter fields. 
In general, the system~\eqref{einstein} must be complemented by equations describing
the evolution of the matter fields. 

For certain matter models, the energy-momentum tensor takes the form of an explicit functional of the metric $g_{ij}$. 
Let us give some examples.
For a (non-tilted) perfect fluid, the energy-momentum tensor reads
$T_{\mu\nu} dx^\mu dx^\nu = \rho \,d t^2 + p\: g_{i j} d x^i d x^j$, 
where the energy density $\rho$ and the pressure $p$ 
are connected via a barotropic equation of state $p(\rho)$. 
When we define $n$ as $n =(\det g)^{-1/2}$, 
the Einstein equations~\eqref{einstein} 
imply $p(\rho)+\rho = n (d\rho/dn)$ (which is a special case of equation~\eqref{Tij} below).
Hence $\rho$ is obtained as a function $\rho = \rho(n)$.
Clearly, prescribing $\rho$ as a function of $n$ is equivalent to
prescribing an equation of state for the fluid (modulo a scaling 
constant).
Likewise, for collisionless matter 
the components of the energy-momentum tensor
are given as functions of the metric $g_{i j}$,
which depend on the initial data for 
the matter (which is described by 
the distribution function of particles in phase space; 
see Section~\ref{mattermodels} for details). 
%
%
We thereby obtain the energy density $\rho$ and 
the principal pressures $p_i$ (which are defined as the
eigenvalues of $T^i_{\weg j}$) as $\rho = \rho(g_{ij})$ and $p_i = p_i(g_{jk})$.
Finally, for elastic matter the constitutive equation of state of the material
determines $\rho$ and $p_i$ as functions of $g_{i j}$;
for details we refer again to Section~\ref{mattermodels}.
Motivated by these examples we make our fundamental assumption on the matter model.

\begin{Assumption}\label{assumptionT}
The components of the energy-momentum tensor 
are represented by smooth (at least $\mathcal{C}^1$)
functions of the metric $g_{ij}$. 
We assume that $\rho=\rho(g_{ij})$ is positive 
(as long as $g_{ij}$ is non-degenerate).
\end{Assumption}

\begin{Remark}
We note that the particular form of the functions that represent 
the components of the energy-momentum tensor
may depend on a number of external parameters or external
functions that describe the properties of the matter, and/or the initial data
of the matter field(s); see Section~\ref{mattermodels} and~\cite{UJR} 
for examples.
\end{Remark}

By Assumption~\ref{assumptionT}, the 
evolution equations of the matter fields in Bianchi type~I are
contained in the Einstein evolution equations~\eqref{evolution} via the 
contracted Bianchi identity.

\begin{Remark}
The regularity of $T^i_{\weg j}$ as a function 
of the spatial metric ensures that the Cauchy problem for the evolution 
equations with initial data at $t=t_0$ is locally well-posed on a time 
interval $(t_-,t_+)$, $t_-<t_0<t_+$. The requirement $\rho >0$ 
and the constraint equation~\eqref{constraints} imply that the mean 
curvature $\tr k$ never vanishes in the interval of existence. 
Without loss of generality we assume that $\tr k<0$ on $(t_-,t_+)$, i.e.,
we consider an expanding spacetime;
the case $\tr k>0$ is obtained from $\tr k<0$ by replacing $t$ with $-t$. 
If the energy momentum tensor satisfies the inequality
$\tr T + 3 \rho > 0$ (which corresponds to  
Assumption~\ref{assumptionwi} in Section~\ref{dynamicalsection}),
then the mean curvature is non-decreasing, the solutions 
of~\eqref{einstein} exist in an interval $(t_-,+\infty)$ with $t_- > -\infty$, 
and the spacetime is future geodesically complete. 
By a time translation we can assume $t_-=0$. 
Under additional conditions on $T^i_{\weg j}$ the limit $t\rightarrow 0$
corresponds to a curvature singularity.
We refer to~\cite{R,R2} for a proof of these statements in the context
of general spatially homogeneous cosmologies;
for the special case we consider these results follow straightforwardly
from our formulation of the equations, cf.~\eqref{dynamicalsystem}.
\end{Remark}

\begin{Lemma}\label{formulaT}
Assumption \ref{assumptionT} implies
\begin{equation}\label{Tij}
T^i_{\ j}=-2\frac{\partial\rho}{\partial g_{il}}\,g_{jl}-\delta^i_{\ j}\,\rho\:.
\end{equation}
In particular, the functional dependence of the energy-momentum tensor on the metric $g_{ij}$ is 
completely determined by the function $\rho(g_{ij})$. 
\end{Lemma}

\begin{proof} 
Differentiating~\eqref{constraints} w.r.t.\ $t$ and using the evolution equation for $k^i_{\ j}$ we obtain
\begin{subequations}
\begin{equation}\label{dtrho1}
\partial_t\rho=k^j_{\ i}\left(T^i_{\ j}+\delta^i_{\ j}\,\rho\right).
\end{equation}
On the other hand, using $\rho=\rho(g_{ij})$ and the evolution equation for the metric we obtain
\begin{equation}\label{dtrho2}
\partial_t\rho=-2k^j_{\ i}\,g_{jl}\,\frac{\partial\rho}{\partial g_{il}}\:.
\end{equation}
\end{subequations}
Equating the r.h.\ sides of~\eqref{dtrho1} and~\eqref{dtrho2} gives an identity between two
polynomials in $k^j_{\ i}$ with coefficients that depend only on the metric $g_{ij}$; 
therefore the corresponding coefficients must be equal, which gives~\eqref{Tij}. 
\end{proof} 

Equation~\eqref{Tij} will play a fundamental role in the following. 
We emphasize that the energy-momentum 
tensor must be independent of the second fundamental form for 
the proof of Lemma~\ref{formulaT} to hold. We refer to Section~\ref{mattermodels}
for a derivation of~\eqref{Tij} within the Lagrangian formalism.

Initial data for the Einstein-matter system are given by
$g_{ij}(t_0)$, $k^i_{\weg j}(t_0)$ and the initial values of the matter fields.
Without loss of generality we can assume that $g_{ij}(t_0)$ and $k^i_{\ j}(t_0)$
are diagonal (by choosing coordinates adapted to an orthogonal basis of eigenvectors of $k^i_{\ j}(t_0)$).

\begin{Assumption}\label{assumptiondiagonal}
We assume that there exists initial data for the matter such that 
$T^i_{\weg j}$ is diagonal in any orthogonal frame.
\end{Assumption}

Uniqueness of solutions of the evolution equations then implies 
that $(g_{ij}, k^i_{\weg j}, T^i_{\weg j})$ remain
diagonal for all times. We refer to solutions of this type as~\textit{diagonal models};
henceforth we restrict ourselves to these models.

\section{Dynamical systems formulation}
\label{dynamicalsection}

We introduce a set of variables for which the Einstein evolution equations~\eqref{evolution} 
decouple.
The reduced system we thereby obtain is a dynamical system on a bounded state space.
Let
\begin{alignat*}{3}
H & = -\frac{\tr k}{3}\:, & \qquad \quad
& \mathrm{x} = g^{11} + g^{22} + g^{33} & \qquad\quad  & \textnormal{(dimensional variables)},\\
\Sigma_i & =-\frac{k^i_{\ i}}{H}-1 \:, & \qquad \quad 
& s_i = \frac{g^{ii}}{\mathrm{x}} & \qquad\quad & \textnormal{(dimensionless variables)}.
\end{alignat*}
There is no summation over the index $i$ in these definitions.
The division by $H$ is not a restriction, since $H>0$ for all solutions, cf.~the remark in Section~\ref{setup}.
The dimensionless variables satisfy the constraints
\[
\Sigma_1+\Sigma_2+\Sigma_3=0 \:,\qquad\qquad s_1+s_2+s_3=1\:.
\]
The transformation from the six variables $(g^{ii}, k^i_{\ i})$ 
to the variables $(H, \mathrm{x},s_i,\Sigma_i)$ that satisfy the constraints 
is one-to-one. The variable $\mathrm{x}$ can be replaced by 
$n =(\det g)^{-1/2}$, since $\mathrm{x} = n^{2/3} (s_1 s_2 s_3)^{-1/3}$. 

We define dimensionless matter quantities by
\[
\Omega=\frac{\rho}{3 H^2}\:,\qquad 
w_i= \frac{p_i}{\rho} = \frac{T^i_{\ i}}{\rho}\:,\qquad 
w=\frac{1}{3}(w_1+w_2+w_3) = \frac{p}{\rho}\:;
\]
there is no summation over $i$.
The quantity  $p$ is the isotropic pressure; it is simply given as the average
of the principal pressures $p_i$ ($i=1,2,3$). 

The quantities $w_i$ ($i=1,2,3$), which are the rescaled principal pressures,
encode the degree of anisotropy of the matter. If $w_i = w$ $\forall i$,
the matter is isotropic. The density is a function of the metric and 
can thus be expressed as $\rho=\rho(n,s_1,s_2,s_3)$. Equation~\eqref{Tij} then entails 
\begin{subequations}\label{wwiintermsof}
\begin{align}
\label{winterms}
w & = \frac{\partial\log \rho}{\partial \log n} - 1 \:, \\
\label{wiinterms}
w_i & = w + 2 \left( \frac{\partial\log \rho}{\partial\log s_i} - 
s_i\,\sum\nolimits_j \frac{\partial\log \rho}{\partial\log s_j} \right) \:.
\end{align}
\end{subequations}
We make the following simplifying assumption:

\begin{Assumption}\label{assumptionwi}
We suppose that the isotropic pressure and the density are proportional, 
i.e., we assume $w = \mathrm{const}$, where
\begin{equation}\label{eqasswi}
w \in (-1,1)\:.
\end{equation}
\end{Assumption}

According to Assumption~\ref{assumptionwi}, the density and the isotropic pressure behave like
those of a perfect fluid with a linear equation of state satisfying the 
dominant energy condition.
It seems natural to focus attention
on anisotropic matter sources satisfying Assumption~\ref{assumptionwi}, 
since they generalize the 
behaviour of perfect fluid models widely used in cosmology.
In the concluding remarks, see Section~\ref{conclusions}, we will indicate how to treat the more general case
where $w \neq \mathrm{const}$.

\begin{Remark} 
In Assumption~\ref{assumptionwi} the cases $w = \pm 1$ are excluded.
The case $w = 1$ leads to different dynamics, which we refrain from discussing here.
The case $w = -1$ is excluded since it comprises as a subcase the de Sitter spacetime which does not
exhibit a singularity. 
\end{Remark}

Taking account of~\eqref{wwiintermsof}, Assumption~\ref{assumptionwi} 
implies that 
\begin{equation}\label{rhopi}
\rho(n,s_1,s_2,s_3) = n^{1 + w} \psi(s_1,s_2,s_3)\:,
\end{equation}
for some function $\psi(s_1,s_2,s_3)$.
Moreover, for all $i$, the matter anisotropies $w_i$ are functions of $(s_1,s_2,s_3)$ alone and
\begin{equation}\label{wiinpsi}
w_i = w_i(s_1,s_2,s_3)  = w + 2 \left( \frac{\partial\log \psi}{\partial\log s_i} - 
s_i\,\sum\nolimits_j \frac{\partial\log \psi}{\partial\log s_j} \right) \:.
\end{equation}

Finally we introduce a dimensionless time variable $\tau$ by
\begin{equation}\label{tandtau}
\frac{d}{d\tau} = H^{-1}\, \frac{d}{d t}\:;
\end{equation}
in the following a prime will denote differentiation w.r.t.\ $\tau$.

Expressed in the new variables, 
the Einstein evolution equations decouple into the dimensional equations
\begin{equation}\label{decoupled}
H' = -3 H \left[1-\frac{\Omega}{2}(1-w)\right]\,,\qquad\quad
\mathrm{x}^\prime = -2\, \mathrm{x} \left(1+ \sum\nolimits_k s_k\Sigma_k\right)\:,
\end{equation}
and an autonomous system of equations
\begin{subequations}\label{dynamicalsystem}
\begin{align}
\label{Sigeq}
& \Sigma_i' = -3\Omega\left[\frac{1}{2}(1-w)\Sigma_i-(w_i-w)\right]&& (i =1,2,3)\:, \\
\label{seq}
& s_i' = -2s_i\left[\Sigma_i-\sum_ks_k\Sigma_k\right]&& (i =1,2,3)\:.
\end{align}
\end{subequations}
The Hamiltonian constraint results in
\begin{equation}
\Omega = 1 - \Sigma^2\:, \qquad\qquad\text{where}\quad \Sigma^2 :=
\textfrac{1}{6} \Big( \Sigma_1^2 +\Sigma_2^2 +\Sigma_3^2 \Big)\:,
\end{equation}
which implies $\Sigma^2 < 1$. 
This constraint is used to substitute for $\Omega$ in~\eqref{dynamicalsystem}.

The reduced dimensionless 
dynamical system~\eqref{dynamicalsystem} encodes the essential 
dynamics of Bianchi type~I anisotropic spacetimes where
the matter is described by the quantities $w_i = w_i(s_1,s_2,s_3)$ and $w=\mathrm{const}$; 
once this system is solved, the 
decoupled equations~\eqref{decoupled} can be integrated and the solution
$(g^{ii}, k^i_{\ i})$ can be constructed. 

There exist useful auxiliary equations in connection with the system~\eqref{dynamicalsystem}.
In particular, the evolution equation for $\Omega$ is given by 
\begin{equation}\label{omega}
\Omega'=\Omega\left[3(1-w)\Sigma^2-\sum\nolimits_k w_k \Sigma_k\right]\,.
\end{equation}
Likewise, for $\rho$ we get $\rho' =  -\rho\, [3(1+w) + \sum_k w_k\,\Sigma_k]$.

\begin{figure}[Ht]
\begin{center}
\subfigure[The Kasner disc $\mathscr{K}$]{%
\psfrag{y1}[cc][cc][1][0]{$\Sigma_1$}
\psfrag{y2}[cc][cc][1][0]{$\Sigma_2$}
\psfrag{y3}[cc][cc][1][0]{$\Sigma_3$}
\psfrag{1-1}[cc][cc][0.8][90]{$\Sigma_1=-1$}
\psfrag{2-1}[cc][cc][0.8][30]{$\Sigma_2=-1$}
\psfrag{3-1}[cc][cc][0.8][-30]{$\Sigma_3=-1$}
\psfrag{11}[cc][cc][0.6][-90]{$\Sigma_1=1$}
\psfrag{21}[cc][cc][0.6][30]{$\Sigma_2=1$}
\psfrag{31}[cc][cc][0.6][-30]{$\Sigma_3=1$}
\psfrag{231}[rt][rt][1][0]{$\langle 231 \rangle$}
\psfrag{213}[tc][tc][1][0]{$\langle 213 \rangle$}
\psfrag{123}[lt][lt][1][0]{$\langle 123 \rangle$}
\psfrag{132}[lb][lb][1][0]{$\langle 132 \rangle$}
\psfrag{312}[bc][bc][1][0]{$\langle 312 \rangle$}
\psfrag{321}[rb][rb][1][0]{$\langle 321 \rangle$}
\psfrag{T1}[cc][cc][1][0]{$\mathrm{T}_1$}
\psfrag{T2}[cc][cc][1][0]{$\mathrm{T}_2$}
\psfrag{T3}[cc][cc][1][0]{$\mathrm{T}_3$}
\psfrag{Q1}[cc][cc][1][0]{$\mathrm{Q}_1$}
\psfrag{Q2}[cc][cc][1][0]{$\mathrm{Q}_2$}
\psfrag{Q3}[cc][cc][1][0]{$\mathrm{Q}_3$}
\includegraphics[width=0.45\textwidth]{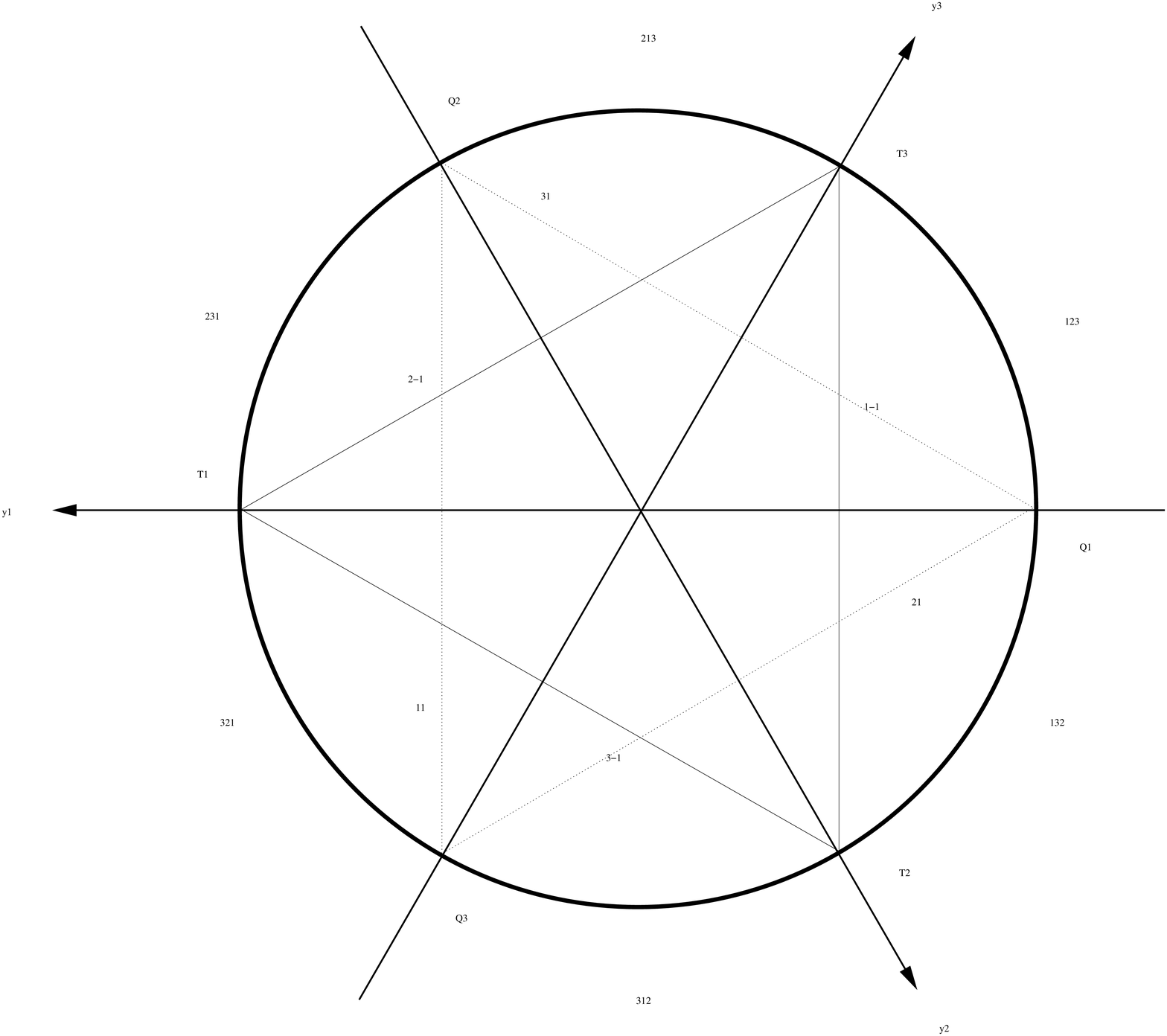}}
\qquad
\subfigure[The space $\mathscr{T}$]{%
\psfrag{s1}[cc][cc][1][0]{$s_1$}
\psfrag{s2}[cc][cc][1][0]{$s_2$}
\psfrag{s3}[cc][cc][1][0]{$s_3$}
\psfrag{0rl}[bc][bc][1][60]{$(0,\rightarrow,\leftarrow)$}
\psfrag{r0l}[bc][bc][1][0]{$(\rightarrow,0,\leftarrow)$}
\psfrag{rl0}[bc][bc][1][-60]{$(\rightarrow,\leftarrow,0)$}
\includegraphics[width=0.45\textwidth]{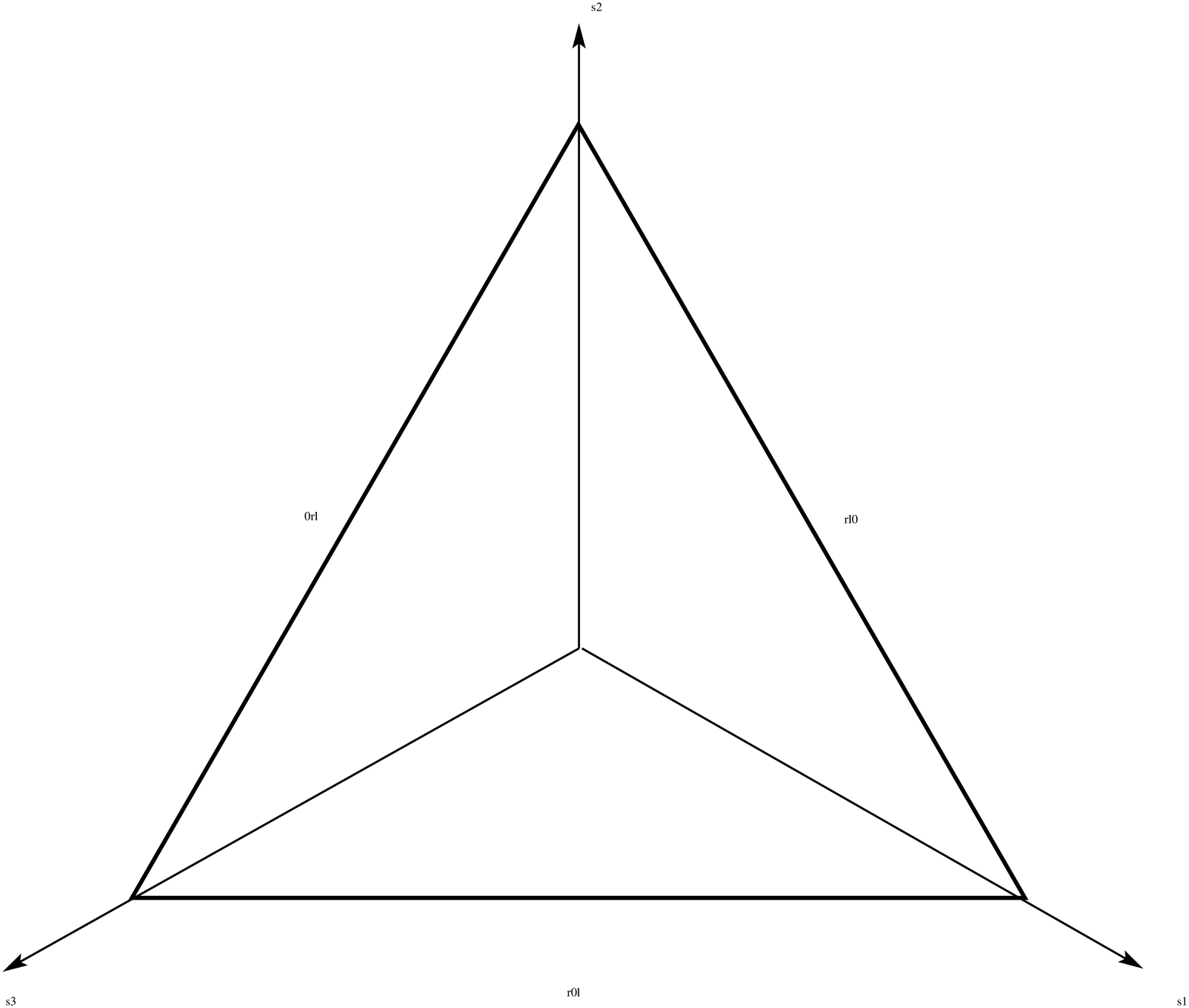}}
\caption{The four-dimensional state space $\mathcal{X}$ is the Cartesian product
of the Kasner disc $\mathscr{K}$ and the set $\mathscr{T}$.
The latter is represented by (the interior of) a triangle; the center of the triangle is the point
$(s_1,s_2,s_3) = (1/3,1/3,1/3)$; for the sides, the values of $(s_1,s_2,s_3)$ are 
given in the figure, where the arrows denote the directions
of increasing values (from $0$ to $1$).}
\label{X}
\end{center}
\end{figure}

The state space of the dimensionless dynamical system~\eqref{dynamicalsystem} 
is the four-dimensional bounded open connected set $\mathcal{X}$ given by
\begin{equation*}
\mathcal{X}=\left\{(\Sigma_1,\Sigma_2,\Sigma_3,s_1,s_2,s_3)\:\Big|\:\left(\Sigma^2<1\right)\wedge 
\left(\sum\nolimits_k\Sigma_k=0\right) 
\wedge (0< s_i < 1 \;\,\forall i) \wedge \left(\sum\nolimits_k s_k = 1\right)\right\}.
\end{equation*}
This set can be written as the Cartesian product $\mathcal{X}=\mathscr{K}\times\mathscr{T}$ of 
two two-dimensional bounded open connected sets,
\begin{subequations}\label{Xdef}
\begin{align}
\mathscr{K} &= \left\{(\Sigma_1,\Sigma_2,\Sigma_3) \:\Big|\: \left(\Sigma^2<1\right)\wedge 
\left(\sum\nolimits_k\Sigma_k=0\right)\right\}\:.\label{KTdef}\\
\mathscr{T} &=\left\{(s_1,s_2,s_3) \:\Big|\: (0 < s_i <1\:\,\forall i\,) \wedge 
\left(\sum\nolimits_k s_k = 1\right) \right\}\:.\label{Tdef}
\end{align}
\end{subequations}
The set $\mathscr{K}$ is the Kasner disc; it is typically 
depicted in a projection onto the plane with conormal $(1,1,1)$, see Figure~\ref{X}.
The boundary of $\mathscr{K}$ is the Kasner circle 
$\partial \mathscr{K} = \{ (\Sigma^2 =1)\wedge (\Sigma_1+ \Sigma_2 + \Sigma_3 =0) \}$.
The Kasner circle contains six special points, which are referred to as LRS points:
The three Taub points $\mathrm{T}_1$, $\mathrm{T}_2$, $\mathrm{T}_3$ 
given by $(\Sigma_1,\Sigma_2,\Sigma_3) = (2, -1,-1)$ and permutations,
and the three non-flat LRS points $\mathrm{Q}_1$, $\mathrm{Q}_2$, $\mathrm{Q}_3$ 
given by $(\Sigma_1,\Sigma_2,\Sigma_3) = (-2, 1,1)$ and permutations.
The six sectors of $\mathscr{K}$ are denoted by permutations of
the triple $\langle 123 \rangle$; by definition, $\Sigma_i < \Sigma_j <\Sigma_k$ holds
in sector $\langle ijk \rangle$.

The set $\mathscr{T}$ is the interior of a triangle contained in the affine plane $s_1+s_2+ s_3 =1$,
see Figure~\ref{X}.
The boundaries form a triangle with the three sides $s_1 = 0$, $s_2 = 0$ and $s_3 = 0$;
the corners are given by $(1,0,0)$, $(0,1,0)$ and $(0,0,1)$.
The functions $\psi(s_1,s_2,s_3)$ and $w_i(s_1,s_2,s_3)$ in~\eqref{rhopi} and~\eqref{wiinpsi}
are understood as functions on the domain $\mathscr{T}$.

We conclude this section with a discussion of the energy conditions.
Our basic assumption is $\rho > 0$.
Then the dominant energy condition 
is expressed in the new matter variables as $|w_i(s_1,s_2,s_3)|\leq 1$, $\forall (s_1,s_2,s_3) \in \mathscr{T}\:$ and  $\forall\,i=1,2,3$;
the weak energy condition is $-1 \leq w_i$ $\forall\, i=1,2,3$;
the strong energy condition is satisfied if the weak energy condition holds and $w\geq -1/3$. 
We are mainly interested in matter models that satisfy the dominant energy condition; however 
we shall also discuss the more general case when the functions $w_i$ are merely bounded: 
\begin{Assumption}\label{assumptiondominant}
We assume that the rescaled principal pressures are bounded functions, 
\begin{equation}\label{wibounded}
\sup_\mathscr{T}|w_i(s_1,s_2,s_3)| <\infty\:,\quad \forall\,i=1,2,3\:.
\end{equation}
\end{Assumption}

\section{The flow on the boundaries of the state space}
\label{boundaries}

The present section is devoted to a detailed analysis of the flow
induced by~\eqref{dynamicalsystem} on the boundary of the state space.
This analysis is rather technical but essential to understand the asymptotic behaviour 
(in particular toward the past) of solutions of~\eqref{dynamicalsystem}.

The state space $\mathcal{X} = \mathscr{K} \times \mathscr{T}$, see~\eqref{Xdef}
and Figure~\ref{X},
is relatively compact.
Its boundary is the union of two three-dimensional compact sets, i.e.,
\begin{equation}\label{boundary}
\partial\mathcal{X}=\left(\partial\mathscr{K}\times\overline{\mathscr{T}}\right)\,\cup\,
\left(\overline{\mathscr{K}}\times\partial\mathscr{T}\right)\,.
\end{equation}
The intersection of these boundary components is the two-dimensional compact set 
$\partial\mathscr{K} \times \partial\mathscr{T}$.

\subsection{The vacuum boundary: $\bm{\partial\mathscr{K}\times\overline{\mathscr{T}}}$}

Consider the boundary component $\partial\mathscr{K}\times\overline{\mathscr{T}}$,
which is characterized by $\Sigma^2 = 1$ (i.e., $\Omega = 0$);
we call this component 
the \textit{vacuum boundary}---the reason for this terminology will become clear through
the remarks in this subsection.
The dynamical system~\eqref{dynamicalsystem} admits a regular extension from
$\mathcal{X}$ to the vacuum boundary;
we simply let $\Omega\rightarrow 0$, so that equation~\eqref{Sigeq} becomes
$\Sigma_i^\prime = 0$ in this limit. 
This is independent of $(s_1,s_2,s_3) \in \overline{\mathscr{T}}$,
since $w_i$ is bounded on $\mathscr{T}$ by Assumption~\ref{assumptiondominant}.
The dynamical system~\eqref{dynamicalsystem} thus induces the system
\begin{equation}\label{dynsysvac}
\Sigma_i^\prime = 0 \:, \qquad  s_i' = -2s_i\left[\Sigma_i-\sum\nolimits_ks_k\Sigma_k\right] \qquad (i =1,2,3)\:,
\end{equation}
on the vacuum boundary. 

\begin{figure}[Ht]
\begin{center}
\subfigure[Vacuum fixed point set]{\label{vacuumfixnet}
\psfrag{TL1}[cc][cc][0.7][60]{$\text{TL}_1$}
\psfrag{TL2}[cc][cc][0.7][0]{$\text{TL}_2$}
\psfrag{TL3}[cc][cc][0.7][-60]{$\text{TL}_3$}
\psfrag{QL1}[cc][cc][0.7][60]{$\text{QL}_1$}
\psfrag{QL2}[cc][cc][0.7][0]{$\text{QL}_2$}
\psfrag{QL3}[cc][cc][0.7][-60]{$\text{QL}_3$}
\psfrag{T21}[cc][cc][0.5][0]{$\text{T}_{21}$}
\psfrag{T22}[cc][cc][0.5][0]{$\text{T}_{22}$}
\psfrag{T23}[cc][cc][0.5][0]{$\text{T}_{23}$}
\psfrag{Q21}[cc][cc][0.5][0]{$\text{Q}_{21}$}
\psfrag{Q22}[cc][cc][0.5][0]{$\text{Q}_{22}$}
\psfrag{Q23}[cc][cc][0.5][0]{$\text{Q}_{23}$}
\psfrag{T11}[cc][cc][0.5][0]{$\text{T}_{11}$}
\psfrag{T12}[cc][cc][0.5][0]{$\text{T}_{12}$}
\psfrag{T13}[cc][cc][0.5][0]{$\text{T}_{13}$}
\psfrag{Q11}[cc][cc][0.5][0]{$\text{Q}_{11}$}
\psfrag{Q12}[cc][cc][0.5][0]{$\text{Q}_{12}$}
\psfrag{Q13}[cc][cc][0.5][0]{$\text{Q}_{13}$}
\psfrag{T31}[cc][cc][0.5][0]{$\text{T}_{31}$}
\psfrag{T32}[cc][cc][0.5][0]{$\text{T}_{32}$}
\psfrag{T33}[cc][cc][0.5][0]{$\text{T}_{33}$}
\psfrag{Q31}[cc][cc][0.5][0]{$\text{Q}_{31}$}
\psfrag{Q32}[cc][cc][0.5][0]{$\text{Q}_{32}$}
\psfrag{Q33}[cc][cc][0.5][0]{$\text{Q}_{33}$}
\psfrag{KC1}[cc][cc][0.8][0]{$\text{KC}_1$}
\psfrag{KC2}[cc][cc][0.8][0]{$\text{KC}_3$}
\psfrag{KC3}[cc][cc][0.8][0]{$\text{KC}_2$}
\psfrag{S1}[cc][cc][0.8][60]{$s$} \psfrag{S2}[cc][cc][0.8][0]{$s$} \psfrag{S3}[cc][cc][0.8][-60]{$s$}
\includegraphics[width=0.45\textwidth]{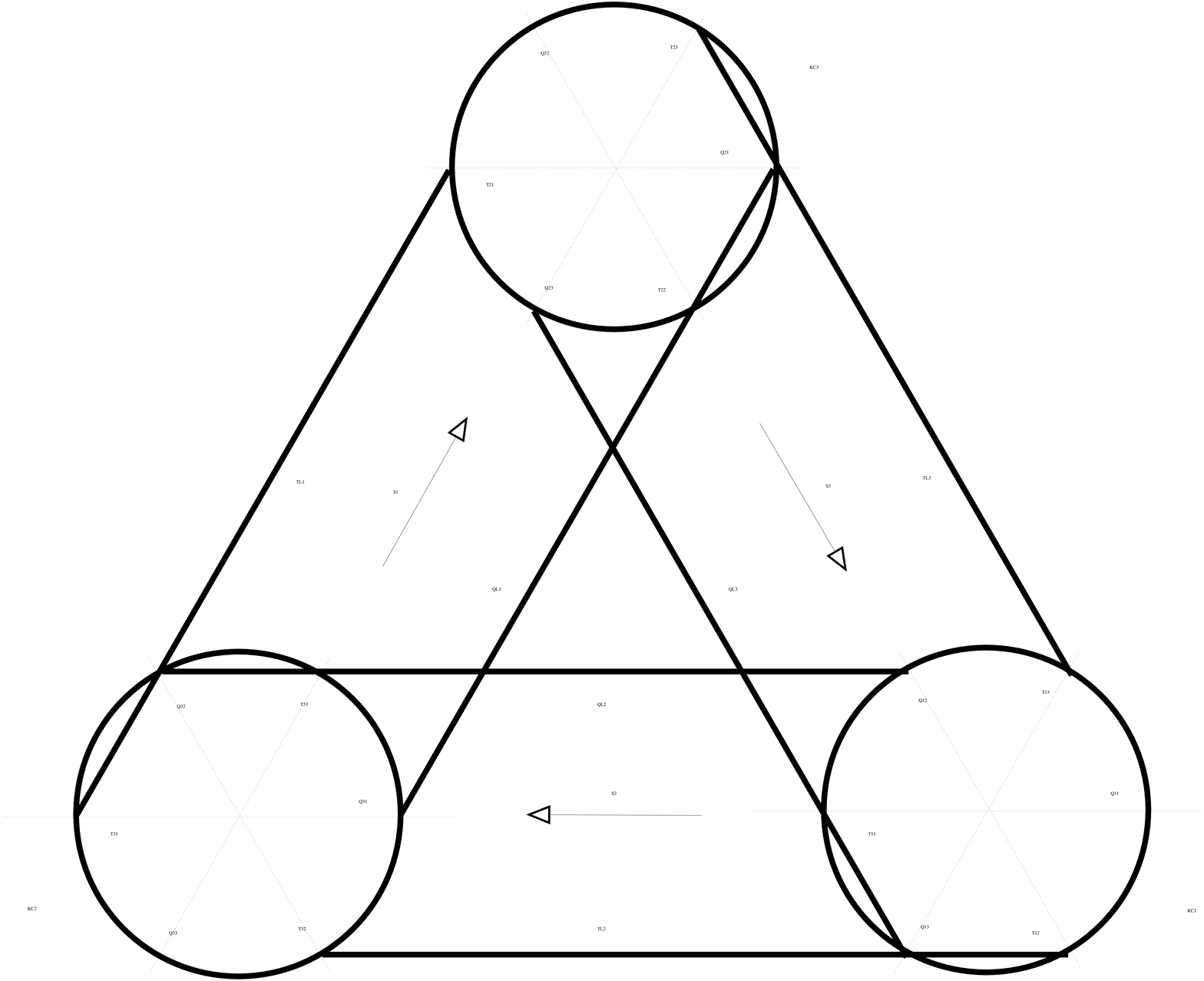}}\qquad
\subfigure[Dynamics on the vacuum boundary]{\label{vacuumflow}
\psfrag{p}[cc][cc][0.6][0]{$\textbf{P}$}
\psfrag{y1}[cc][cc][0.7][0]{$\Sigma_1$}
\psfrag{y2}[cc][cc][0.7][0]{$\Sigma_2$}
\psfrag{y3}[cc][cc][0.7][0]{$\Sigma_3$}
\psfrag{231}[lt][lt][0.7][0]{$\langle 231 \rangle$}
\psfrag{213}[tc][tc][0.7][0]{$\langle 213 \rangle$}
\psfrag{123}[rt][rt][0.7][0]{$\langle 123 \rangle$}
\psfrag{132}[rb][rb][0.7][0]{$\langle 132 \rangle$}
\psfrag{312}[bc][bc][0.7][0]{$\langle 312 \rangle$}
\psfrag{321}[lb][lb][0.7][0]{$\langle 321 \rangle$}
\includegraphics[width=0.45\textwidth]{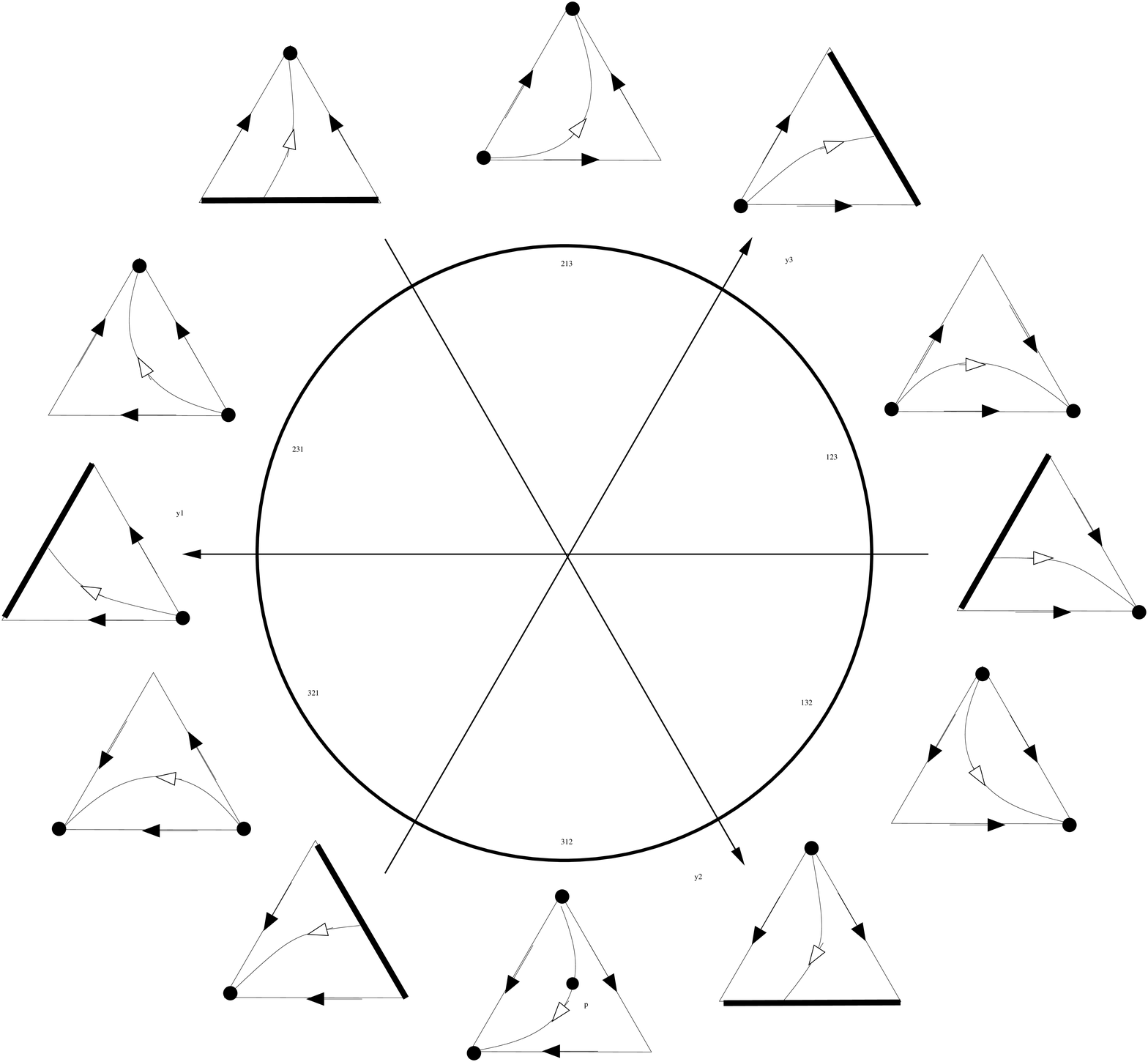}}
\caption{A schematic depiction of (a) the network of fixed points
and (b) the flow of the dynamical system on the vacuum
boundary $\partial\mathscr{K}\times\overline{\mathscr{T}}$.
Orbits in the interior of $\partial\mathscr{K}\times\overline{\mathscr{T}}$ connect a transversally hyperbolic 
source with a transversally hyperbolic sink;
for instance, the $\alpha$-limit ($\omega$-limit) 
of the orbit through the point $\mathrm{P}\in\partial\mathscr{K}\times\overline{\mathscr{T}}$ in 
Figure~\ref{vacuumflow} is a point in sector $\langle 312\rangle$ of $\mathrm{KC}_2$ ($\mathrm{KC}_3$).}
\label{vacuumpic}
\end{center}
\end{figure}

The vacuum boundary $\partial\mathscr{K}\times\overline{\mathscr{T}}$ 
is a solid torus whose cross
section is $\overline{\mathscr{T}}$; each cross section corresponds
to $(\Sigma_1,\Sigma_2,\Sigma_3) = \mathrm{const}$ and is an invariant
subspace of~\eqref{dynsysvac}.
The fixed points of the system~\eqref{dynsysvac} are transversally hyperbolic and form a connected network of lines
on $\partial\mathscr{K} \times \partial\mathscr{T}$:
\begin{itemize}
\item \textit{Kasner circles}: There exist three circles of fixed points
that can be interpreted as Kasner circles.
The Kasner circles are located at the vertices of $\overline{\mathscr{T}}$, i.e., 
let $(ijk)$ be a cyclic permutation of $(123)$, then $\mathrm{KC}_i$ is given by
\begin{equation*}
\mathrm{KC}_i :\quad \Sigma^2 = 1\,,\quad (s_i, s_j,s_k) = (1,0,0) \:.
\end{equation*}
\item \textit{Taub lines}: These are three lines of fixed points given by
\begin{equation*}
\mathrm{TL}_i : \quad (\Sigma_i,\Sigma_j,\Sigma_k) = (2,-1,-1)\,,\quad  (s_i,s_j,s_k) = (0, s, 1-s), \,s \in [0,1]\:.
\end{equation*}
\item \textit{Non-flat LRS lines}: These are three lines of fixed points given by
\begin{equation*}
\mathrm{QL}_i : \quad (\Sigma_i,\Sigma_j,\Sigma_k) = (-2,1,1)\,, \quad (s_i,s_j,s_k) = (0, s, 1-s), \,s \in [0,1]\:.
\end{equation*}
\end{itemize}
Let $(ijk)$ be a cyclic permutation of $(123)$. 
On the Kasner circle $\mathrm{KC}_i$ there exist three `Taub points': $\mathrm{T}_{ii}$,
$\mathrm{T}_{ij}$, and $\mathrm{T}_{ik}$, which are given by
$(\Sigma_i,\Sigma_j,\Sigma_k) = (2,-1,-1)$, $(-1,2,-1)$, and $(-1,-1,2)$, respectively.
The point $\mathrm{T}_{ij}$ ($\mathrm{T}_{ik}$) 
is the point of intersection of $\mathrm{KC}_i$ with $\mathrm{TL}_j$ ($\mathrm{TL}_k$);
$\mathrm{T}_{ii}$ is an isolated Taub point since it does not lie on any of the Taub lines.
Analogously, there exist three `non-flat LRS points' on $\mathrm{KC}_i$: $\mathrm{Q}_{ii}$,
$\mathrm{Q}_{ij}$, and $\mathrm{Q}_{ik}$, which are given by
$(\Sigma_i,\Sigma_j,\Sigma_k) = (-2,1,1)$, $(1,-2,1)$, and $(1, 1,-2)$, respectively;
the analogous comments apply.
In Figure~\ref{vacuumfixnet} we give a schematic depiction of the
network of fixed points.

\begin{Remark}
Each of the fixed points on the Kasner circles is associated with a Kasner solution (Bianchi type~I vacuum
solution) of the Einstein equations,
\begin{equation}\label{Kasnersol}
d s^2 = -d t^2 + a_1 \,t^{2 p_1} d x^1\otimes dx^1 + a_2\, t^{2 p_2} d x^2\otimes d x^2 + a_3 \,t^{2 p_3} d x^3\otimes dx^3\:,
\end{equation}
where $a_1$, $a_2$, $a_3$ are positive constants, and $p_1$, $p_2$, $p_3$ are
the so-called Kasner exponents, which are constants that satisfy $p_1 + p_2 + p_3 = p_1^2 +p_2^2 +p_3^2 = 1$.
The relation between $(\Sigma_1, \Sigma_2,\Sigma_3)$ and $(p_1, p_2,p_3)$ for 
a Kasner fixed point is $3 p_i = \Sigma_i +1$, $i=1,2,3$. 
The fixed points on $\mathrm{TL}_i$ and $\mathrm{QL}_i$ represent the
flat LRS Kasner solution (Taub solution) and the non-flat LRS Kasner solution,
respectively; in these cases the Kasner exponents are $(1,0,0)$ and permutations and
$(2/3,-1/3,-1/3)$ and permutations, respectively .
\end{Remark}

\begin{Lemma}\label{flowvacuum}
The solutions of the dynamical system~\eqref{dynsysvac} on the vacuum boundary 
are heteroclinic orbits, 
i.e., the $\alpha$- and the $\omega$-limit sets consist of one fixed point each;
see Figure~\ref{vacuumflow}.
\end{Lemma}

\begin{proof}
Since $\Sigma_i = \mathrm{const}$ $\forall i$, the result follows by studying the sign of $s'_i$ 
for each sector $\langle ijk \rangle$ (and at the special points) of the Kasner circle separately.
\end{proof}

\begin{Remark}
Not only the Kasner fixed points themselves, but each 
solution of the dynamical system~\eqref{dynsysvac} on the vacuum boundary 
can be interpreted as a Kasner solution.
This is simply because $(\Sigma_1,\Sigma_2,\Sigma_3) \equiv \mathrm{const}$
and $\Sigma^2 = 1$.
\end{Remark}


\subsection{The cylindrical boundary: $\bm{\overline{\mathscr{K}}\times\partial\mathscr{T}}$}

The second component of $\partial \mathcal{X}$ is the set $\overline{\mathscr{K}}\times\partial\mathscr{T}$;
the set $\partial\mathscr{T}$ is a triangle and consists of the three sides $\{s_i = 0\}$ ($i=1,2,3$), hence
\begin{equation}\label{cylbou}
\overline{\mathscr{K}}\times\partial\mathscr{T} = \overline{\mathscr{C}}_1 \cup 
\overline{\mathscr{C}}_2 \cup \overline{\mathscr{C}}_3 \:,
\qquad \quad\text{where}\quad \overline{\mathscr{C}}_i := \overline{\mathscr{K}} \times \{s_i = 0\}\:.
\end{equation}
Each set $\overline{\mathscr{C}}_i$ is compact; since it 
has the form of a cylinder, we call $\overline{\mathscr{K}}\times\partial\mathscr{T}$ 
the \textit{cylindrical boundary}.
By $\mathscr{C}_i$ we denote the interior of $\overline{\mathscr{C}}_i$.

\subsubsection*{Asymptotic properties of matter models}
The dynamical system~\eqref{dynamicalsystem} admits
a regular extension from $\mathcal{X}$ to the lateral surfaces of the cylinders $\overline{\mathscr{C}}_i$,
cf.~the discussion of the vacuum boundary; however, in general,
the system is \textit{not} extendible to the interior $\mathscr{C}_i$, nor to the top/base surfaces.
The simple reason for this is that, 
although $w_i(s_1,s_2,s_3)$, $i=1,2,3$, are bounded on $\mathscr{T}$,
cf.~\eqref{wibounded},
in general these functions need not
possess limits as $(s_1,s_2,s_3)$ converges to a point on $\partial \mathscr{T}$.
However, for reasonable matter models 
(such as collisionless matter and elastic matter),
the matter anisotropies $w_i(s_1,s_2,s_3)$ are functions on $\mathscr{T}$
that admit a unique extension to $\overline{\mathscr{T}}$. 
We make this a general assumption.

\begin{Assumption}\label{a7}
The matter anisotropies $w_i(s_1,s_2,s_3)$ ($i=1,2,3$)
admit unique extensions from $\mathscr{T}$ to $\overline{\mathscr{T}}$.
The extended functions are assumed to be 
sufficiently smooth on $\partial\mathscr{T}$.
(For the majority of our future purposes, continuity is
a sufficient requirement; however, we will assume the functions to be $\mathcal{C}^1$
to facilitate our analysis.)
\end{Assumption}

Assumption~\ref{a7} ensures that the dynamical system~\eqref{dynamicalsystem}
can be extended to $\overline{\mathscr{C}}_i$ ($i=1,2,3$) and thus to 
the entire boundary $\partial\mathcal{X}$ of 
the state space $\mathcal{X}$.
The next assumption is a simplifying assumption on the functions $w_i$ ($i=1,2,3$)
on $\partial \mathscr{T}$.

\begin{Assumption}\label{a9}
We assume that 
$w_i(s_1,s_2,s_3) \equiv v_- = \mathrm{constant}$ on $\overline{\mathscr{C}}_i$
(i.e., when $s_i = 0$) for all $i=1,2,3$.
\end{Assumption}

\begin{Remark}
Assumption~\ref{a9} stems from basic physical considerations; in particular, it
is satisfied for collisionless matter and for elastic matter, see Section~\ref{mattermodels}.
Loosely speaking, Assumption~\ref{a9} means that the (rescaled) principal pressure in a direction $i$
becomes independent of the values of $g_{jj}$ and $g_{kk}$ ($i\neq j\neq k \neq i$) 
in the limit $g_{ii} \rightarrow \infty$ ($\Leftrightarrow g^{ii} \rightarrow 0$), 
provided that $g_{jj}$, $g_{kk}$ remain bounded.
(The complementary statement is implicit in 
our previous assumptions:
The fact that $w_i$ is well-defined for $s_i = 1$ 
means that the rescaled principal pressure in a direction $i$
converges to a limit as $g_{ii} \rightarrow 0$, when
$g_{jj}$ and $g_{kk}$ remain bounded from below.)
\end{Remark}

For some matter models like elastic matter, there exists a function $v(z_1,z_2,z_3)$, 
which is defined on $\overline{\mathscr{T}}\ni(z_1,z_2,z_3)$, 
sufficiently smooth on $\partial\mathscr{T}$, and symmetric in the arguments
$z_2$ and $z_3$, such that
\begin{subequations}
\begin{equation}\label{wiandv}
w_1(s_1,s_2,s_3) = v(s_1,s_2,s_3)\:,\quad
w_2(s_1,s_2,s_3) = v(s_2,s_3,s_1)\:,\quad
w_3(s_1,s_2,s_3) = v(s_3,s_1,s_2)\:
\end{equation}
for all $(s_1,s_2,s_3) \in \overline{\mathscr{T}}$.
Equation~\eqref{wiandv} follows from basic physical considerations;
it is a consequence of the fact that there is no
distinguished direction and reflects the freedom of permuting the axes.
Assumption~\ref{a9} is equivalent to the requirement that
$v(0,\zeta_1,\zeta_2) = v_-$ for all $\zeta_1$, $\zeta_2$.

\begin{Definition}\label{v(s)def}
For matter models satisfying the symmetry property~\eqref{wiandv},
in slight abuse of notation we define a function $v(s)$ on
the interval $[0,1]$ by
\begin{equation}\label{vdef}
[0,1]\ni s \mapsto v(s) := v(s,1-s,0) \:.
\end{equation}
The value of $v$ at the endpoint $s=0$ is 
$v(0) = v_-$, cf.~Assumption~\ref{a9}; 
the value at $s=1$ we denote by $v_+$, i.e., $v(1) = v_+$.
\end{Definition}
\end{subequations}

For more general matter models, the initial data for the matter
fields might break the symmetry~\eqref{wiandv}, an example being
collisionless matter.
Let us denote by $I$ the initial data of the matter fields
and by $\mathcal{I}$ the space of possible initial data.
Let $\sigma$ be a permutation of the triple $(123)$,
which induces the permutation 
$(s_1, s_2, s_3) \mapsto (s_{\sigma(1)}, s_{\sigma(2)},s_{\sigma(3)})$ 
on any ordered triple $(s_1,s_2,s_3)$ in $\mathbb{R}^3$; 
in slight abuse of notation
we use the symbol $\sigma$ for the induced permutation as well.
Finally, let $I_{(\sigma)}$ denote
the initial data arising from $I$ by the (induced) permutation.
(For instance, in the case of collisionless matter, $I$ represents the distribution function $f_0$, 
and $I_{(\sigma)} = f_0 \circ \sigma$, cf.~Section~\ref{mattermodels}).
Then there exists $u: \mathcal{I} \times \overline{\mathscr{T}} \rightarrow \mathbb{R}$,
$I \times (z_1,z_2,z_3) \mapsto u[I](z_1,z_2,z_3)$, sufficiently smooth,
such that
\begin{equation}\label{wiandvI}
w_i = u[I_{(\sigma_i^{-1})}]\circ\sigma_i\:,
\tag{\ref{wiandv}${}^\prime$}
\end{equation}
where $\sigma_i$ is a permutation with $\sigma_i(1) = i$.
Equation~\eqref{wiandvI} reduces to~\eqref{wiandv} if there is no dependence on~$I\in\mathcal{I}$;
then $u[I]$ is simply replaced by $v$. 
(The symmetry of $v$ in the second and third argument corresponds to
the freedom of choosing even or odd permutations in~\eqref{wiandvI}.)
Assumption~\ref{a9} is equivalent to the requirement that 
$u[I](0,\zeta_1,\zeta_2) = v_-$ 
for all $I\in\mathcal{I}$, for all $\zeta_1$, $\zeta_2$.
For matter models of this type Definition~\ref{v(s)def} is replaced by

\renewcommand{\theDefinition}{\textbf{\ref{v(s)def}}$\bm{{}^\prime}$}
\begin{Definition}
For $i=1,2,3$, let $j$, $k$ be such that\/ $\sgn(ijk) = {+1}$. 
We define a function $u_i(s)$ on the interval $[0,1]$ by
\begin{equation}\label{uidef}
[0,1] \ni s \mapsto u_i(s) := u[I_{(\sigma_j^{-1})}](s,1-s,0)\:.
\tag{\ref{vdef}${}^\prime$}
\end{equation}
The value of $u_i$ at the endpoint $s=0$ is
$u_i(0) = v_-$, cf.~Assumption~\ref{a9};
the value at $s=1$ we denote by $v_+$, i.e., $u_i(1) = v_+$.
Equation~\eqref{uidef} reduces to~\eqref{vdef} if there is no dependence on $\mathcal{I}$
and $u_i$ is replaced by $v$. 
\end{Definition}%
\renewcommand{\theDefinition}{\arabic{Definition}}%
\setcounter{Definition}{1}%

\begin{Remark}
The important fact that $u_i(1)$ is independent of $i$ is a direct consequence of~\eqref{wiandvI}
and the identity $w_1 + w_2 + w_3 = 3 w$; see~\eqref{vpmvspez} and the discussion below.
\end{Remark}

\subsubsection*{The dynamical system on $\bm{\overline{\mathscr{C}}_i}$}

In the following let $(ijk)$ be a cyclic permutation of $(123)$.
We begin our analysis of the flow of the dynamical system 
on the cylindrical boundary~\eqref{cylbou} by
fixing the notation.
Consider the cylinder $\overline{\mathscr{C}}_i$.
Since $s_i = 0$ on $\overline{\mathscr{C}}_i$, we may set $s_j =s$, $s_k = 1-s$, $s\in[0,1]$;
the cylinder $\overline{\mathscr{C}}_i$ is then given as the Cartesian product
$\overline{\mathscr{K}}\times \{ 0\leq s \leq 1\}$.
We call the set $\overline{\mathscr{K}}\times \{ s = 0\}$
the base, and $\overline{\mathscr{K}}\times \{ s = 1\}$ the top of the cylinder.
The boundary of the top is the Kasner circle $\mathrm{KC}_j$, the boundary of the base is $\mathrm{KC}_k$, 
see Figure \ref{cylinder}. 

The matter anisotropies on the cylinder $\overline{\mathscr{C}}_i$ 
are represented by $w_i$, $w_j$, and $w_k$.
Assumption~\ref{a9} implies that $w_i \equiv v_-$ on $\overline{\mathscr{C}}_i$.
When regarded as a function of $s\in [0,1]$, $w_j$
coincides with the function $u_i(s)$ of Definition~\ref{v(s)def}${}^\prime$, i.e.,
$w_j(s_1,s_2,s_3) = u[I_{(\sigma_j^{-1})}](s_j,s_k,s_i) =
u[I_{(\sigma_j^{-1})}](s,1-s,0) = u_i(s)$ on $\overline{\mathscr{C}}_i$.
Finally, $w_k = 3 w - w_i - w_j$, since $w_i + w_j + w_k = 3 w$.
Evaluation of this identity at $s = 1$ (i.e., for $(s_i,s_j,s_k) = (0,1,0)$) 
yields $v_- + u_i(1) + v_- = 3 w$, because $w_k = v_-$ for $(s_i,s_j,s_k) = (0,1,0)$
by Assumption~\ref{a9}.
Since, by definition, $u_i(1) = v_+$, we obtain
\begin{equation}\label{vpmvspez}
2 v_- + v_+ = 3 w \:.
\end{equation}
Using this equation in the form $3 w - v_- = v_- + v_+$, the
anisotropies on the cylinder $\overline{\mathscr{C}}_i$ take the form
\begin{equation}\label{wijku}
w_i = v_- \:,\qquad
w_j = u_i(s)\:, \qquad
w_k = v_+ + v_- - u_i(s)\:,
\tag{\ref{wijkv}${}^\prime$}
\end{equation}
where $u_i(s)$ satisfies $u_i(0) = v_-$ and $u_i(1) = v_+$
(independently of $i$).

In the simpler case~\eqref{wiandv} we have
\begin{equation}\label{wijkv}
w_i = v_- \:,\qquad
w_j = v(s)\:, \qquad
w_k = v_+ + v_- - v(s)\:,
\end{equation}
where $v(s)$ satisfies $v(0) = v_-$ and $v(1) = v_+$,
and again~\eqref{vpmvspez}.
However, by~\eqref{wiandv}, on $\overline{\mathscr{C}}_i$, 
$w_k$ can also be written as $w_k = v(1-s,0,s) = v(1-s)$.
Therefore, 
$v(s) + v(1-s) = v_+ + v_-$;
this identity can be written in an alternative form,
\begin{equation}\label{vantisymm}
\left( v(s) - \frac{v_-+v_+}{2} \right) = -\left( v(1-s) - \frac{v_-+v_+}{2} \right) \:,
\end{equation}
which states that $v(s) - (v_-+v_+)/2$ is antisymmetric around $s=1/2$.

Using~\eqref{wijkv} we can write the dynamical system that is induced on $\overline{\mathscr{C}}_i$ 
by~\eqref{dynamicalsystem} as
\begin{subequations}\label{dynoncyl}
\begin{alignat}{2}
& s' = -2 s (1-s) (\Sigma_j-\Sigma_k)\:, \qquad
& &  \Sigma_i'  = -3\Omega\left(\frac{1}{2}(1-w)\Sigma_i-[v_--w]\right),\\
& \Sigma_j' = -3\Omega\left(\frac{1}{2}(1-w)\Sigma_j-[v(s)-w]\right),\qquad
& & \Sigma_k' = -3\Omega\left(\frac{1}{2}(1-w)\Sigma_k-[v_+ + v_- - v(s) -w]\right),
\end{alignat}
\end{subequations}
where we recall that $\Omega = 1-\Sigma^2$.
The remainder of this section is devoted to a detailed analysis of the
flow of the dynamical system~\eqref{dynoncyl}.

In the case~\eqref{wijku} the function $v(s)$ in~\eqref{dynoncyl} 
is simply replaced by $u_i(s)$.
The qualitative dynamics of the resulting system remains unchanged
because the key quantities in our analysis 
will turn out to be
$v_-$ and $v_+$, which are the same for~\eqref{wijkv} and~\eqref{wijku}.

In the following we analyze in detail
the dynamical system~\eqref{dynoncyl} which is connected with
the matter models satisfying~\eqref{wiandv},~\eqref{vdef} and~\eqref{wijkv}.
The more general case, represented by~\eqref{wiandvI},~\eqref{uidef}
and~\eqref{wijku}, leads to the same conclusions
and will only be commented on sporadically.

\subsubsection*{Flow on the lateral boundary of $\bm{\overline{\mathscr{C}}_i}$}

\begin{figure}[Ht]
\begin{center}
\psfrag{sigj}[tc][tc][1][0]{$\mathrm{\Sigma}_i$}
\psfrag{sigk}[tc][tc][1][0]{$\mathrm{\Sigma}_j$}
\psfrag{sigi}[tc][tc][1][0]{$\mathrm{\Sigma}_k$}
\psfrag{yj}[tc][tc][1][0]{$s_j$}
\psfrag{maggiore}[tc][tc][0.8][0]{$\mathrm{\Sigma}_j-\mathrm{\Sigma}_k>0$}
\psfrag{minore}[tc][tc][0.8][0]{$\mathrm{\Sigma}_j-\mathrm{\Sigma}_k<0$}
\psfrag{t}[rc][rc][1.2][0]{$\mathrm{TL}_i$}
\psfrag{q}[lc][lc][1.2][0]{$\mathrm{QL}_i$}
\psfrag{KC1}[lt][lt][1.2][0]{$\mathrm{KC}_k$}
\psfrag{KC2}[lb][lb][1.2][0]{$\mathrm{KC}_j$}
\psfrag{y1}[cr][cr][0.7][0]{$\begin{pmatrix} s_i = 0 \\ s_j =1 \\ s_k = 0 \end{pmatrix}$}
\psfrag{y2}[cr][cr][0.7][0]{$\begin{pmatrix} s_i = 0 \\ s_j = 0\\ s_k = 1 \end{pmatrix}$}
\psfrag{y}[cr][cr][0.7][0]{$\begin{pmatrix} s_i = 0 \\ s_j \: \uparrow \\ s_k  \downarrow \end{pmatrix}$}
\includegraphics[width=0.8\textwidth]{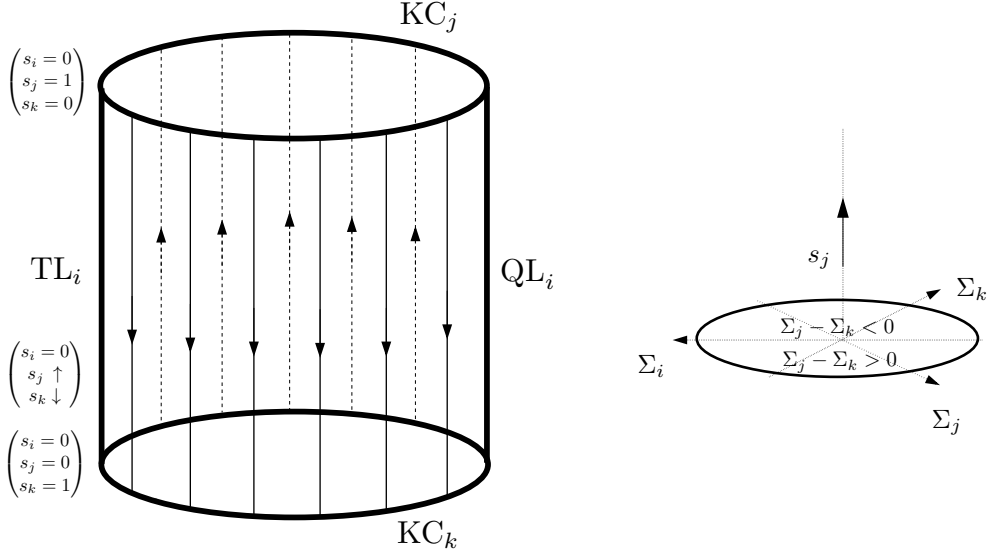}
\caption{The boundary component $\overline{\mathscr{K}}\times\partial\mathscr{T}$
consists of the three cylinders $\overline{\mathscr{C}}_1 \cup \overline{\mathscr{C}}_2 \cup \overline{\mathscr{C}}_3$. 
In this figure, $\overline{\mathscr{C}}_i$ 
is depicted together with the flow of
the dynamical system on the lateral boundary and the vacuum fixed points.}
\label{cylinder}
\end{center}
\end{figure}

The flow on the lateral surface of $\overline{\mathscr{C}}_i$ 
(which is given by $\Sigma^2 = 1$)
is independent of the matter quantities, since $\Omega=0$.
Orbits on the lateral surface satisfy
\[
s'\lessgtr 0\, \:\Leftrightarrow\: \Sigma_j - \Sigma_k \gtrless 0\:,
\]
while $s'=0$ on the lines of fixed points $\mathrm{TL}_i$ and $\mathrm{QL}_i$, see Figure~\ref{cylinder}.

\subsubsection*{Flow on the base/top of $\bm{\overline{\mathscr{C}}_i}$ and definition of the various cases}

On the base of the cylinder $\overline{\mathscr{C}}_i$, where $s=0$, we have
$w_i = w_j = v_-$ and $w_k = v_+$, see~\eqref{wijkv} or~\eqref{wijku}.
The dynamical system 
induced on the base of $\overline{\mathscr{C}}_i$ is thus
\begin{equation}\label{dynonbottom}
\left\{ \begin{matrix} \Sigma_{i}^\prime\\\Sigma_{j}^\prime \end{matrix}\right\} 
=  -3\Omega\left[\frac{1}{2}(1-w)\left\{ \begin{matrix} \Sigma_{i} \\\Sigma_{j} \end{matrix}\right\} - (v_- - w)\right]\,,
\qquad
\Sigma_k^\prime =  -3\Omega\left[\frac{1}{2}(1-w)\Sigma_{k}-(v_+ - w)\right]\,;
\end{equation}
since $2 v_- + v_+ = 3 w$ we have $v_+ -w = 2 (w -v_-)$.

\begin{Definition}
We define the quantity $\bm{\beta}$ as
\begin{equation}\label{betadef}
\beta :=  2 \,\frac{w -v_-}{1-w} =  \frac{v_+ -w}{1-w}\:.
\end{equation}
\end{Definition}
This quantity will play an essential role in our analysis.

\begin{figure}[Ht]
\begin{center}
\subfigure[\Cplus]{
\label{case1}
\psfrag{rk}[cc][cc][0.6][0]{$\mathbf{\mathrm{R}_k}$}
\psfrag{y1}[cc][cc][1][0]{$\Sigma_i$}
\psfrag{y3}[cc][cc][1][0]{$\Sigma_k$}
\psfrag{y2}[cc][cc][1][0]{$\Sigma_j$}
\psfrag{beta}[cc][cc][0.6][150]{$\Sigma_k=2/\beta$}
\includegraphics[width=0.25\textwidth]{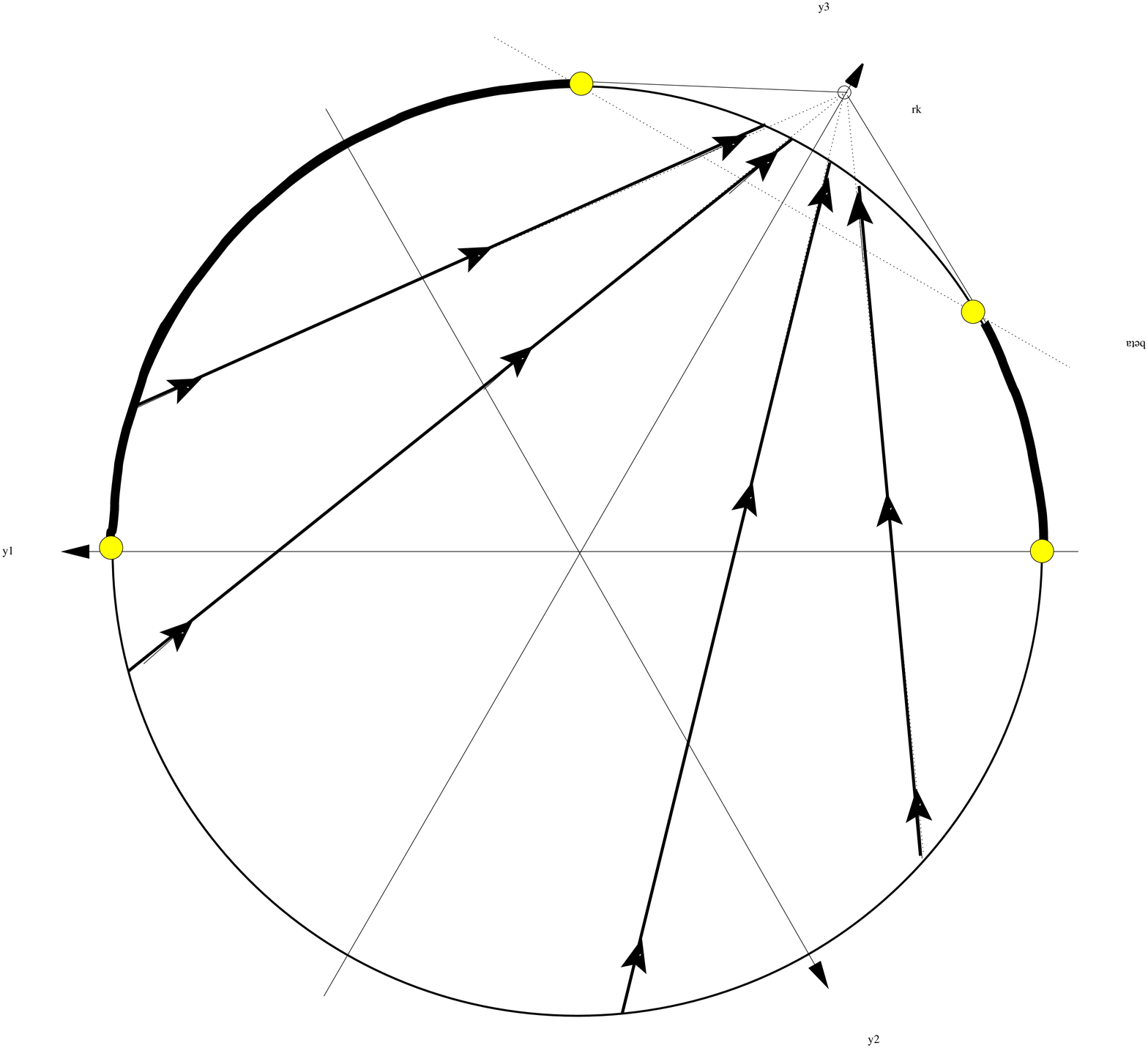}}\qquad
\subfigure[\Bplus]{
\label{case2}
\psfrag{tkk}[lc][lc][0.6][0]{$\mathbf{\mathrm{T}_{kk}}\equiv\mathbf{\mathrm{R}_k}$}
\psfrag{y1}[cc][cc][1][0]{$\Sigma_i$}
\psfrag{y3}[cc][cc][1][0]{$\Sigma_k$}
\psfrag{y2}[cc][cc][1][0]{$\Sigma_j$}
\includegraphics[width=0.25\textwidth]{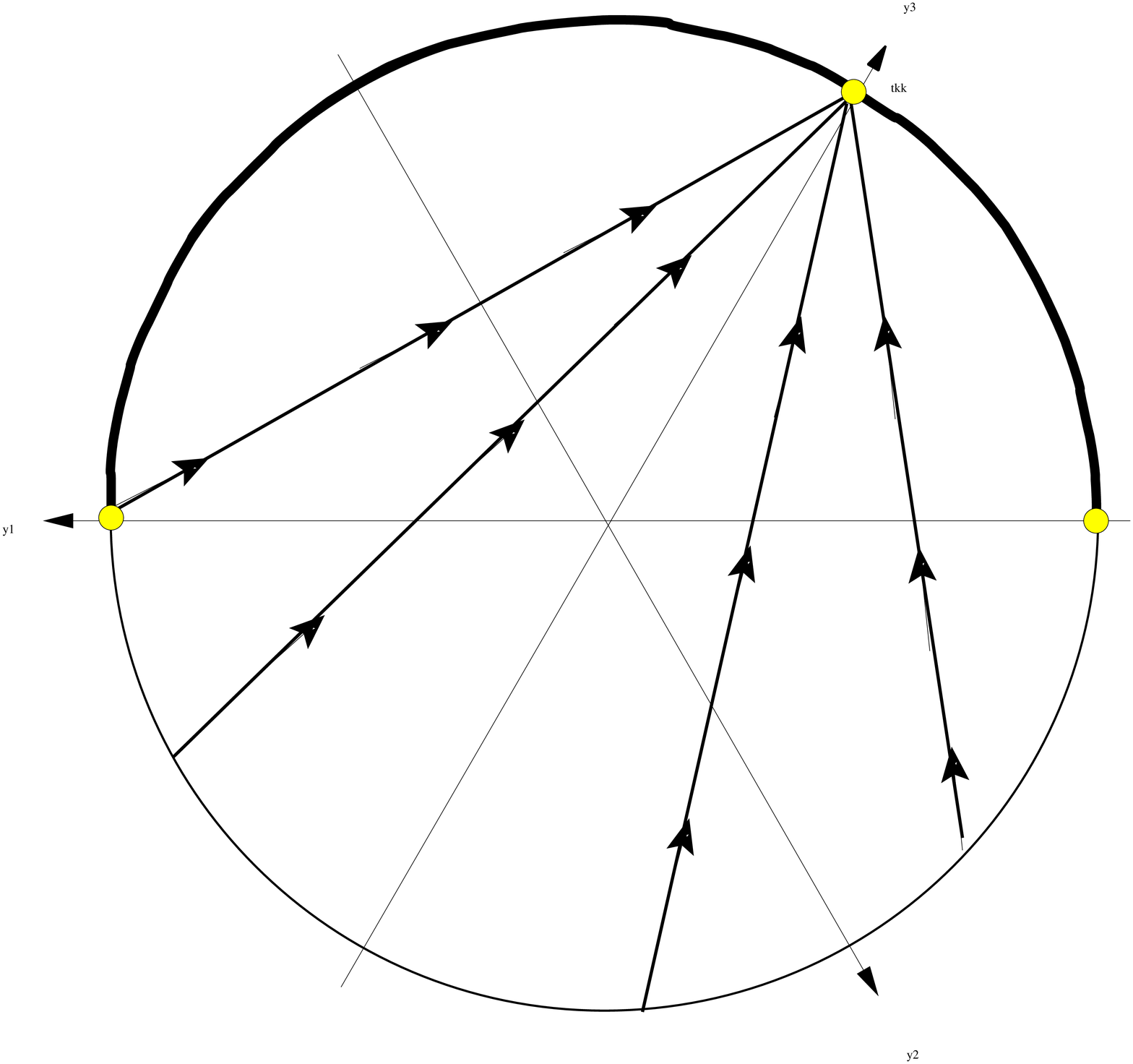}}
\qquad
\subfigure[\Aplus]{
\label{case3}
\psfrag{rk}[cc][cc][0.6][0]{$\mathbf{\mathrm{R}_k}$}
\psfrag{tkk}[lc][lc][0.6][0]{$\mathbf{\mathrm{T}_{kk}}$}
\psfrag{y1}[cc][cc][1][0]{$\Sigma_i$}
\psfrag{y3}[cc][cc][1][0]{$\Sigma_k$}
\psfrag{y2}[cc][cc][1][0]{$\Sigma_j$}
\includegraphics[width=0.25\textwidth]{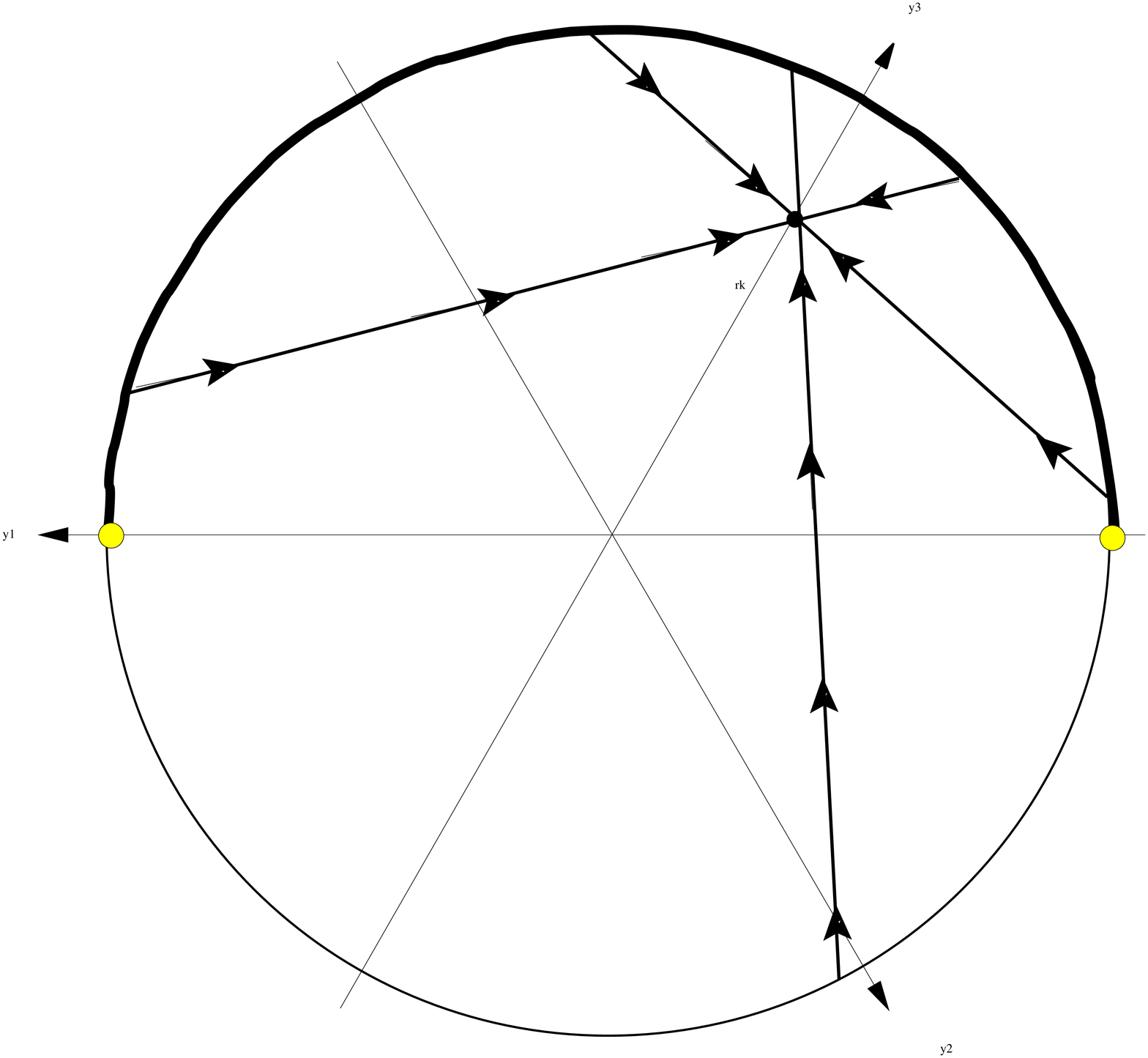}}\\
\subfigure[\Aminus]{
\label{case4}
\psfrag{rk}[cc][cc][0.6][180]{$\mathbf{\mathrm{R}_k}$}
\psfrag{y1}[cc][cc][1][180]{$\Sigma_i$}
\psfrag{y3}[cc][cc][1][180]{$\Sigma_k$}
\psfrag{y2}[cc][cc][1][180]{$\Sigma_j$}
\includegraphics[width=0.25\textwidth]{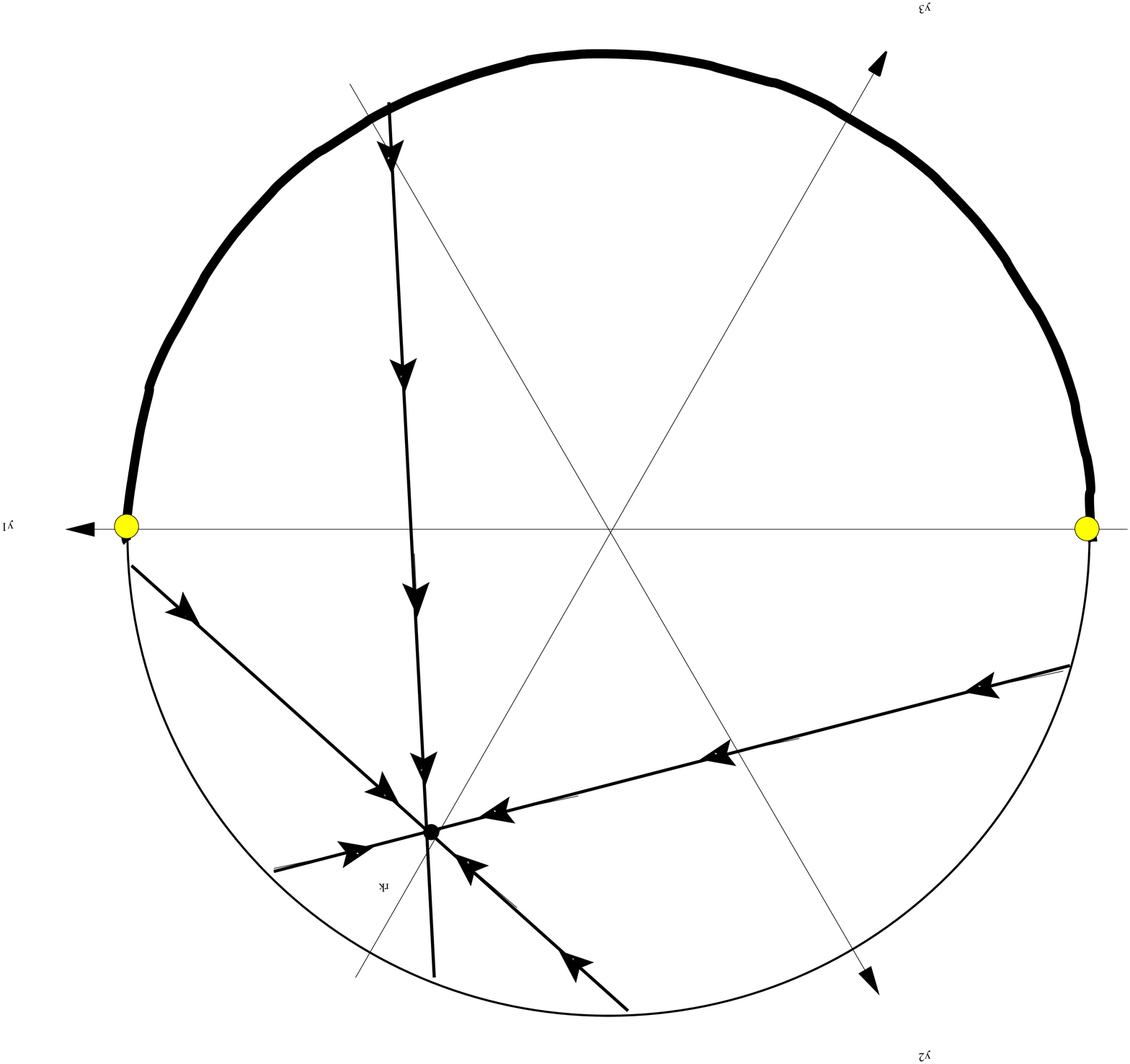}}
\qquad
\subfigure[\Bminus]{
\label{case5}
\psfrag{tkk}[rr][rr][0.6][180]{$\mathbf{\mathrm{Q}_{kk}}\equiv\mathbf{\mathrm{R}_k}$}
\psfrag{y1}[cc][cc][1][180]{$\Sigma_i$}
\psfrag{y3}[cc][cc][1][180]{$\Sigma_k$}
\psfrag{y2}[cc][cc][1][180]{$\Sigma_j$}
\includegraphics[width=0.25\textwidth]{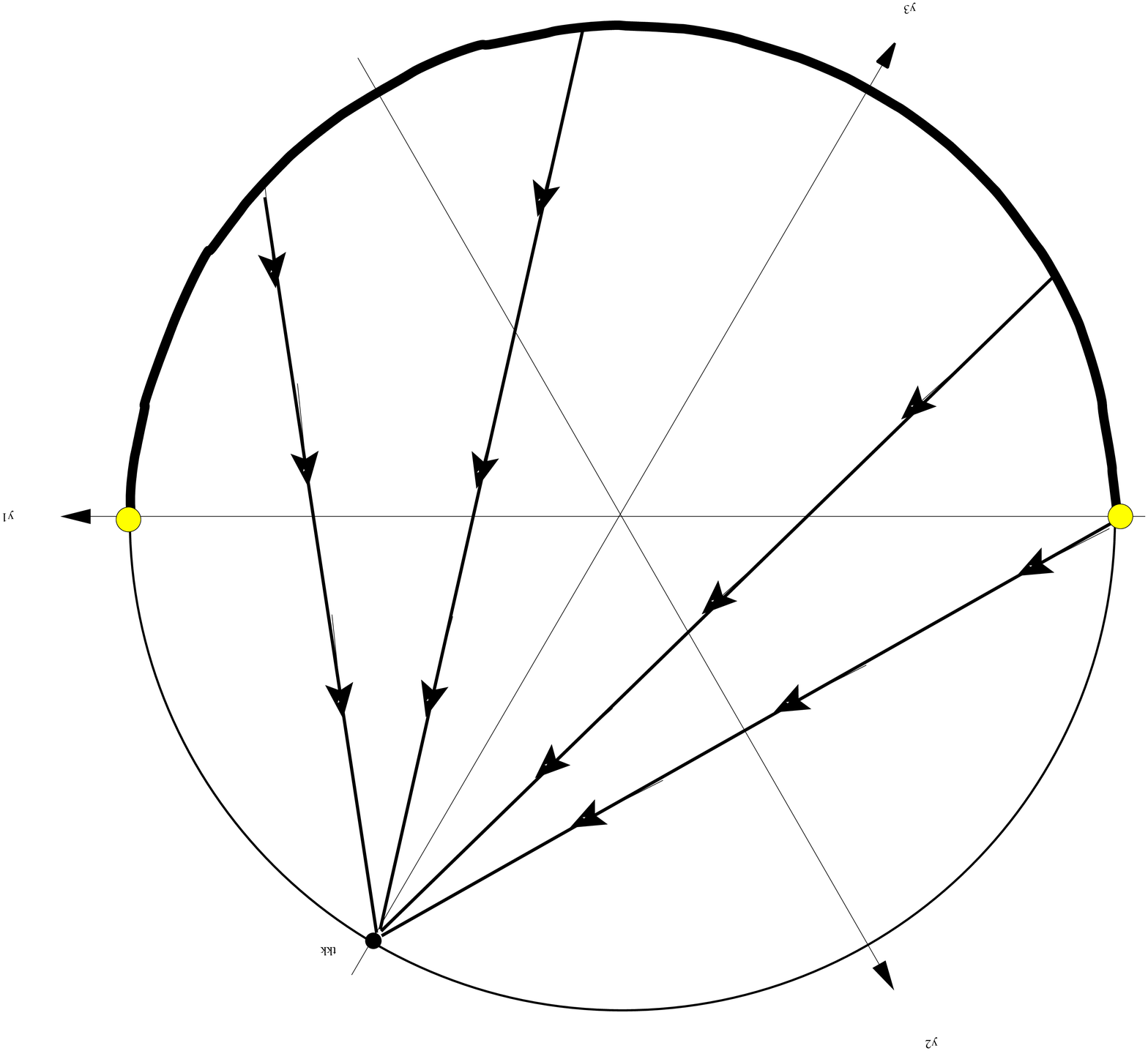}}
\qquad
\subfigure[\Cminus]{
\label{case6}
\psfrag{rk}[cc][cc][0.7][0]{$\mathbf{\mathrm{R}_k}$}
\psfrag{tkk}[lc][lc][0.7][0]{$\mathbf{\mathrm{T}_{kk}}$}
\psfrag{y1}[cc][cc][1][180]{$\Sigma_i$}
\psfrag{y3}[cc][cc][1][180]{$\Sigma_k$}
\psfrag{y2}[cc][cc][1][180]{$\Sigma_j$}
\psfrag{beta}[cc][cc][0.6][-30]{$\Sigma_k=2/\beta$}
\includegraphics[width=0.25\textwidth]{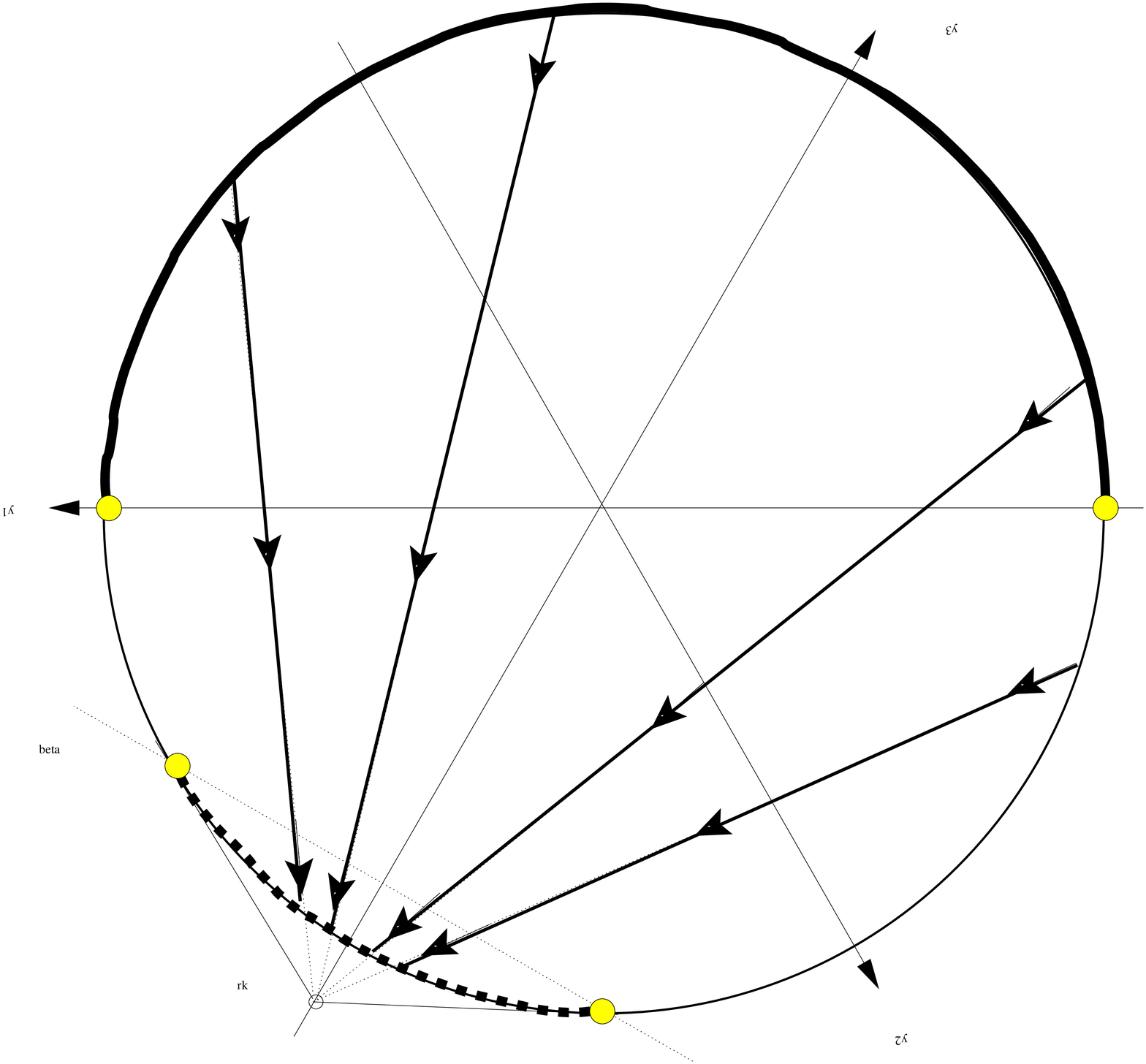}}
\caption{Flow on the base of the cylinder $\overline{\mathscr{C}}_i$;
the orbits are straight lines focusing at $\mathrm{R}_k$.
The Kasner circle $\mathrm{KC}_k$ consists of fixed points that can act as sources, saddles, or sinks
for orbits in the interior of the cylinder $\mathscr{C}_i$; in the figures,
bold continuous [dashed] lines denote transversally 
hyperbolic sources [sinks].
In the cases \Bplus\ and \Bminus, $\mathrm{R}_k$ is not transversally hyperbolic.
In the cases \Cplus\ and \Cminus\ the point $\mathrm{R}_k$ is close to the Kasner circle 
since $|\beta|<2$; increasing the value of $|\beta|$ amounts to increasing the distance
between $\mathrm{R}_k$ and the Kasner circle; in the case $|\beta|>2$ 
the distance is greater than the Kasner radius.
Accordingly, if $\beta > 2$
there appear transversally hyperbolic sinks
in the lower half of the Kasner circle;
if $\beta < -2$, some of the transversally
hyperbolic sources in the upper half vanish.
The flow on the top of the cylinder and the stability properties of the
Kasner fixed points on $\mathrm{KC}_j$ are obtained by a reflection w.r.t.\ the $\Sigma_i$-axis.}
\label{base}
\end{center}
\end{figure}

It is not difficult to show that the solutions of the system~\eqref{dynonbottom} form a 
family of straight lines that are attracted by a common focal point,
the fixed point 
\begin{equation}\label{R3}
\mathrm{R}_k :\quad (\Sigma_i,\Sigma_j,\Sigma_k)= \beta\, (-1,-1,2) \:,
\end{equation}
see Figure~\ref{base}.
Analogously, on the top of the cylinder $\overline{\mathscr{C}}_i$ (which is given by $s_i =0$, $s_j =1$, $s_k =0$) 
there is a fixed point $\mathrm{R}_j$, where $(\Sigma_i,\Sigma_j,\Sigma_k)= \beta\, (-1,2,-1)$.
The flow of the dynamical system that is induced on the top of $\overline{\mathscr{C}}_i$
is represented by a family of straight lines focusing at $\mathrm{R}_j$.

Depending on the value of $\beta$, see~\eqref{betadef}, 
we distinguish several scenarios
whose main characteristic is the position of the fixed point $\mathrm{R}_k$ in relation
to the Kasner circle $\mathrm{KC}_k$ (on the base of $\overline{\mathscr{C}}_i$).
In the following we discuss these scenarios in detail, 
in particular in view of their
compatibility with the energy conditions.
\renewcommand{\labelenumi}{\textbf{\Roman{enumi}}}
\begin{enumerate}
\item[\Cplus] $\beta>1$ (and $\beta <2$, see below).
The point $\mathrm{R}_k$ lies beyond the Taub point on the Kasner circle (which is the 
point $\mathrm{T}_{kk}$ on $\mathrm{KC}_k$), i.e., $\Sigma_k|_{\mathrm{R}_k} > 2$; see Figure~\ref{case1}.
This is the case if and only if $v_+ > 1$, which is not compatible with the dominant energy
condition. 
\item[\Bplus] $\beta=1$. 
The point $\mathrm{R}_k$ coincides with the Taub point $\mathrm{T}_{kk}$, i.e., $\Sigma_k|_{\mathrm{R}_k} = 2$;
see Figure~\ref{case2}. This is the case if and only if $v_+ = 1$;
accordingly, $v_- =  (3 w -1)/2$ by~\eqref{vpmvspez}, i.e., we have
the chain $(3 w -1)/2 = v_- < w < v_+ =1$.
Imposing the dominant 
energy condition on $v_-$ leads to the restriction
\[
w \geq -\textfrac{1}{3}\:.
\]
At the same time the strong energy condition is satisfied. 
\item[\Aplus] 
$\beta \in (0,1)$. The point
$\mathrm{R}_k$ lies between the center of the Kasner disk and the Taub point with
a value $\Sigma_k|_{\mathrm{R}_k} \in (0,2)$;
see Figure~\ref{case3}. 
This is the case if and only if $(3w-1)/2 < v_- < w < v_+ < 1$.
If $w \geq -1/3$, the dominant (and strong) energy condition are
automatically satisfied for $v_\pm$ and $\beta$ can take any value in $(0,1)$.
If $w < -1/3$, the condition
$v_- \geq -1$ is stronger than the condition $v_- > (3 w -1)/2$;
the range of $\beta$ is then restricted to $\beta \in (0, 2 (1+w)/(1-w)]$.
\item[\Azero] 
$\beta = 0$. The point $\mathrm{R}_k$ lies at the center of the Kasner disc.
This is the case iff $v_- = v_+ = w$.
\item[\Aminus] 
$\beta\in (-1,0)$. 
The point
$\mathrm{R}_k$ lies between the center of the Kasner disk and the non-flat LRS point (the point $\mathrm{Q}_{kk}$)
with a value $\Sigma_k|_{\mathrm{R}_k} \in (-2,0)$; see Figure~\ref{case4}.
This is the case iff $-1 +2 w < v_+ < w < v_- < (1 +w)/2$.
If $w \geq 0$, the dominant (and strong) energy condition are
automatically satisfied for $v_\pm$ and $\beta$ can take any value in $(-1,0)$.
If $w <0$, the condition $v_+ \geq -1$ is stronger than the condition $v_+> -1 +2 w$;
the range of $\beta$ is then restricted to $\beta \in [-(1+w)/(1-w),0)$.
\item[\Bminus] 
$\beta=-1$. 
The point $\mathrm{R}_k$ coincides with the non-flat LRS point $\mathrm{Q}_{kk}$, 
i.e., $\Sigma_k|_{\mathrm{R}_k} = -2$; see Figure~\ref{case5}. 
This is the case iff $-1 + 2 w = v_+ < w < v_- = (1+w)/2$.
The dominant energy condition requires
\[
w \geq 0 
\]
and at the same time the strong energy condition is satisfied.
\item[\Cminus] 
$\beta < -1$ (and $\beta > -2$, see below).
The point $\mathrm{R}_k$ lies beyond the non-flat LRS point $\mathrm{Q}_{kk}$, 
i.e., $\Sigma_k|_{\mathrm{R}_k} < -2$; see Figure~\ref{case6}. 
This is the case if and only if $v_+ < -1 +2 w < w < (1+w)/2 < v_-$.
To ensure compatibility with the dominant energy condition we must presuppose
\[
w > 0\:.
\]
The quantity $\beta$ cannot assume arbitrary values less than $-1$ without a violation
of the energy conditions.
Imposing the energy conditions on $v_\pm$ restricts the possible range;
we obtain that $\beta$ must be greater than or equal to the maximum
$\max\{-(1+w)/(1-w),-2\}$.
The extremal case would be $\beta=-2$, 
which corresponds to $v_-=1$ ($\Leftrightarrow v_+ = 3 w -2$). 
In this extremal case, the dominant energy condition requires $w \geq 1/3$. 
\end{enumerate}

\begin{Remark}
In a nutshell: Iff $\beta \in [-2, 1]$, then compatibility with the energy conditions
is possible; in general it is required that $w$ be sufficiently large ($w \geq 1/3$ is always 
sufficient).  
\end{Remark}

\begin{Remark}
For the function $v(s)$ (where we recall $v(0) = v_-$ and $v(1) = v_+$) 
we have the fundamental inequalities
\begin{equation}\label{vwv}
\begin{array}{cl} 
v_- < w < v_+ & \quad \textnormal{in the $\boldsymbol{+}$ cases } (\beta > 0) \\ 
v_- = w = v_+ &\quad \textnormal{in the case \Azero\ } \!(\beta =0)\\ 
v_- > w > v_+ &\quad \textnormal{in the $\boldsymbol{-}$ cases }(\beta< 0)\end{array}
\end{equation}
which will be used several times in the following.
\end{Remark}

\begin{Remark}
As we will see below, several characteristic properties of 
the flow of the dynamical system on the cylinder $\mathscr{C}_i$
change when $|\beta| \geq 2$. Therefore, in the following,
\Cplus\ will denote the case $\beta \in (1,2)$ only,
and \Cminus\ will denote the case $\beta \in (-2,-1)$.
The cases $|\beta| \geq 2$ will be denoted by extra symbols:
\renewcommand{\labelenumi}{\textbf{\Roman{enumi}}}
\begin{enumerate}
\item[\Dplus] $\beta\geq 2$.
The point $\mathrm{R}_k$ lies far beyond the Taub point on the Kasner circle, i.e., $\Sigma_k|_{\mathrm{R}_k} \geq 4$.
Like case \Cplus\ this case is not compatible with the dominant energy condition. 
\item[\Dminus] $\beta\leq -2$. 
The point $\mathrm{R}_k$ lies far beyond the non-flat LRS point $\mathrm{Q}_{kk}$, 
i.e., $\Sigma_k|_{\mathrm{R}_k} \leq -4$.
The only subcase that is compatible with the dominant energy condition 
is $\beta=-2$, which corresponds to $v_-=1$ ($\Leftrightarrow v_+ = 3 w -2$). 
In this case, the dominant energy condition requires $w \geq 1/3$. 
\end{enumerate}
\end{Remark}

\subsubsection*{Stability of the fixed points on the boundary of $\bm{\mathscr{C}_i}$}

Combining the analysis of the flow on the lateral surface 
and on the base/top surface we obtain the stability properties of the 
fixed points on the Kasner circles $\mathrm{KC}_j$ and $\mathrm{KC}_k$. 
(We simply put Figures~\ref{cylinder} and~\ref{base} on top of each other.)
Depending on the case there exist 
(transversally hyperbolic) 
sources and sinks on $\mathrm{KC}_j$/$\mathrm{KC}_k$; the location of 
these sources and sinks on $\mathrm{KC}_k$ is depicted in Figure~\ref{base}
(the location of sources/sinks on $\mathrm{KC}_j$ is obtained by a reflection w.r.t.\ the $\Sigma_i$-axis). 
In cases \Bplus\ and \Bminus, $\mathrm{R}_k$  (which coincides with $\mathrm{T}_{kk}$ and $\mathrm{Q}_{kk}$, respectively) 
is not transversally hyperbolic; the center manifold reduction theorem
implies that, in case \Bplus, the fixed point is a center saddle;
in case \Bminus, it acts as a sink, i.e., it is the $\omega$-limit for a two-parameter set of orbits in $\mathscr{C}_i$;
the analog holds for $\mathrm{R}_j$.

The point $\mathrm{R}_j$ [$\mathrm{R}_k$] lies in the interior of the top [base] of the cylinder,
iff we are in one of the \A\ scenarios.
By~\eqref{dynoncyl} we obtain
\begin{equation}
(1-s)^{-1} (1-s)^\prime\,\big|_{\mathrm{R}_j} = 6 \beta \:,\qquad
s^{-1} s^\prime\,\big|_{\mathrm{R}_k} = 6 \beta \:.
\end{equation}
In combination with the results on the flow on the base/top, 
see~Figure~\ref{base}, we obtain that $\mathrm{R}_j$ [$\mathrm{R}_k$] 
is a saddle in case \Aplus, while
it is a sink in case \Aminus.

\begin{Remark}
For completeness we briefly discuss the case \Azero\ as well.
In this case, $\beta = 0$ and the fixed points $\mathrm{R}_j$/$\mathrm{R}_k$
are not hyperbolic.
Center manifold analysis reveals that
the character of these points depends on the derivative of $v(s)$ at $s=0$ and $s=1$, 
i.e., on $v'(0)$ and $v'(1)$.
As the antisymmetry property~\eqref{vantisymm} of the function $v(s)$
guarantees that $v'(0) = v'(1)$,
there exist two subcases:
\begin{itemize}\setlength{\itemindent}{2em}
\item[\Azeroplus] $v'(0) = v'(1) < 0$. $\mathrm{R}_j$/$\mathrm{R}_k$ are center saddles. This case resembles the case \Aplus.
\item[\Azerominus] $v'(0) = v'(1) > 0$. $\mathrm{R}_j$/$\mathrm{R}_k$ act as sinks. This case resembles the case \Aminus.
\end{itemize}
Apart from a special subcase ($v(s) \equiv w$), the degenerate case $v'(0)=v'(1) = 0$ is
excluded by Assumption~\ref{vassum} below.
(If we analyze the system~\eqref{dynoncyl} with the function $u_i(s)$ instead of $v(s)$, cf.~\eqref{wijku},
there might exist more subcases of \Azero, because the signs of $u_i'(0)$ and $u_i'(1)$ might be different;
however, since the focus of our analysis lies on the $\boldsymbol{+}$ and $\boldsymbol{-}$ cases and not on \Azero,
we refrain from discussing these additional subcases further.)
\end{Remark}

The remaining fixed points on the boundary of $\overline{\mathscr{C}}_i$
are located on the lateral boundary: $\mathrm{TL}_i$ and $\mathrm{QL}_i$.
Using~\eqref{omega},~\eqref{vpmvspez}, and~\eqref{betadef} we find
\begin{subequations}\label{TLQLlocal1}
\begin{align}
\Omega^{-1} \Omega'\,\big|_{\mathrm{TL}_i} & = 3 (1 - v_-) = \textfrac{3}{2} (1-w) (2 + \beta) \:,\\
\Omega^{-1} \Omega'\,\big|_{\mathrm{QL}_i} & = 2(1-v_+) + (1-v_-) = 
\textfrac{3}{2} (1-w) + \textfrac{3}{2} (1-v_+) = \textfrac{3}{2} (1-w) (2 -\beta) \:.
\end{align}
\end{subequations}
Therefore, if $\beta < 2$
each fixed point on $\mathrm{QL}_i$ 
acts as the source for exactly one orbit in the interior of $\mathscr{C}_i$; 
if $\beta > 2$, each point on $\mathrm{QL}_i$ is the $\omega$-limit set
for one interior orbit. 
The analog is true for $\mathrm{TL}_i$: If $\beta > -2$,
then each fixed point on $\mathrm{TL}_i$ acts as the source for
one interior orbit; if $\beta < -2$, each fixed point on $\mathrm{TL}_i$ 
attracts one interior orbit as $\tau\to \infty$.
(The proof of these statements is based on the center manifold reduction theorem, where we
use that the lateral boundary is the center manifold of $\mathrm{QL}_i$.)
The borderline cases $\beta=\pm 2$ cannot be dealt with by using local methods; these cases
are discussed in Lemma~\ref{alphalimitcyl} and Lemma~\ref{omegalimitcyl1} below.

\subsubsection*{Dynamics in the interior of $\bm{\mathscr{C}_i}$}

The dynamical system~\eqref{dynoncyl} on the cylinder $\mathscr{C}_i$ admits a non-negative 
function,
\begin{equation}\label{monMi}
M_{(i)} := \big( \Sigma_i  + \beta  \big)^2 \:,
\qquad
M_{(i)}^\prime = -3 \Omega (1- w) M_{(i)} \:,
\end{equation}
which is strictly monotonically decreasing on $\mathscr{C}_i$ whenever $\Sigma_i \neq -\beta$.
The plane $\Sigma_i = -\beta$ itself, i.e.,
\begin{equation}
\mathscr{D}_i = \Big\{ (s, \Sigma_i,\Sigma_j,\Sigma_k)\in \mathscr{C}_i \:\Big|\: \Sigma_i = - \beta \Big\} 
\end{equation}
is an invariant subset. 
It can be characterized as the unique plane, orthogonal to the base/top of the
cylinder, whose boundary contains the points $\mathrm{R}_j$ and $\mathrm{R}_k$.
The plane $\mathscr{D}_i$ intersects the cylinder $\mathscr{C}_i$ 
whenever $-2 < \beta < 2$. (When $\beta = \pm 2$, 
$\mathscr{D}_i$ is tangent to $\overline{\mathscr{C}}_i$; 
the intersection of the two is $\mathrm{TL}_i$ or $\mathrm{QL}_i$.
When $|\beta| > 2$, $\mathscr{D}_i$ does not intersect $\overline{\mathscr{C}}_i$.)

\begin{Lemma}\label{alphalimitcyl}
Let $\gamma$ be an orbit in the (interior of the) cylinder $\mathscr{C}_i$ such
that $\gamma\not\subset \mathscr{D}_i$.
The $\alpha$-limit set of $\gamma$ is 
\begin{itemize}
\item one of the transversally hyperbolic sources
on the Kasner circles $\mathrm{KC}_j$/$\mathrm{KC}_k$ (where the points
on $\partial\mathscr{D}_i$ are excluded), or
\item a fixed point on $\mathrm{QL}_i$ (when $\beta < 2$), or 
\item a fixed point on $\mathrm{TL}_i$ (when $\beta > -2$).
\end{itemize}
Each transversally hyperbolic source on $\mathrm{KC}_j$/$\mathrm{KC}_k$ 
is the $\alpha$-limit set for a one-parameter family of orbits; 
each fixed point on $\mathrm{QL}_i$ and $\mathrm{TL}_i$ is the 
$\alpha$-limit set for one orbit in $\mathscr{C}_i$ (under the given assumption on $\beta$).
\end{Lemma}

\begin{proof}
Assume first that $|\beta| < 2$. 
The orbit $\gamma$ is either contained in the invariant subset $\mathscr{S}^+_i = \{\Sigma_i > -\beta\}$ 
or in the invariant subset $\mathscr{S}^-_i = \{\Sigma_i < -\beta\}$ of $\mathscr{C}_i$.
Since the function $M_{(i)}$ is monotonically decreasing on the invariant sets $\mathscr{S}^\pm_i$,
the monotonicity principle implies that the $\alpha$-limit set of $\gamma$ 
is contained on the boundary of $\mathscr{S}^\pm_i$; however, $\overline{\mathscr{D}}_i$ 
is excluded, since $M_{(i)}$ attains its minimum there.
This leaves the lateral boundary of $\mathscr{C}_i$
as the only possible superset of the $\alpha$-limit set of $\gamma$.
From the structure of the flow on the boundary of $\mathscr{C}_i$, 
which is depicted in Figure~\ref{cylinder}, we conclude
that it is only the fixed points that come into question as possible $\alpha$-limit sets.
Combining these observations with the local analysis of the fixed points, 
which we performed in the previous subsection, leads to the statement of the lemma.
In the case $\beta \geq 2$, the set $\mathscr{S}^-_i$ 
is empty and $\mathscr{S}^+_i$ corresponds to the interior of the cylinder.
The monotonicity principle, when applied to the function $M_{(i)}$, implies
that $\alpha(\gamma)$ is located on the lateral boundary of $\mathscr{C}_i$,
where $\mathrm{QL}_i$ is excluded.
The case $\beta \leq -2$ is analogous and thus the claim of the
lemma is proved.
\end{proof}

\begin{Lemma}\label{omegalimitcyl1}
Let $\gamma \subset \mathscr{C}_i$,  $\gamma\not\subset \mathscr{D}_i$.
Then the $\omega$-limit set $\omega(\gamma)$ of $\gamma$ depends on the actual case under
consideration (as characterized by the quantity $\beta$):
\begin{itemize}
\item $\beta > 2$ \textnormal{($\rightarrow$ \Dplus)}. $\omega(\gamma)$ is one of the transversally hyperbolic sinks on 
$\mathrm{KC}_j$/$\mathrm{KC}_k$ or a point on $\mathrm{QL}_i$.
\item $\beta =2$ \textnormal{($\rightarrow$ \Dplus)}. $\omega(\gamma) \subseteq \mathrm{QL}_i$.
\item $\beta \in [-1,2)$ \textnormal{($\leftrightarrow$ \A, \B, \Cplus)}. 
$\omega(\gamma) \subseteq \overline{\mathscr{D}}_i$.
\item $\beta \in (-2,-1)$ \textnormal{($\leftrightarrow$ \Cminus)}. 
$\omega(\gamma)$ is one of the transversally hyperbolic sinks on 
$\mathrm{KC}_j$/$\mathrm{KC}_k$ or it lies on  $\overline{\mathscr{D}}_i$.
\item $\beta \leq -2$ \textnormal{($\leftrightarrow$ \Dminus)}. 
$\omega(\gamma)$ is one of the transversally hyperbolic sinks on 
$\mathrm{KC}_j$/$\mathrm{KC}_k$ or it lies on  $\mathrm{TL}_i$.
\end{itemize}
\end{Lemma}

\begin{Remark}
In analogy to the statement of Lemma~\ref{alphalimitcyl}, each transversally hyperbolic sink on $\mathrm{KC}_j$/$\mathrm{KC}_k$
is the $\omega$-limit set for a one-parameter set of orbits; for $\beta > 2$,
each fixed point on $\mathrm{QL}_i$ attracts exactly one interior orbit, so that
the set $\mathrm{QL}_i$, when regarded as a whole, attracts a one-parameter set of interior orbits;
see~\eqref{TLQLlocal1};
in contrast, for $\beta = 2$, $\mathrm{QL}_i$ attracts every orbit in $\mathscr{C}_i$, i.e., a two-parameter set.
\end{Remark}

\begin{proof}
The proof is analogous to the proof of Lemma~\ref{alphalimitcyl}. 
Consider first the case $|\beta| < 2$. 
On $\overline{\mathscr{S}}^+_i$, which is the closure of the invariant subset defined 
in the proof of Lemma~\ref{alphalimitcyl}, the function $M_{(i)}$
attains its maximum on $\mathrm{TL}_i$; on $\overline{\mathscr{S}}^-_i$,
the maximum is attained on $\mathrm{QL}_i$; therefore, these sets are
excluded as possible $\omega$-limits sets. Using the analysis of
the flow on the boundary of the cylinder, the statement of the lemma ensues.
The cases $|\beta| \geq 2$ are analogous.
\end{proof}

\subsubsection*{Dynamics on the invariant plane $\bm{\overline{\mathscr{D}}_i}$}

Lemma~\ref{omegalimitcyl1} states that
the future dynamics of orbits in $\mathscr{C}_i$
is connected with the properties of the flow on $\overline{\mathscr{D}}_i$.
(This is for $|\beta| < 2$, i.e., for all cases \A, \B, \C.)
We thus proceed by investigating the flow of the dynamical system on 
this invariant plane.

Let $\mathrm{D}_i$ denote the set of fixed points on (the interior of) $\mathscr{D}_i$ 
for the dynamical system~\eqref{dynoncyl}.
The coordinates $(\bar{s},\bar{\Sigma}_j,\bar{\Sigma}_k)$ 
of any fixed point $\mathrm{P}\in \mathrm{D}_i$ must satisfy $\bar{\Sigma}_j = \bar{\Sigma}_k = \beta/2$
and 
\begin{equation}\label{diex}
v(\bar{s}) = 
\frac{1}{2} ( v_- + v_+)\:,
\end{equation}
where $\bar{s}\in (0,1)$.
In the $\boldsymbol{+}$ and $\boldsymbol{-}$
cases (i.e., $\beta \lessgtr 0$) 
we have $v_- \gtrless v_+$; therefore
there exists at least one solution of equation~\eqref{diex}.
(Analogously, in the cases \Azeroplus\ and \Azerominus, 
existence of a solution is guaranteed by the fact 
that the derivatives of $v(s)$, or $u_i(s)$, at $s=0$ and $s=1$ 
have the same sign.)

\begin{Remark}
Because $v(s) - (v_-+v_+)/2$ is antisymmetric around $s=1/2$, cf.~\eqref{vantisymm},
one solution of equation~\eqref{diex} is $\bar{s} = 1/2$.
Consequently, the point $\mathrm{P}_i^{(0)}$ with coordinates
$(\bar{s}, \bar{\Sigma}_j,\bar{\Sigma}_k) = (1/2)(1,\beta,\beta)$ 
is always a fixed point on $\mathscr{D}_i$.
(If we analyze the system~\eqref{dynoncyl} with the 
function $u_i(s)$ instead of $v(s)$, cf.~\eqref{wijku},
$\bar{s} = 1/2$ will in general not be a solution.)
\end{Remark}

We make the simplifying 
assumption that the zeros of the function $v(s)- (v_-+v_+)/2$
are simple zeros:

\begin{Assumption}\label{vassum}
We assume that the function $v(s)$ either satisfies 
\begin{equation}\label{vassumeq}
v^\prime(\bar{s}) \neq 0\:, \qquad \text{for all }\bar{s}\in [0,1] \,\text{ such that }
v(\bar{s}) = \frac{v_-+v_+}{2}\:,
\end{equation}
or $v(s) \equiv w$ for all $s$. 
\end{Assumption}

Assumption~\ref{vassum} is automatically satisfied for collisionless matter
and elastic matter and
helps to avoid unnecessary clutter in the analysis of the flow on $\mathscr{D}_i$;
requiring~\eqref{vassumeq} is equivalent to assuming that the set of fixed points $\mathrm{D}_i$
on $\mathscr{D}_i$ is a discrete set of hyperbolic fixed points.
In addition, it follows that
the number of fixed points is always odd, i.e.,
$\#\mathrm{D}_i = 2 d +1$, $d \in \mathbb{N}$.
Accordingly, we can write 
$\mathrm{D}_i = 
\{ \mathrm{P}_i^{(-d)}, \ldots, \mathrm{P}_i^{(-1)}, \mathrm{P}_i^{(0)}, \mathrm{P}_i^{(1)}, \ldots, \mathrm{P}_i^{(d)} \}$,
where the coordinates of these fixed points satisfy 
$0< \bar{s}_{(-d)} < \ldots < \bar{s}_{(-1)} < \bar{s}_{(0)} < \bar{s}_{(1)} < \ldots < \bar{s}_{(d)} <1$.
The only exception we admit is the case $v(s) \equiv w$, which is
a subcase of \Azero; in this special case,
the set $\mathrm{D}_i$ of fixed points is not a discrete set but 
coincides with the line $\Sigma_j = \Sigma_k = 0$
in $\mathscr{D}_i$; this special case will be briefly commented on 
at the end of this section.
%

\begin{Remark}
The antisymmetry property~\eqref{vantisymm} of $v(s)$
guarantees that $\bar{s}_n+\bar{s}_{-n}=1$, $n=0,1,\ldots,d$.
This is not true in general, if we analyze the system~\eqref{dynoncyl} with the 
function $u_i(s)$ instead of $v(s)$, cf.~\eqref{wijku}.
\end{Remark}

Let $\mathrm{P}\in\mathrm{D}_i$ be a fixed point in $\mathscr{D}_i$ with coordinates $(\bar{s},\beta/2,\beta/2)$.
The eigenvalues of the linearization of the 
dynamical system~\eqref{dynoncyl} at $\mathrm{P}$ are given by
\begin{subequations}\label{ewofDi}
\begin{equation}
-\frac{9}{16} (1-w) (2+\beta) (2-\beta) \: \left[ 1 \pm \sqrt{1 - 
\frac{64 \bar{s} (1-\bar{s})}{3 (1-w)^2 (2-\beta)}\: v'(\bar{s})} \:\right]
\end{equation}
and 
\begin{equation}
-\frac{3}{2} (1-w)\: \Omega\big|_{\mathrm{P}} \:.
\end{equation}
\end{subequations}
The former are associated with eigenvectors tangential to the plane $\mathscr{D}_i$,
the latter is associated with an eigenvector transversal to $\mathscr{D}_i$.
It follows that $\mathrm{P}$ is a sink if $v'(\bar{s}) > 0$,
while $\mathrm{P}$ is a saddle if $v'(\bar{s}) < 0$.
Therefore, by Assumption~\ref{vassum}, the set $\mathrm{D}_i = \{\mathrm{P}_i^{(-d)}, \ldots, \mathrm{P}_i^{(d)}\}$
of fixed points can be regarded as an alternating sequence of saddles and sinks.
In the $\boldsymbol{+}$ cases, inequality~\eqref{vwv} implies the structure
$\mathrm{D}_i = 
\{\text{sink}$, $\text{saddle}$, $\text{sink}$, $\ldots$, $\text{sink}$, $\text{saddle}$, $\text{sink}\}$;
in particular, $d+1$ fixed points are sinks and $d$ points are saddles.
In the $\boldsymbol{-}$ cases, inequality~\eqref{vwv} implies the structure
$\mathrm{D}_i = \{\text{saddle}$, $\text{sink}$, $\text{saddle}$, $\ldots$, $\text{saddle}$, $\text{sink}$, $\text{saddle}\}$;
in particular, $d+1$ fixed points are saddles and $d$ points are sinks.

\begin{Remark}
In the cases \Aplusminus\ the fixed points $\mathrm{R}_j$ and $\mathrm{R}_k$ can be added
to the alternating sequence $\{\mathrm{P}_i^{(-d)}, \ldots, \mathrm{P}_i^{(d)}\}$
of saddles and sinks; 
this follows from the local analysis of the fixed points $\mathrm{R}_j$/$\mathrm{R}_k$.
Accordingly, in the case \Aplus, the sequence 
$\{\mathrm{R}_k, \mathrm{P}_i^{(-d)}, \ldots, \mathrm{P}_i^{(d)},\mathrm{R}_j\}$
is of the type $\{\text{saddle}$, $\text{sink}$, $\text{saddle}$, $\ldots$,
$\text{saddle}$, $\text{sink}$, $\text{saddle}\}$,
while in the case \Aminus, it is of the type
$\{\text{sink}$, $\text{saddle}$, $\text{sink}$, $\ldots$, $\text{sink}$, $\text{saddle}$, $\text{sink}\}$.
The cases \Azeroplus\ and \Azerominus\ can be subsumed under the cases \Aplus\ and \Aminus, respectively,
the only difference being that $\mathrm{R}_j$/$\mathrm{R}_k$ are non-hyperbolic.
\end{Remark}

\begin{figure}[Ht!]
\begin{center}
\subfigure[\Bplus, \Cplus]{%
\label{Dc1}
\includegraphics[width=0.2\textwidth]{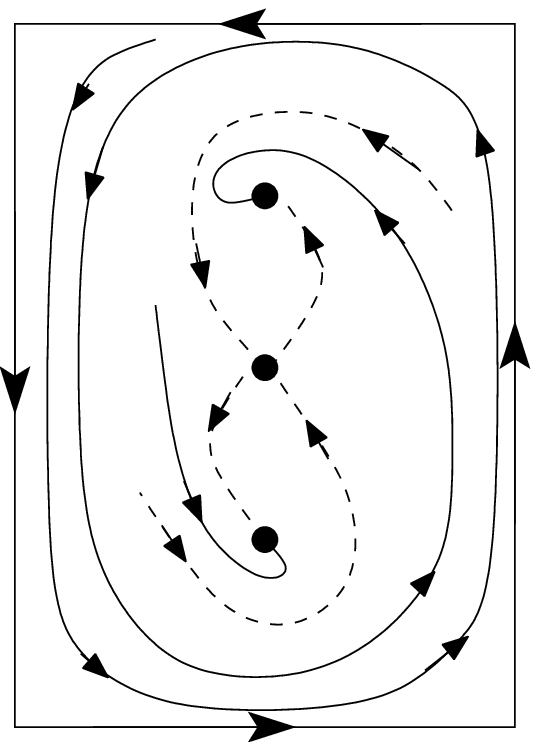}}\qquad
\subfigure[\Aplus\ (and \Azeroplus)]{%
\label{Dc2}
\psfrag{rk}[cc][cc][0.7][0]{$\mathbf{\mathrm{R}_k}$}
\psfrag{rj}[cc][cc][0.7][0]{$\mathbf{\mathrm{R}_j}$}
\includegraphics[width=0.2\textwidth]{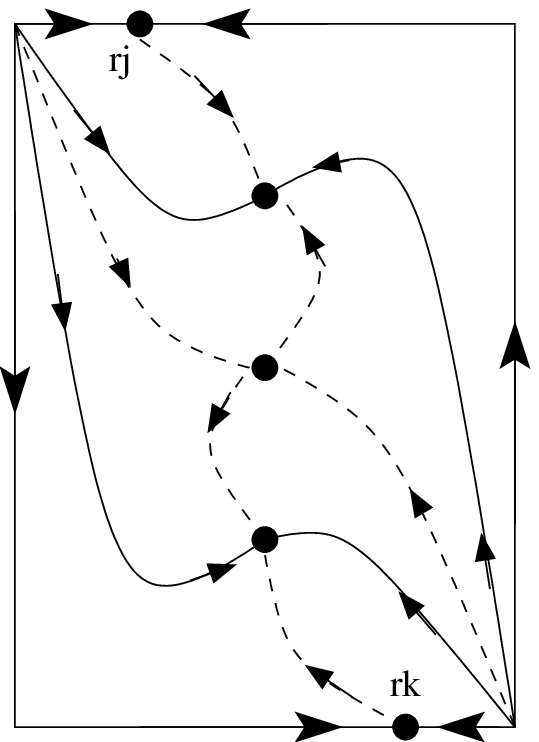}}\qquad
\subfigure[\Aminus\ (and \Azerominus)]{%
\label{Dc5}
\psfrag{rk}[cc][cc][0.7][0]{$\mathbf{\mathrm{R}_k}$}
\psfrag{rj}[cc][cc][0.7][0]{$\mathbf{\mathrm{R}_j}$}
\includegraphics[width=0.2\textwidth]{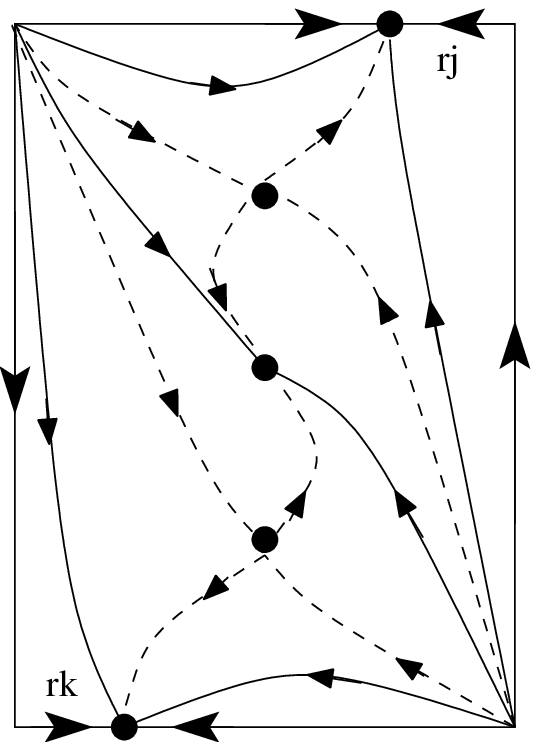}}\qquad
\subfigure[\Bminus, \Cminus]{%
\label{Dc6}
\includegraphics[width=0.2\textwidth]{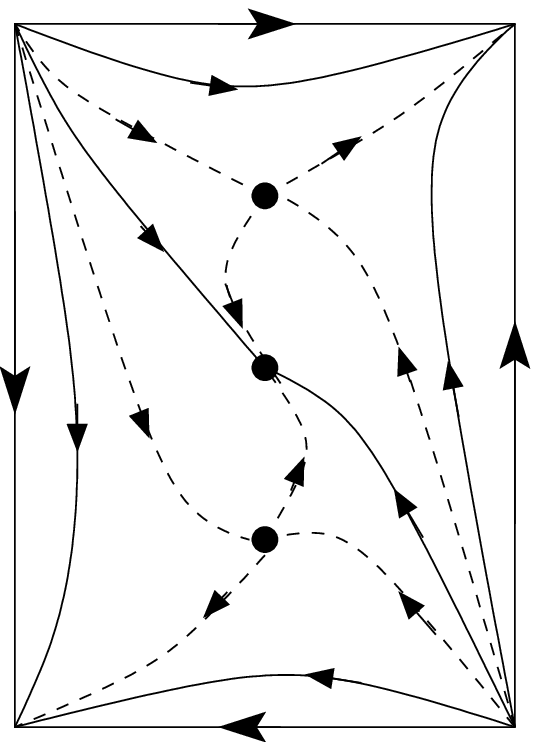}}
\caption{The figures display the flow of the dynamical system on the plane $\mathscr{D}_i$. 
We assume $|\beta| <2$ (cases \A, \B, \C) 
so that $\mathscr{D}_i$ 
is an invariant subset lying in $\mathscr{C}_i$. 
The number of fixed points 
in the interior of $\mathscr{D}_i$ is necessarily odd,
i.e., $\#\mathrm{D}_i = 2 d +1$, $d\in\mathbb{N}$; we depict the case $d=1$, i.e., $\#\mathrm{D}_i = 3$.
[The case $\#\mathrm{D}_i = 1$, which is not depicted, is commented on in square brackets.] 
Non-generic orbits are represented by dashed lines.
(a) The boundary $\partial\mathscr{D}_i$ forms a heteroclinic cycle in cases \Bplus\ and \Cplus.
Generic orbits converge to this heteroclinic cycle as $\tau\rightarrow -\infty$. The central
fixed point $\mathrm{P}_i^{(0)}$ is a saddle; the two other points $\mathrm{P}_i^{(\pm 1)}$ 
are sinks.
[In the case $\#\mathrm{D}_i =1$, $\partial\mathscr{D}_i$ is the $\alpha$-limit set,
and the (unique) fixed point $\mathrm{P}_i^{(0)}$ is the $\omega$-limit set for all (non-trivial) solutions.]
(b) In case \Aplus, the properties of the three fixed points $\mathrm{D}_i$ are analogous. 
The points $\mathrm{R}_j$/$\mathrm{R}_k$ are saddles; one interior orbit converges to $\mathrm{R}_j$,
one to $\mathrm{R}_k$ as $\tau\rightarrow -\infty$. (Whether these two orbits
converge to the sinks as $\tau\rightarrow \infty$, as depicted, 
or the to central fixed point, depends on the function $v(s)$.)
The past attractor for generic orbits consists of the Kasner points in the upper left/lower right corner.
[In the case $\#\mathrm{D}_i =1$, 
the fixed point  $\mathrm{P}_i^{(0)}$ is the $\omega$-limit set for all (non-trivial) solutions.]
The flow in the case \Azeroplus\ resembles the flow in the case \Aplus; however, $\mathrm{R}_j$/$\mathrm{R}_k$ are
center saddles.
(c) In case \Aminus, the central fixed point $\mathrm{P}_i^{(0)}$ is a sink, the two other interior fixed points $\mathrm{P}_i^{(\pm 1)}$ are
saddles; the points $\mathrm{R}_j$/$\mathrm{R}_k$ are sinks. The Kasner points in the upper left/lower right corner
are the $\alpha$-limit for generic orbits. [In the case $\#\mathrm{D}_i =1$, 
the fixed point $\mathrm{P}_i^{(0)}$ is a saddle, which leaves $\mathrm{R}_j$/$\mathrm{R}_k$ as
the exclusive future attractor for generic orbits.] The flow in the case \Azero$_{\!\boldsymbol{-}}$ 
resembles the flow in the case \Aminus; 
however, $\mathrm{R}_j$/$\mathrm{R}_k$ are not hyperbolic.
(d) The cases \Bminus\ and \Cminus\ are similar to \Aminus; however, 
the Kasner points in the lower left/upper right corner are sinks. 
[In the case $\#\mathrm{D}_i =1$,  
the fixed point $\mathrm{P}_i^{(0)}$ is a saddle, which leaves these Kasner points as
the generic future attractor.]
}
\label{Diflow}
\end{center}
\end{figure}

The next step in our study of the flow of the dynamical
system on the plane $\mathscr{D}_i$ is to investigate the global dynamics. 
To this end we consider the function
\begin{equation}\label{monN}
N = (1-\Sigma^2)^{-1} \kappa(s)\:,
\end{equation}
where $\kappa(s)$ is positive and satisfies the differential equation
\begin{equation}
s (1-s) \frac{d\kappa(s)}{d s} = \frac{1}{2} \Big( v(s) - \frac{v_+ + v_-}{2} \Big) \kappa(s)\:.
\end{equation}
By~\eqref{vwv}, in the $\boldsymbol{+}$ cases the function $\kappa(s)$ goes to $\infty$ 
as $s\rightarrow 0$ and $s\rightarrow 1$; in the $\boldsymbol{-}$ cases, 
$\kappa(s)\rightarrow 0$ in the limit $s\rightarrow 0$ and $s\rightarrow 1$.
A straightforward computation shows that $N$ is strictly monotonically decreasing 
on $\mathscr{D}_i$ except on the fixed point set $\mathrm{D}_i$, i.e.,
\begin{equation}
N^\prime\Big|_{\mathscr{D}_i} = -\frac{1-w}{2} \left[ \Big(\Sigma_j - \frac{\beta}{2}\Big)^2 +  
\Big(\Sigma_k - \frac{\beta}{2}\Big)^2\right]\: N \:,
\end{equation}
and 
\begin{equation*}
N^{\prime\prime\prime}\Big|_{\Sigma_j =\Sigma_k = \beta/2} =
-\frac{9}{16} (2+\beta)^2 (2-\beta)^2 (1-w)  \left[ \Big( v(s)- \frac{v_-+ v_+}{2}\Big)^2
+\Big( v(1-s)- \frac{v_-+ v_+}{2}\Big)^2\right]\: N\:.
\end{equation*}
The monotone function $N$ on $\mathscr{D}_i$ excludes
the existence of periodic orbits, homoclinic orbits and heteroclinic
cycles in (the interior of) $\mathscr{D}_i$.
Using the monotone function in combination with the local analysis 
of the fixed points, it is possible to solve the
global dynamics of the flow on $\mathscr{D}_i$.
Since the relevant arguments are rather standard (because the theory of
planar flows is well-developed) we omit the derivation 
here 
and merely summarize 
the results in Figure~\ref{Diflow}.
Note that in the cases \Bplus, \Cplus, the flow on $\partial\mathscr{D}_i$ 
is given by a heteroclinic cycle;
in the case \Bplus\ it is represented as follows: 
\begin{equation*}
\begin{CD}
\mathrm{T}_{jj} @ <<< \mathrm{T}_{jk} \\
@VVV @AAA \\
\mathrm{T}_{kj} @>>> \mathrm{T}_{kk} 
\end{CD}
\end{equation*}
Having solved the dynamics on $\overline{\mathscr{D}}_i$
we are now in a position to complete the analysis of the future dynamics of orbits in $\mathscr{C}_i$.

\subsubsection*{Dynamics in the interior of $\bm{\mathscr{C}_i}$ (cont.)}

While Lemma~\ref{alphalimitcyl} describes the past asymptotics of
orbits in $\mathscr{C}_i \backslash \mathscr{D}_i$ completely,
the description of the future asymptotics by 
Lemma~\ref{omegalimitcyl1} is incomplete in the cases where $|\beta| < 2$.
Exploiting the results of the previous subsection we are able to
complete the discussion.

\begin{Lemma}\label{DLemma}
Consider either of the $\boldsymbol{+}$ cases (with $\beta < 2$), i.e.,
either of the cases \textnormal{\Aplus, \Bplus, \Cplus} (or \textnormal{\Azeroplus}).
Let $\gamma$ be an orbit such that $\gamma \not\subset \mathscr{D}_i$.
Then $\omega(\gamma)$ is one of the fixed points of\/ $\mathrm{D}_i$.
More specifically, of the $\# \mathrm{D}_i = 2 d+1$ fixed points, there are $d+1$ sinks that
attract a two-parameter set of orbits each, while 
each of the $d$ saddles attracts a one-parameter family.   
\end{Lemma}

\begin{proof}
The flow on $\mathscr{D}_i$ is known in detail from the analysis of the preceding subsection.
We see that the only cases that require a careful analysis are \Bplus\ and \Cplus, since
in these cases, the boundary $\partial\mathscr{D}_i$ of the
plane $\mathscr{D}_i$ is a heteroclinic cycle, which has to be excluded
as a possible $\omega$-limit of $\gamma$.
To this end we use the function $N$ given by~\eqref{monN}.
Suppose that $\gamma$ converges to the heteroclinic cycle $\partial\mathscr{D}_i$.
A straightforward computation shows that 
\begin{equation}\label{NNp}
N^{-1} N^\prime = -\frac{1-w}{2} \left[ \Big(\Sigma_j +\frac{\Sigma_i}{2} \Big)^2 +
\Big(\Sigma_k +\frac{\Sigma_i}{2} \Big)^2  \right] - \frac{3}{4} (1-w) (\Sigma_i^2 - \beta^2) 
\end{equation}
along $\gamma$. The term in square brackets on the r.h.s.~of~\eqref{NNp} is a positive function,
which does not converge to zero along $\gamma$; on the contrary, for sufficiently large
times $\tau$ this function is approximately equal to $(3/2)(4-\beta^2)$ for most of the time
(which is because the orbit
$\gamma$ spends most of its time in a neighborhood of the fixed points on $\partial\mathscr{D}_i$).
In contrast, the second term on the r.h.s.~of~\eqref{NNp} converges to zero as $\tau\rightarrow \infty$.
It follows that $N\rightarrow 0$ along $\gamma$. However, this is a contradiction, because
$N$ is infinite on $\partial\mathscr{D}_i$; therefore, the heteroclinic
cycle $\partial\mathscr{D}_i$ is excluded as a $\omega$-limit of $\gamma$.
\end{proof}

\begin{Lemma}\label{lemmaCminus}
Consider either of the cases\/ \textnormal{\Aminus, \Bminus} (or \textnormal{\Azerominus}).
Let $\gamma$ be an orbit such that $\gamma \not\subset \mathscr{D}_i$.
Then $\omega(\gamma)$ is one of the fixed points on\/ $\overline{\mathscr{D}}_i$,
i.e., one of the points\/ $\mathrm{R}_j$, $\mathrm{R}_k$, or one of the points of the 
set\/ $\mathrm{D}_i$.
More specifically, of the $\# \mathrm{D}_i = 2 d+1$ fixed points, there are $d$ sinks that
attract a two-parameter set of orbits each, while 
each of the $d+1$ saddles attracts a one-parameter family.
The fixed points $\mathrm{R}_j$, $\mathrm{R}_k$ 
(which coincide with $\mathrm{Q}_{jj}$, $\mathrm{Q}_{kk}$ in \textnormal{\Bminus})
attract a two-parameter family of orbits each.
\end{Lemma}

\begin{Lemma}\label{lemmaAminus}
Consider the case\/ \textnormal{\Cminus}.
Let $\gamma$ be an orbit such that $\gamma \not\subset \mathscr{D}_i$.
Then $\omega(\gamma)$ is one of the fixed points of the set\/ $\mathrm{D}_i$
or one of the transversally hyperbolic sinks on the Kasner circles $\mathrm{KC}_j$/$\mathrm{KC}_k$.
More specifically, of the $\# \mathrm{D}_i = 2 d+1$ fixed points, there are $d$ sinks that
attract a two-parameter set of orbits each, while 
each of the $d+1$ saddles attracts a one-parameter family.
Each sink on the Kasner circles attracts a one-parameter family of orbits.
\end{Lemma}

\begin{proof}
The statements of the two previous lemmas follow from Lemma~\ref{omegalimitcyl1}, the analysis of
the flow on $\overline{\mathscr{D}}_i$, and the local analysis of the fixed points.
\end{proof}

\begin{figure}[Ht]
\begin{center}
\subfigure[\Cplus]{
\psfrag{Pi}[cc][cc][0.6][0]{$\mathrm{D}_i$}
\includegraphics[width=0.25\textwidth]{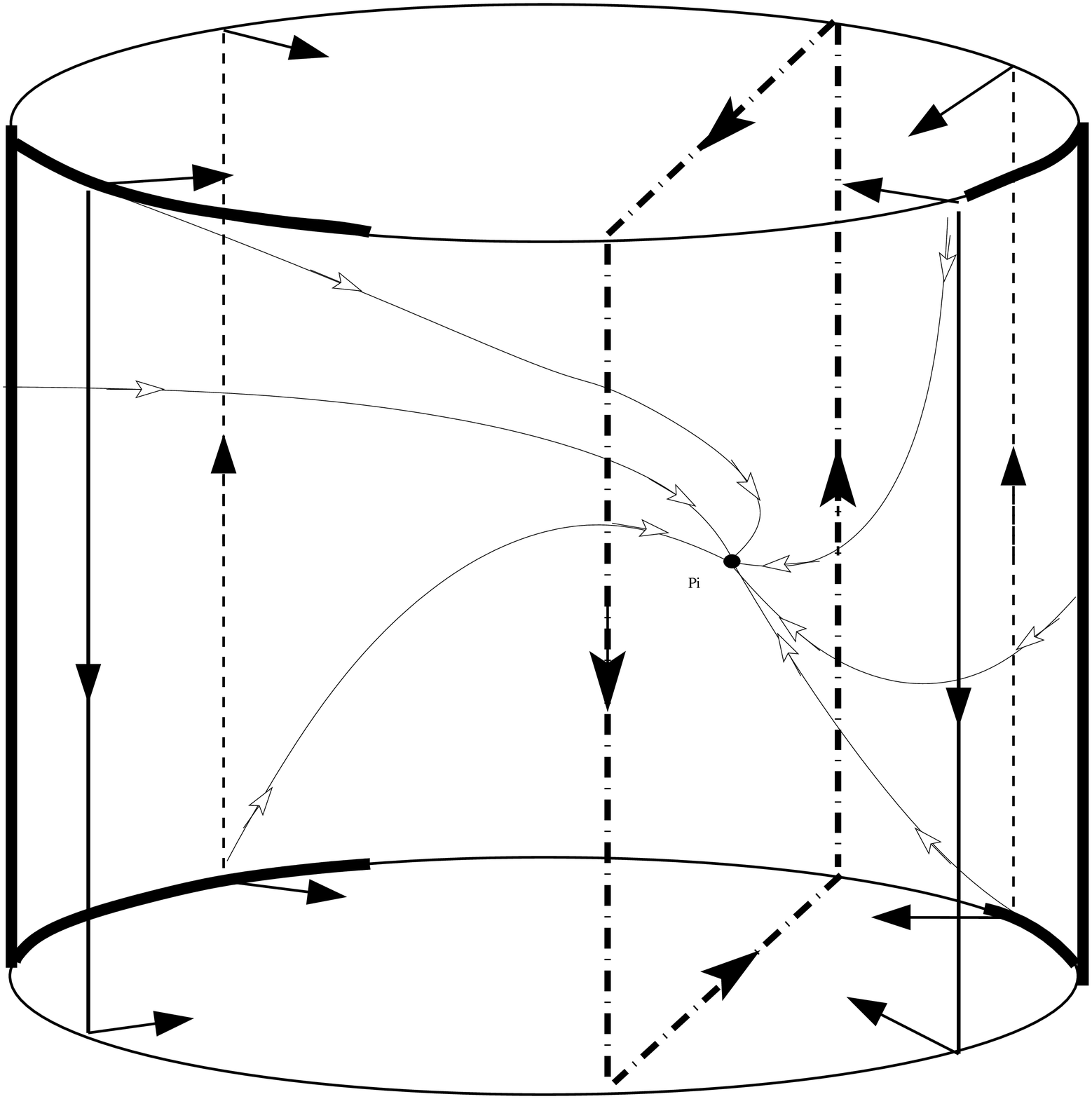}}
\qquad\quad
\subfigure[\Bplus]{
\psfrag{pijp}[cc][cc][0.6][0]{$\mathrm{T}_{jj}$}
\psfrag{pijm}[cc][cc][0.6][0]{$\mathrm{T}_{jk}$}
\psfrag{pikp}[cc][cc][0.6][0]{$\mathrm{T}_{kk}$}
\psfrag{pikm}[cc][cc][0.6][0]{$\mathrm{T}_{kj}$}
\psfrag{pi}[cc][cc][0.6][0]{$\mathrm{D}_i$}
\includegraphics[width=0.25\textwidth]{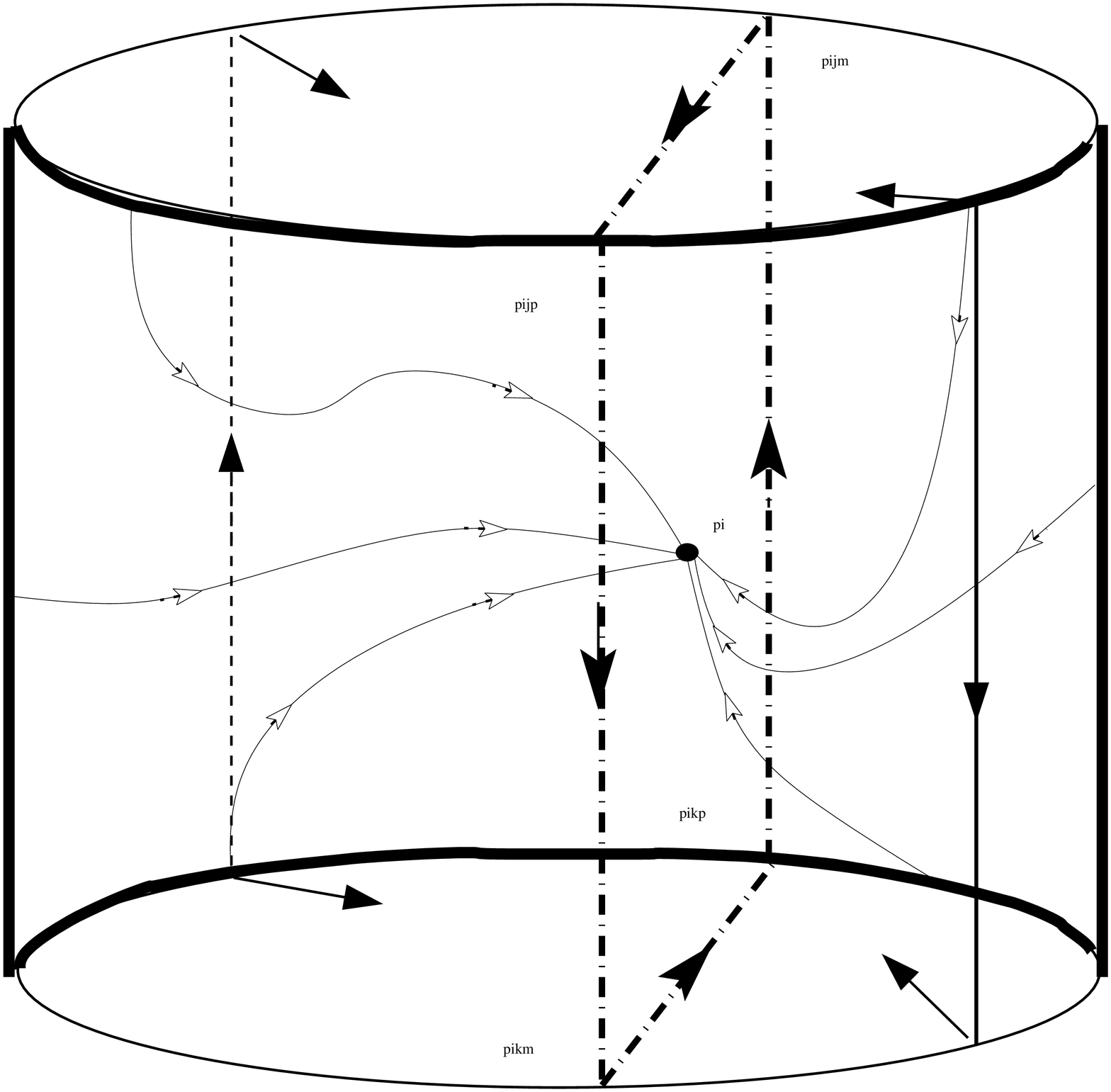}}\qquad\quad
\subfigure[\Aplus\ (and \Azeroplus)]{\label{cylApl}
\psfrag{pi}[cc][cc][0.6][0]{$\mathrm{D}_i$}
\psfrag{rj}[cc][cc][0.6][0]{$\mathrm{R}_j$}
\psfrag{rk}[cc][cc][0.6][0]{$\mathrm{R}_k$}
\includegraphics[width=0.25\textwidth]{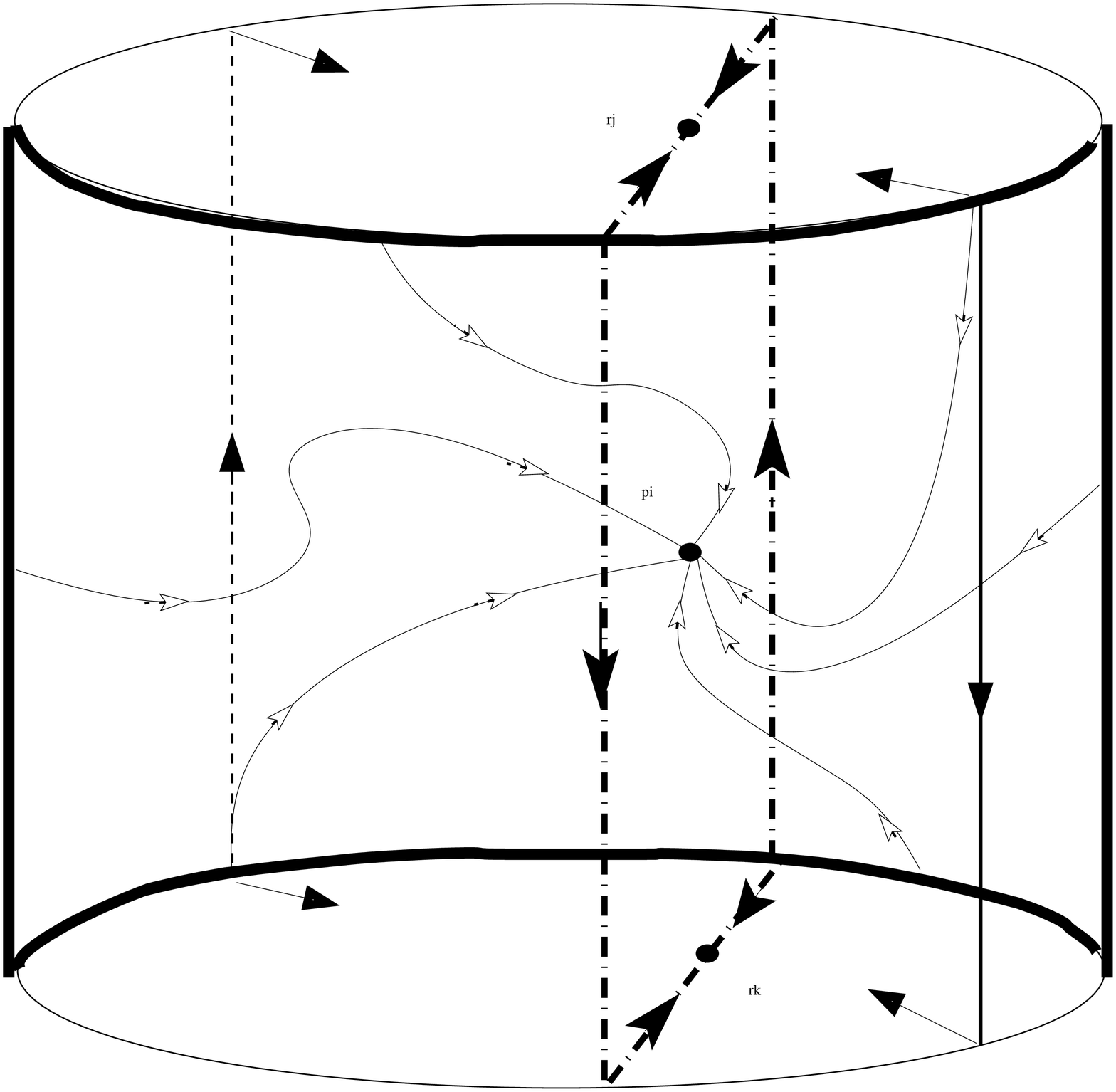}}\\
\subfigure[\Aminus\ (and \Azerominus)]{
\psfrag{pi}[cc][cc][0.6][0]{$\mathrm{D}_i$}
\psfrag{rj}[cc][cc][0.6][0]{$\mathrm{R}_j$}
\psfrag{rk}[cc][cc][0.6][0]{$\mathrm{R}_k$}
\includegraphics[width=0.25\textwidth]{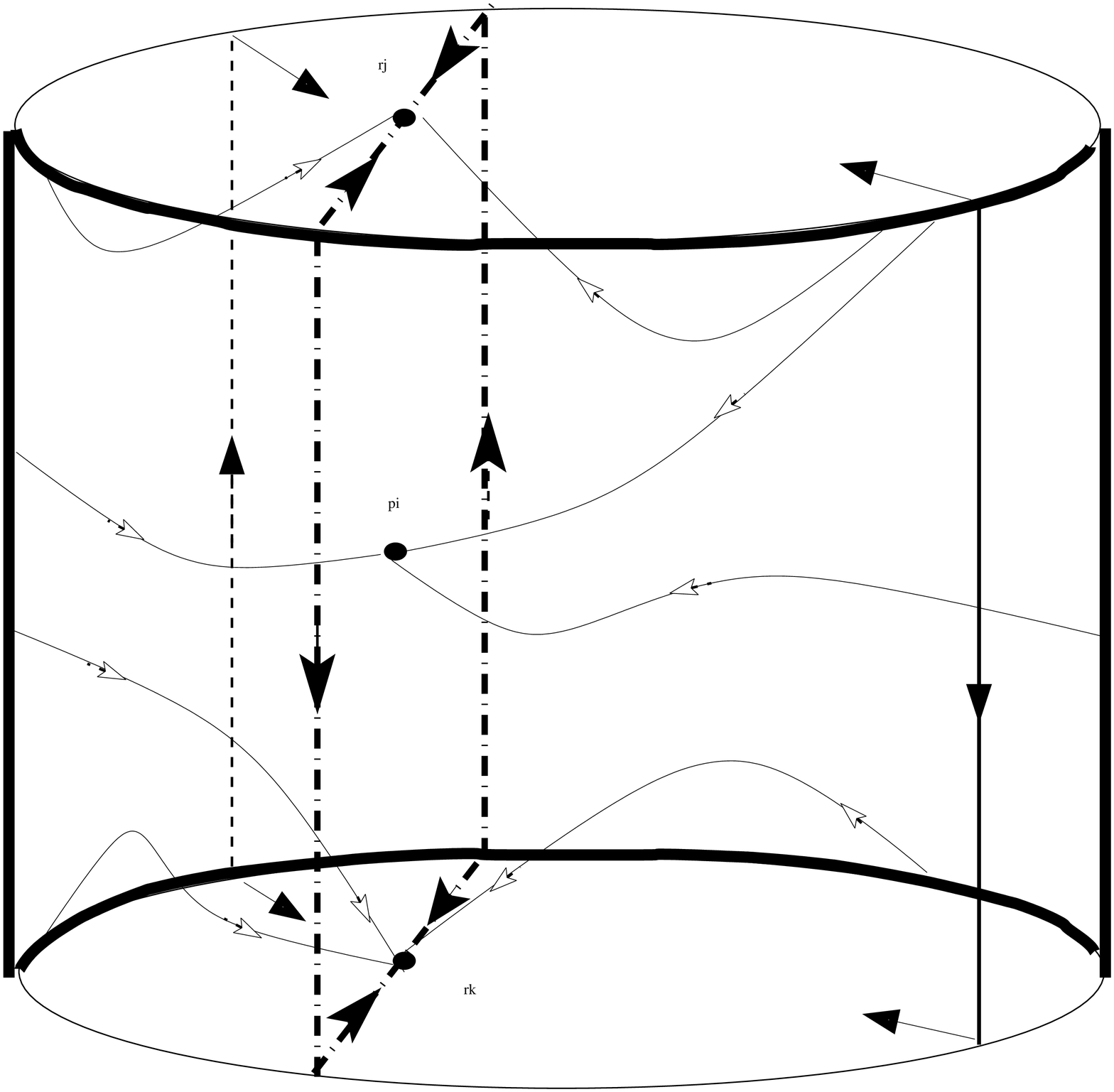}}
\qquad\quad
\subfigure[\Bminus]{
\psfrag{pi}[cc][cc][0.6][0]{$\mathrm{D}_i$}
\psfrag{qkk}[cc][cc][0.6][0]{$\mathrm{Q}_{kk}$}
\psfrag{qjj}[cc][cc][0.6][0]{$\mathrm{Q}_{jj}$}
\includegraphics[width=0.25\textwidth]{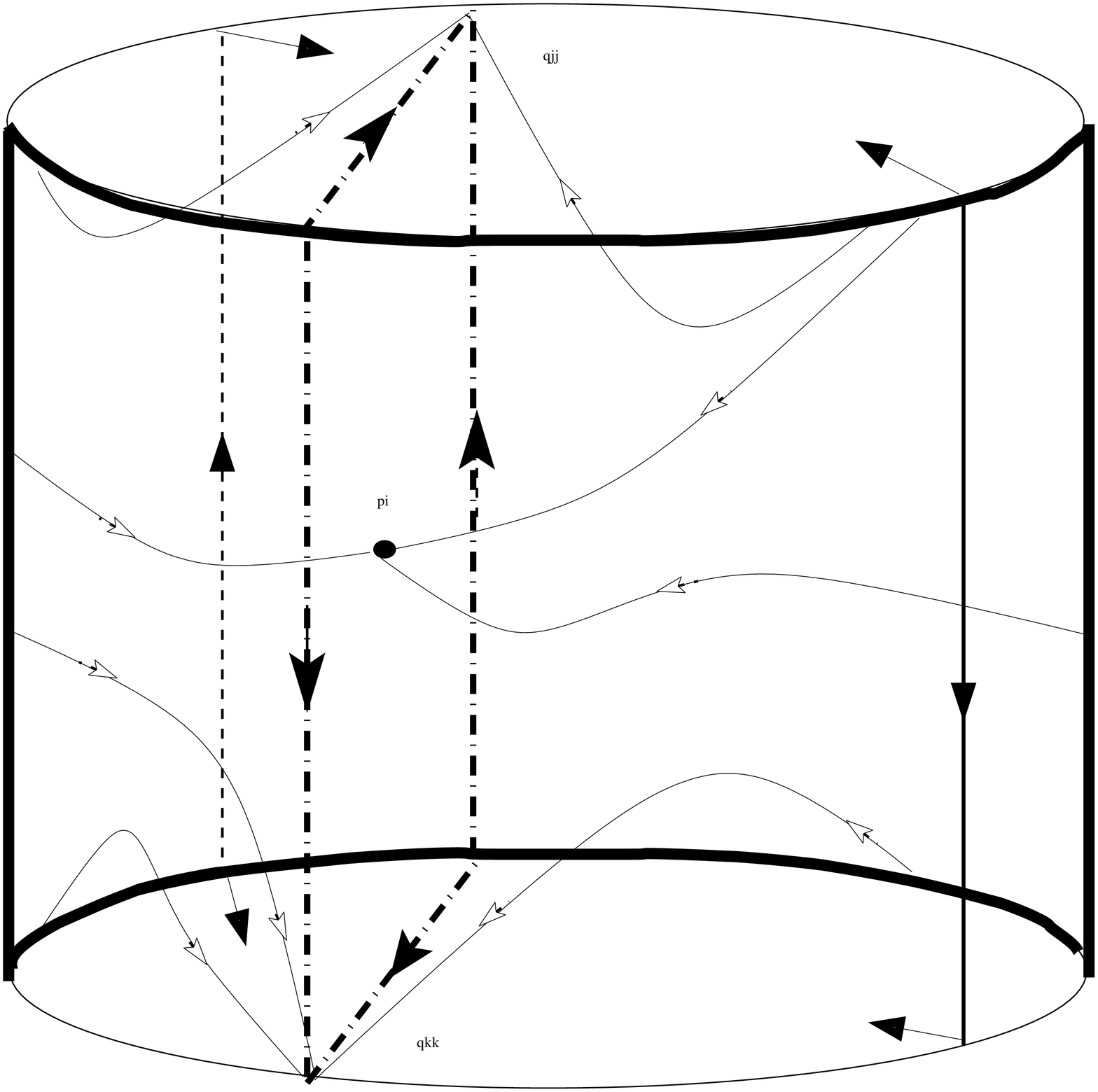}}
\qquad\quad
\subfigure[\Cminus]{
\psfrag{pi}[cc][cc][0.5][0]{$\mathrm{D}_i$}
\psfrag{tkk}[cc][cc][0.5][0]{$\mathrm{T}_{kk}$}
\psfrag{tjj}[cc][cc][0.5][0]{$\mathrm{T}_{jj}$}
\includegraphics[width=0.25\textwidth]{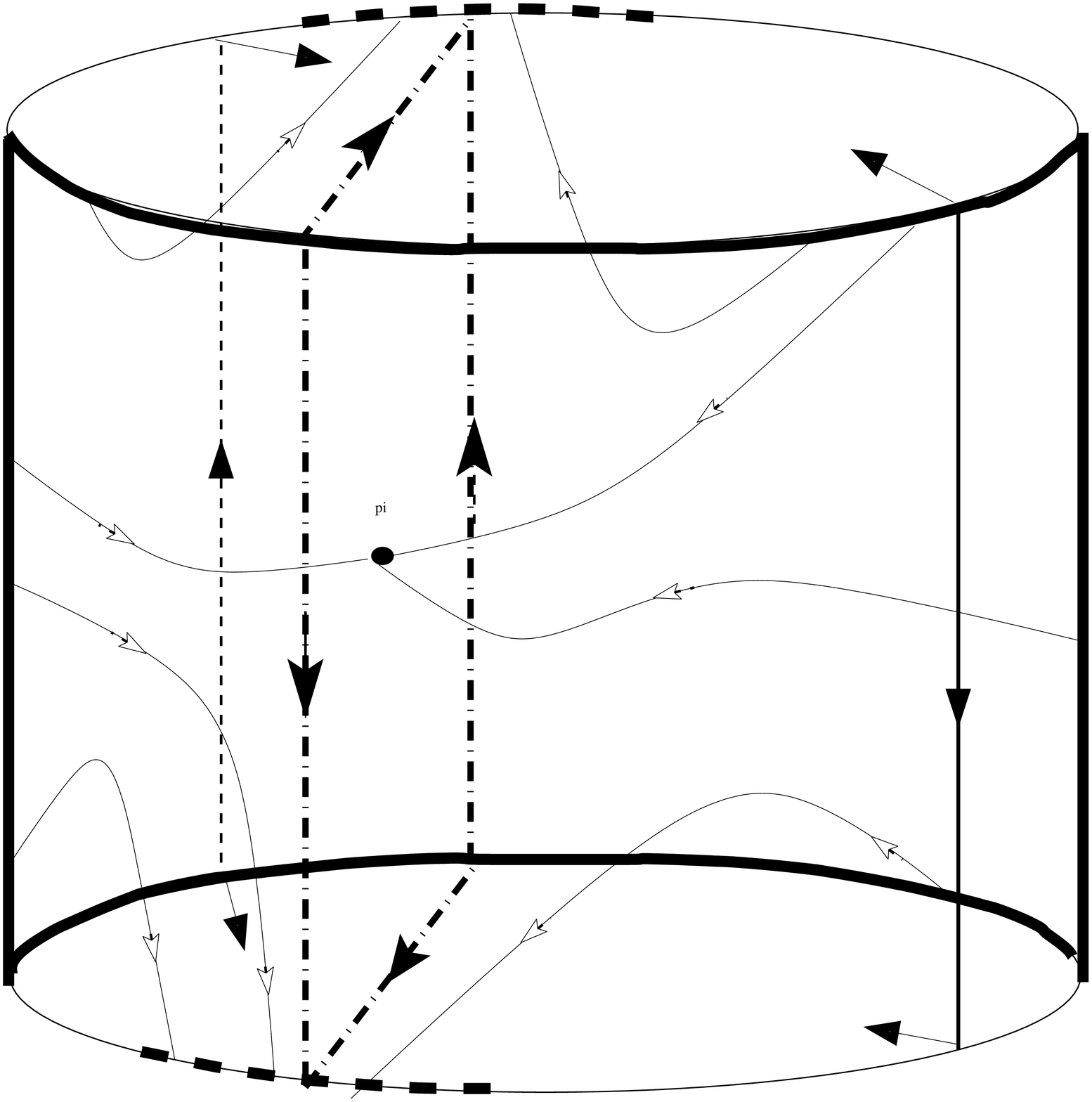}}
\caption{A schematic depiction of the flow on the cylinder $\overline{\mathscr{C}}_i$. 
In contrast to Figure~\ref{Diflow}, to simplify the illustration we assume in this figure 
that the function $v(s)$ 
is such that there is only one fixed point $\mathrm{D}_i$ 
in the plane $\mathscr{D}_i$.}
\label{cylinder2}
\end{center}
\end{figure}

Lemma~\ref{alphalimitcyl} describes the past dynamics, 
Lemma~\ref{omegalimitcyl1} (in the cases \D) 
and Lemmas~\ref{DLemma}--\ref{lemmaAminus} (in the cases \A, \B, \C)
describe the future dynamics of orbits in $\mathscr{C}_i$.
The qualitative behaviour of orbits is depicted in Figure~\ref{cylinder2}.

\begin{Remark}
In Assumption~\ref{vassum} we allow for a special case, defined by
the requirement $v(s) \equiv w$ for all $s$.
In this special case, $\mathrm{D}_i$ is not a discrete set of fixed points but
a line of fixed points, the line $\Sigma_j = \Sigma_k$. It follows from~\eqref{ewofDi}
that the fixed points of $\mathrm{D}_i$ are transversally hyperbolic sinks.
Using the monotone function~\eqref{monN} and the simple structure of the flow
on $\mathscr{D}_i$ it is not difficult to show that
$\omega(\gamma) \in \mathrm{D}_i$ for all orbits $\gamma\subset \mathscr{C}_i\backslash \mathrm{D}_i$.
\end{Remark}

\section{Local dynamics}
\label{local}

In this section we perform a stability analysis of the fixed points in the (closure of
the) state space $\overline{\mathcal{X}}$.
This local dynamical systems analysis is an essential ingredient
in order to understand the global dynamics of solutions in $\mathcal{X}$.

\subsubsection*{Fixed points on the boundary $\bm{\partial\mathcal{X}}$}

In the previous section we have analyzed in detail the
dynamical system~\eqref{dynamicalsystem} on $\partial\mathcal{X}$.
The fixed points of this system reside on the cylindrical boundary 
$\overline{\mathscr{C}}_1 \cup 
\overline{\mathscr{C}}_2 \cup \overline{\mathscr{C}}_3 \subset \partial\mathcal{X}$.
Since $\overline{\mathscr{C}}_i$ is given as $s_i = 0$, 
the direction orthogonal to $\overline{\mathscr{C}}_i$ is given by the variable $s_i$.
From~\eqref{dynamicalsystem} we obtain 
\begin{equation}
s_i^{-1} s_i^\prime\,\big|_{s_i = 0} = -2 \left(\Sigma_i - s \Sigma_j - (1-s) \Sigma_k \right)\:,
\end{equation}
where we again use the convention $(s_j, s_k) = (s, 1-s)$.
Evaluated at the fixed points we get
\begin{subequations}\label{fixoffbound}
\begin{alignat}{4}
& \mathrm{TL}_i :\: & & s_i^{-1} s_i^\prime\,\big|_{s_i = 0} = -6\,,
& \qquad\qquad\quad
& \mathrm{QL}_i :\: & & s_i^{-1} s_i^\prime\,\big|_{s_i = 0} = 6 \,,\\[0.3ex]
& \mathrm{D}_i: \: & & s_i^{-1} s_i^\prime\,\big|_{s_i = 0} = 3 \beta\,, & & & & \\[0.3ex]
\label{RjkKCjk}
& \mathrm{R}_{j}/\mathrm{R}_{k}:\:\: & & s_i^{-1} s_i^\prime\,\big|_{s_i = 0} = 6 \beta\,,
& \qquad
& \mathrm{KC}_{j}/\mathrm{KC}_k :\:\: & & s_i^{-1} s_i^\prime\,\big|_{s_i = 0} = -2 (\Sigma_i - \Sigma_{j/k})\:,
\end{alignat}
\end{subequations}
where $\beta$ is given by~\eqref{betadef}; $\beta$ encodes the dependence
on the rescaled matter anisotropies, which are represented by $v_\pm$, cf.~\eqref{wijkv}.
The results of~\eqref{fixoffbound} can be combined with the results of Section~\ref{boundaries} 
to complete the stability analysis of the fixed points on the boundary
of the state space.
The results are summarized in Table~\ref{tab1}, Table~\ref{tab2}, and Table~\ref{tab3}.

\begin{Example} 
Take, for instance, the fixed point $\mathrm{R}_k$ in the case \Aplus:
From Figure~\ref{case3} we see that $\mathrm{R}_k$ is a sink on
the base of the cylinder $\overline{\mathscr{C}}_i$; see also Figure~\ref{cylApl}.
Figure~\ref{Dc2} shows that there exists exactly one orbit in $\mathscr{D}_i$
(and thus in $\mathscr{C}_i$) that emerges from $\mathrm{R}_k$.
Together with~\eqref{RjkKCjk} it thus follows that $\mathrm{R}_k$ possesses
a two-dimensional unstable manifold (which lies in the interior of the state space 
$\mathcal{X}$), so that there exists a one-parameter family of orbits (in $\mathcal{X}$)
that converge to $\mathrm{R}_k$ as $\tau\rightarrow -\infty$; cf.~Table~\ref{tab1}.
\end{Example}

\begin{table}
\begin{center}
\begin{tabular}{|c|ccc|}
\hline  & & & \\[-2ex]
$\text{point} \in \overline{\mathscr{C}}_i \subset \partial\mathcal{X}$ & \Aplus & \Bplus & \Cplus  \\ \hline & & & \\[-2ex] 
$\text{point} \in \mathrm{QL}_i$ & 1-parameter family & 1-parameter family & 1-parameter family \\
$d+1$ points $\in \mathrm{D}_i$ & one orbit & one orbit & one orbit \\
$d$ points $\in \mathrm{D}_i$ &  1-parameter family & 1-parameter family & 1-parameter family \\
$\mathrm{R}_{j}/\mathrm{R}_k$ &  1-parameter family & part of $\partial\mathscr{D}_i$ & --- \\
point $\mathrm{P} \in \mathrm{KC}_j/\mathrm{KC}_k$ &  2-parameter family & 2-parameter family & 2-parameter family \\
\protect[condition on $\mathrm{P}$] &  [$1 < \Sigma_{j/k}|_{\mathrm{P}}$]
& [$1 < \Sigma_{j/k}|_{\mathrm{P}} < 2$] & [$1< \Sigma_{j/k}|_{\mathrm{P}} < 2/\beta$]  \\ 
$\text{point}\in \mathrm{TL}_i$ & \multicolumn{3}{c|}{does not attract any interior orbit as $\tau\rightarrow-\infty$} \\ 
\hline 
\end{tabular}
\caption{In the $\boldsymbol{+}$ cases, the fixed points on $\partial\mathcal{X}$ are 
$\alpha$-limits for orbits of
the interior of the state space $\mathcal{X}$; in this table we list how many interior orbits
converge to a particular fixed point as $\tau\rightarrow -\infty$.
The fixed points on the Kasner circles that satisfy the condition in the table
are transversally hyperbolic sources and thus act as the $\alpha$-limit for a 
two-parameter set of orbits each. (The points that do not fulfill the condition do not attract any orbits.)
The fixed points on $\mathrm{QL}_i$ (where the end points $\mathrm{Q}_{ij}$ and 
$\mathrm{Q}_{ik}$ are excluded) act as the $\alpha$-limit for
a one-parameter set of interior orbits; hence, in total, a two-parameter 
set of orbits has an $\alpha$-limit on $\mathrm{QL}_i$.
In case \Bplus, the fixed points $\mathrm{R}_j/\mathrm{R}_k$ (which coincide with $\mathrm{T}_{jj}/\mathrm{T}_{kk}$) 
lie on $\partial\mathscr{D}_i$,
which is a heteroclinic cycle. The collection of these cycles, 
$\partial\mathscr{D}_1 \cup \partial\mathscr{D}_2 \cup \partial\mathscr{D}_3$,
forms a so-called heteroclinic network.}
\label{tab1}
\end{center}
\end{table}

\begin{table}
\begin{center}
\begin{tabular}{|c|c|ccc|}
\hline &  & & & \\[-2ex]
$\text{point} \in \overline{\mathscr{C}}_i \subset \partial\mathcal{X}$ & & \Aminus & \Bminus & \Cminus  \\ \hline & & & & \\[-2ex] 
$\text{point} \in \mathrm{QL}_i$ & $\bm{\alpha}$ & 1-parameter family & 1-parameter family & 1-parameter family \\
point $\mathrm{P} \in \mathrm{KC}_j/\mathrm{KC}_k$ & $\bm{\alpha}$ & 
2-parameter family & 2-parameter family & 2-parameter family \\
\protect[condition on $\mathrm{P}$] &  &  [$1 < \Sigma_{j/k}|_{\mathrm{P}}$] 
& [$1< \Sigma_{j/k}|_{\mathrm{P}}$]  & [$1 < \Sigma_{j/k}|_{\mathrm{P}}$] \\[0.3ex]
$\text{point}\in \mathrm{TL}_i$ &  & \multicolumn{3}{c|}{does not attract any interior orbit as $\tau\rightarrow-\infty$ or $\tau\rightarrow \infty$} \\
$d+1$ points $\in \mathrm{D}_i$ & $\bm{\omega}$ & 2-parameter family & 2-parameter family & 2-parameter family \\
$d$ points $\in \mathrm{D}_i$ &   $\bm{\omega}$ & 3-parameter family & 3-parameter family & 3-parameter family \\
$\mathrm{R}_{j}/\mathrm{R}_k$ & $\bm{\omega}$ & 3-parameter family & 3-parameter family & --- \\
point $\mathrm{P} \in \mathrm{KC}_j/\mathrm{KC}_k$ &  $\bm{\omega}$ & \multicolumn{2}{c}{does not act as $\omega$-limit} & 2-parameter family \\
\protect[condition on $\mathrm{P}$] &  & \multicolumn{2}{c}{[---]}  & [$\Sigma_{j/k}|_{\mathrm{P}} < 2/\beta$]  \\ \hline 
\end{tabular}
\caption{In the $\boldsymbol{-}$ cases, some fixed points on $\partial\mathcal{X}$ are $\alpha$-limits, some are
$\omega$-limits for orbits of the interior of the state space $\mathcal{X}$; in this table we list how 
many interior orbits converge to a particular fixed point as $\tau\rightarrow -\infty$ (denoted by $\bm{\alpha}$)
or $\tau\rightarrow\infty$ (denoted by $\bm{\omega}$).
In case \Bminus, the fixed points $\mathrm{R}_j/\mathrm{R}_k$ coincide with $\mathrm{Q}_{jj}/\mathrm{Q}_{kk}$.
The proof that these points attract a three-parameter set of orbits requires center manifold analysis.}
\label{tab2}
\end{center}
\end{table}

\begin{table}
\begin{center}
\begin{tabular}{|c|c|cc|}
\hline &  & &  \\[-2ex]
$\text{point} \in \overline{\mathscr{C}}_i \subset \partial\mathcal{X}$ & & \Dplus & \Dminus  \\ \hline & & & \\[-2ex] 
$\text{point}\in \mathrm{QL}_i$ & $\bm{\alpha}$  & not attractive & 1-parameter family \\
point $\mathrm{P} \in \mathrm{KC}_j/\mathrm{KC}_k$ & $\bm{\alpha}$ &  not attractive
& 2-parameter family \\
\protect[condition on $\mathrm{P}$] &
& [---]  & [$1 < \Sigma_{j/k}|_{\mathrm{P}}$] \\[0.3ex]
$\text{point} \in \mathrm{TL}_i$ & $\bm{\omega}$ &  not attractive & 1-parameter family \\
point $\mathrm{P} \in \mathrm{KC}_j/\mathrm{KC}_k$ &  $\bm{\omega}$ &  not attractive & 2-parameter family  \\
\protect[condition on $\mathrm{P}$] &  & [---] & [$\Sigma_{j/k}|_{\mathrm{P}} < -1$] \\ \hline 
\end{tabular}
\caption{If $|\beta| \geq 2$, the properties of the fixed points change considerably.
If $\beta \geq 2$ (\Dplus), neither of the fixed points can be the $\alpha$/$\omega$-limit set of an interior orbit.
(However, a fixed point can be \textit{contained} in the $\alpha$-limit
set of some orbit $\gamma \subseteq \mathcal{X}$.)
If $\beta \leq -2$ (\Dminus), then each fixed point $\mathrm{P}$ on
$\mathrm{KC}_j/\mathrm{KC}_k$ with $1 < \Sigma_{j/k}|_{\mathrm{P}}$
is a transversally hyperbolic source, while $\Sigma_{j/k}|_{\mathrm{P}} < -1$
yields a transversally hyperbolic sink.}
\label{tab3}
\end{center}
\end{table}

\subsubsection*{Fixed points in the interior of $\bm{\mathcal{X}}$}

The dynamical system~\eqref{dynamicalsystem} possesses 
a set of fixed points in the interior of $\mathcal{X}$ ($= \mathscr{K} \times \mathscr{T}$),
which is given as the set of solutions of the equations
\begin{equation}\label{Fdef}
\Sigma_1=\Sigma_2=\Sigma_3=0 \:, \qquad  w_1=w_2=w_3= w\:;
\end{equation}
the first condition defines a point in $\mathscr{K}$; in general,
the second condition defines a subset of $\mathscr{T}$.
Since the principal pressures coincide,
each of these fixed points represents the flat isotropic 
Friedmann-Robertson-Walker (FRW) perfect fluid solution 
associated with the equation of state $p = w \rho$. 

\begin{Assumption}\label{F}
For simplicity, we assume that the matter model is such
that $w_1 = w_2 = w_3 = w$ defines a \textit{point} in $\mathscr{T}$;
hence~\eqref{Fdef}
defines an isolated fixed point in $\mathcal{X}$,
which we denote by $\mathrm{F}$.
\end{Assumption}

\begin{Remark}
This is an assumption out of convenience. (It is automatically satisfied
for the matter models we consider in detail 
in Section~\ref{mattermodels}---collisionless matter and elastic matter.)
The analysis of this section 
is equally valid when~\eqref{Fdef} has more than one solution.
In that case, $\mathrm{F}$ denotes any of the multiple fixed points in $\mathcal{X}$.
\end{Remark}

To study the stability properties of $\mathrm{F}$,
it is useful to define a different set of coordinates on $\mathscr{T}$.
Let $(ijk)$ be a cyclic permutation of the triple $(123)$
and set
\begin{equation}\label{stcoords}
s_i = \frac{e^{t_i}}{1 + e^{t_i} + e^{t_j}} \:,\qquad
s_j = \frac{e^{t_j}}{1 + e^{t_i} + e^{t_j}} \:,\qquad
s_k = \frac{1}{1 + e^{t_i} + e^{t_j}} \:.
\end{equation}
This defines a coordinate system $\mathbb{R}^2 \ni (t_i,t_j) \mapsto \mathscr{T}$, such that 
the point $(t_i,t_j) = (0,0)$
corresponds to the center $(s_1,s_2,s_3) = (1/3,1/3,1/3)$ of $\mathscr{T}$;
the limit $s_i\rightarrow 0$ corresponds to $t_i\rightarrow -\infty$, $s_j\rightarrow 0$ to 
$t_j\rightarrow -\infty$; the limit $s_i\rightarrow 1$ corresponds to $t_i\rightarrow \infty$,
$s_j\rightarrow 1$ to $t_j\rightarrow \infty$; finally,
$s_k \rightarrow 0$ corresponds to a combined limit 
$t_i\rightarrow \infty$, $t_j\rightarrow \infty$.
In these coordinates the fixed point $\mathrm{F}$ is given by
\begin{equation}\label{Fint}
\mathrm{F}: \:(\Sigma_1,\Sigma_2,\Sigma_3)_{|\mathrm{F}} = (0,0,0), \;(t_i,t_j)_{|\mathrm{F}} := (\bar{t}_i,\bar{t}_j)\:.
\end{equation}
Using $(t_i,t_j)$ on $\mathscr{T}$, 
the rescaled pressures~\eqref{wiinpsi} can be expressed as
\begin{equation}\label{wiinpsi2}
w_i = w + 2\, \frac{\partial}{\partial t_i}\log\psi \:,
\qquad
w_j = w + 2\, \frac{\partial}{\partial t_j}\log\psi \:,
\qquad
w_k = w - 2 \left(\frac{\partial}{\partial t_i} +  \frac{\partial}{\partial t_j}\right)\log\psi\:,
\end{equation}
where $\psi$ is regarded as a function of the two variables $(t_i, t_j)$.
Since $w_1 = w_2 =w_3$ at $\mathrm{F}$, it follows that 
$(\bar{t}_i,\bar{t}_j)\in\mathscr{T}$ is a critical point of $\psi$;
we set $\bar{\psi} = \psi(\bar{t}_i,\bar{t}_j)$.

The dynamical system~\eqref{dynamicalsystem} can be represented in the four 
variables $(\Sigma_i,\Sigma_j,t_i,t_j)$ as
\begin{equation*}
\Sigma_n^\prime = -3\Omega\left[\frac{1}{2}(1-w)\Sigma_n-(w_n -w )\right],\:\quad
t_n^\prime = -2(2\Sigma_n+\Sigma_m)\,,\qquad
(n,m) \in \big\{(i,j),(j,i)\big\}\:,
\end{equation*}
where $\Omega = 1- \frac{1}{3}(\Sigma_i^2+\Sigma_j^2+\Sigma_i\Sigma_j)$.
The matrix representing the linearization of~\eqref{dynamicalsystem} at the point $\mathrm{F}$ 
contains the Hessian $\mathfrak{h}$ of $\psi$ evaluated at $(\bar{t}_i,\bar{t}_j)$.
The four eigenvalues of this matrix are given by
\begin{equation*}
\lambda_{\pm\pm} = \frac{3}{4}(1-w)\left[-1\pm\sqrt{1-\Lambda_\pm}\right]\:,
\end{equation*}
where $\Lambda_{\pm}$ are given by
$\Lambda_{\pm}=64\,\bar{\psi}\left[\tr\mathfrak{h}+\mathfrak{h}_{ij}\pm\sqrt{(\tr\mathfrak{h}+\mathfrak{h}_{ij})^2-3\det\mathfrak{h}}\right]/[3 (1-w)^2]$.

\begin{Lemma}\label{stabilityF}
If \textnormal{(i)} $\mathfrak{h}$ is positive definite, i.e., $\det\mathfrak{h}>0$ and $\tr\mathfrak{h}>0$, then
all eigenvalues $\lambda_{\pm\pm}$ have negative real part and\/ $\mathrm{F}$ is a hyperbolic sink. 
If \textnormal{(ii)} $\mathfrak{h}$ is negative definite, i.e., $\det\mathfrak{h}>0$ and $\tr\mathfrak{h}< 0$,
then the eigenvalues $\lambda_{+\pm}$ are positive and $\lambda_{-\pm}$ negative so that\/ $\mathrm{F}$ 
is a hyperbolic saddle with a two-dimensional stable and a two-dimensional unstable manifold. 
If \textnormal{(iii)} $\det\mathfrak{h}<0$, then $\lambda_{\pm+}$ and $\lambda_{--}$
have negative real part, whereas $\lambda_{+-}$ has positive real part, 
so that\/ $\mathrm{F}$ is a hyperbolic saddle with a three-dimensional stable and one-dimensional unstable manifold. 
In the exceptional case \textnormal{(iv)} $\det\mathfrak{h}=0$, there exists at least one zero eigenvalue.
\end{Lemma}

\begin{proof}
From the inequality
\begin{equation}\label{inequalitydet}
\det\mathfrak{h}\leq\frac{1}{4}\tr\mathfrak{h}^2-\mathfrak{h}_{ij}^2\:, 
\end{equation}
we obtain $(\tr\mathfrak{h}+\mathfrak{h}_{ij})^2-3\det\mathfrak{h}\geq 0$, 
hence $\Lambda_\pm$ are real numbers. Studying the sign of $\Lambda_\pm$ in the different cases, 
the claim of the lemma follows immediately.
\end{proof}

\section{Global dynamics}
\label{global}

On the state space $\mathcal{X}$ we define the positive function 
\begin{equation}\label{Mdef}
M = (1-\Sigma^2)^{-1}\, \psi(s_1,s_2,s_3) \:,
\end{equation}
where $\psi(s_1,s_2,s_3)$, $(s_1,s_2,s_3) \in \mathscr{T}$, 
is defined by~\eqref{rhopi}. 
By a straightforward computation, where we use that 
$M$ can be written as $M=3 H^2/n^{1+w}$ (which makes it possible to apply~\eqref{decoupled}
and the equation $n' = -3 n$, which follows from $d\sqrt{\det g}/d t =3 H \sqrt{\det g}$), we obtain
\begin{subequations}\label{M1properties}
\begin{equation}\label{M1prime}
M^\prime = -3 (1 - w) \Sigma^2 M\:.
\end{equation}
The computation of higher derivatives reveals that
\begin{equation}
M^{\prime\prime\prime} \Big|_{\Sigma^2 = 0} = -9 (1-w) M \sum_k (w_k-w)^2
\end{equation}
\end{subequations}
on the subset $\Sigma^2 =0$ of the state space.
Therefore, since $w < 1$ by Assumption~\ref{assumptionwi}, 
we obtain $M^\prime < 0$ when $\Sigma^2 \neq 0$
and $M^{\prime\prime\prime}|_{\Sigma^2 = 0} < 0$ 
except at the point $\mathrm{F}$, cf.~Assumption~\ref{F}.
Hence we have proved
that 
$M$ is a strictly monotonically decreasing function 
along the flow of the dynamical system~\eqref{dynamicalsystem} on $\mathcal{X} \backslash \mathrm{F}$. 

\begin{Lemma}\label{limitonboundary}
Let $\gamma$ be an orbit in $\mathcal{X}\backslash \mathrm{F}$.
Then the $\alpha$-limit set $\alpha(\gamma)$ and the $\omega$-limit set $\omega(\gamma)$
satisfy 
\begin{itemize}
\item 
either $\alpha(\gamma) = \mathrm{F}$ or $\alpha(\gamma) \subseteq \partial\mathcal{X}$, and
\item
either $\omega(\gamma) = \mathrm{F}$ or $\omega(\gamma) \subseteq \overline{\mathscr{C}}_1 \cup 
\overline{\mathscr{C}}_2 \cup \overline{\mathscr{C}}_3 \subset \partial\mathcal{X}$, cf.~\eqref{cylbou}.
\end{itemize}
\end{Lemma}

\begin{proof}
The lemma is a direct consequence of the monotonicity principle. 
Since $M$ is strictly monotone along the flow of the dynamical system~\eqref{dynamicalsystem} 
on $\mathcal{X}\backslash \mathrm{F}$, the $\alpha$/$\omega$-limit set of $\gamma$
must be contained in $\mathrm{F} \cup \partial \mathcal{X}$. 
Because these limit sets are necessarily connected, we find that 
either $\alpha(\gamma) = \mathrm{F}$ or $\alpha(\gamma) \subseteq \partial \mathcal{X}$,
and likewise for $\omega(\gamma)$. Moreover, 
$M|_{\partial\mathscr{K} \times \mathscr{T}} = \infty$ (i.e.,
$M \rightarrow \infty$  
along every sequence converging to a point on $\partial\mathscr{K} \times \mathscr{T}$),
hence this subset of $\partial\mathcal{X}$ is excluded for $\omega(\gamma)$,
which leaves 
$\omega(\gamma) \subseteq \overline{\mathscr{K}}\times \partial\mathscr{T} = \overline{\mathscr{C}}_1 \cup 
\overline{\mathscr{C}}_2 \cup \overline{\mathscr{C}}_3$.
\end{proof}

It is the quantity $\beta$, see~\eqref{betadef}, that determines the
details of the asymptotics. 
Therefore we discuss the cases $\beta >0$
and $\beta < 0$ separately; the former 
includes the $\boldsymbol{+}$ cases \Aplus, \Bplus, \Cplus, and the extreme case \Dplus, 
the latter includes the $\boldsymbol{-}$ 
cases \Aminus, \Bminus, \Cminus, and the extreme case \Dminus. 

\subsection*{The $\boldsymbol{+}$ cases}

\begin{Theorem}[\textbf{Future asymptotics}]\label{futurethm}
In the $\boldsymbol{+}$ cases\/ \textnormal{\Aplus, \Bplus, \Cplus}, and \textnormal{\Dplus}, 
the fixed point\/ $\mathrm{F}$ is the future attractor of the dynamical system~\eqref{dynamicalsystem};
every orbit in the state space $\mathcal{X}$ converges to the fixed point\/ $\mathrm{F}$
as $\tau\rightarrow +\infty$.
\end{Theorem}

\textit{Interpretation of Theorem~\ref{futurethm}}.
Since the fixed point $\mathrm{F}$ corresponds to a FRW perfect fluid solution
associated with the equation of state $p = w \rho$,
the theorem states that each Bianchi type~I model with anisotropic 
matter that satisfies $v_- < w < v_+$, cf.~\eqref{vwv},
isotropizes toward the future and behaves, to first order, 
like an (infinitely diluted) isotropic perfect fluid solution.

\begin{Remark}
Theorem~\ref{futurethm} can be proved under assumptions that are considerably 
weaker than the ones we made. (Assumptions~\ref{a7}--\ref{vassum} are rather irrelevant
for the future asymptotics---they are tailored to
the past asymptotics.)
This fact will become obvious from the proof and is thus not further commented on.
\end{Remark}

\begin{proof}
To prove Theorem~\ref{futurethm} we must show that 
$\omega(\gamma) \subseteq \overline{\mathscr{C}}_1 \cup \overline{\mathscr{C}}_2 \cup \overline{\mathscr{C}}_3$ 
is impossible, see Lemma~\ref{limitonboundary}.
To this end we use that $\psi(s_1,s_2,s_3)\rightarrow \infty$ as $(s_1,s_2,s_3) \rightarrow \partial\mathscr{T}$
(which we prove below).
It then follows that $M = \sup_{\mathcal{X}} M = \infty$ 
on the cylindrical boundary and consequently on the 
entire boundary $\partial\mathcal{X}$ of the state space, cf.~\eqref{Mdef}.
To finish the proof we apply 
the monotonicity principle~\cite{WE} for the function $M$ on $\mathcal{X}\backslash \{F\}$:
The monotonicity principle
precludes the possibility that $\omega(\gamma)$ 
might be contained on $\partial\mathcal{X}$ for any orbit $\gamma \subseteq \mathcal{X}$
and the theorem is established.
To show that $\psi\rightarrow \infty$ as $\partial\mathscr{T}$ is approached,
we use that $v_- < w < v_+$, which holds by assumption, cf.~\eqref{vwv}.
Employing the coordinates~\eqref{stcoords} on $\mathscr{T}$ and the
representation~\eqref{wiinpsi2} of the rescaled matter quantities,
we find that there exists some $\epsilon > 0$ such that 
for all $t_j$ there exists $T>0$ and
\[
\frac{\partial}{\partial t_i} \log \psi = \frac{w_i -w}{2} \leq -\epsilon\,,\qquad
\frac{\partial}{\partial t_i} \log \psi = \frac{w_i -w}{2} \geq \epsilon
\]
for all $t_i < -|T|$ and $t_i > T$ respectively.
Consequently, $\log\psi \rightarrow \infty$ and thus $\psi\rightarrow \infty$
as $t_i \rightarrow \pm\infty$.
(Likewise, $\psi\rightarrow \infty$ as $t_j \rightarrow \pm\infty$.)
We thus conclude that $\psi = \infty$ on $\partial\mathscr{T}$ and the
theorem is proved.
\end{proof}

\begin{Theorem}[\textbf{Past asymptotics}]\label{pastthm1}
Let $\gamma\subset \mathcal{X}\backslash\mathrm{F}$ be an orbit of the dynamical system~\eqref{dynamicalsystem}.
\vspace{-2ex}
\begin{itemize}
\item
In the $\boldsymbol{+}$ cases\/ \textnormal{\Aplus, \Bplus, \Cplus},
the $\alpha$-limit set of $\gamma$ is 
\begin{itemize}
\item[\textnormal{\Aplus}]
one of the fixed points
on 
$\partial\mathcal{X}$;
\item[\textnormal{\Bplus}]
one of the fixed points
on 
$\partial\mathcal{X}$ or, possibly, 
the heteroclinic network $\partial\mathscr{D}_1 \cup\partial\mathscr{D}_2\cup\partial\mathscr{D}_3$;
\item[\textnormal{\Cplus}] one of the fixed points
on 
$\partial\mathcal{X}$
or one of the heteroclinic cycles $\partial\mathscr{D}_1$, $\partial\mathscr{D}_2$, $\partial\mathscr{D}_3$.
\end{itemize}
Convergence to a fixed point is the generic scenario, i.e., the set of orbits that 
converge to the heteroclinic structures is a set of measure zero in $\mathcal{X}$.
Table~\ref{tab1} gives a complete list of the fixed points on $\partial\mathcal{X}$
together with the number of orbits that converge to the particular points.
\item
In the case \textnormal{\Dplus}, 
the $\alpha$-limit set of $\gamma$ is represented either by
a heteroclinic cycle or by a heteroclinic sequence on $\partial\mathcal{X}$ and thus contains
a sequence of Kasner points;
in particular, $\alpha(\gamma)$ cannot be a fixed point.
\end{itemize}
\end{Theorem}

\textit{Interpretation of Theorem~\ref{pastthm1} and conclusions}.
Theorem~\ref{pastthm1} (in conjunction with the results of
Table~\ref{tab1}) entails that
in each of the $\boldsymbol{+}$ cases, \textit{generic} solutions approach the fixed points
on the Kasner circle(s), 
which implies that the past attractor
of the dynamical system~\eqref{dynamicalsystem} is located on the Kasner circle(s).
Therefore, each generic Bianchi type~I model with
anisotropic matter, where the matter is supposed to satisfy the requirements of the $\boldsymbol{+}$ 
cases, approaches a Kasner solution toward the singularity.
However, there are interesting differences among the cases.
In case \Aplus\ each Kasner solution is a possible past asymptotic state, 
i.e., given an arbitrary Kasner solution~\eqref{Kasnersol}, represented by the Kasner exponents 
$(p_1,p_2,p_3) = (1/3)(\Sigma_1+1,\Sigma_2+1,\Sigma_3+1)$, $p_1 + p_2 + p_3 = p_1^2+p_2^2+p_3^2 = 1$,
there exist anisotropic Bianchi type~I solutions that converge to this Kasner solution as $t\rightarrow 0$.
In particular, each of the three Taub solutions, which are characterized by $(p_i,p_j,p_k) = (1,0,0)$, 
represents a possible past asymptotic state.
In case \Bplus\ this is no longer correct. Since the Taub points are (center) saddles there
does not exist any solution that converges to a Taub solution as $t\rightarrow 0$. However, the Taub points
are part of a larger structure, the heteroclinic network
$\partial\mathscr{D}_1 \cup \partial\mathscr{D}_2 \cup \partial\mathscr{D}_3$,
which represents a potential $\alpha$-limit set for (a probably two-parameter set)
of solutions. Accordingly, we conjecture that there exists a two-parameter set of 
solutions whose asymptotic behaviour is characterized by oscillations between
different representations of the Taub solution.
In case \Cplus\ (which is characterized by $1 < \beta < 2$) 
there is an entire
one-parameter set of Kasner solutions (including the Taub states) 
that are excluded as $\alpha$-limit sets.
The condition of Table~\ref{tab1} yields that a Kasner solution is 
excluded as a past asymptotic state for Bianchi type~I models with anisotropic matter iff
$\max_l \Sigma_l \geq 2/\beta$, or, equivalently, $\max_l p_l \geq (2+\beta)/(3 \beta)$.
In other words, generic Bianchi type~I models behave asymptotically like 
Kasner solutions as the singularity is approached; however, the set of possible
Kasner limits is restricted to those that are ``sufficiently different'' from the Taub solutions.
If $\beta \geq 2$, i.e., in case \Dplus, this set is the empty set. Each point on the Kasner circles
takes the role of a saddle point. This induces generic \textit{oscillatory asymptotic behaviour}
of Bianchi type~I models. The $\alpha$-limit of a generic solution contains
a sequence of Kasner states, hence the solution
undergoes a sequence of phases (``Kasner epochs''), 
in each of which the behaviour of the solution is approximately described by
a Kasner solution, and ``transitions'' between these epochs.

\begin{Remark}
Theorem~\ref{pastthm1} (in conjunction with the results of
Table~\ref{tab1}) also implies that there exist (non-generic) Bianchi type~I models
whose asymptotic behaviour is not connected to the dynamics of any Kasner solution.
There exist solutions whose asymptotic behaviour is characterized
by $(\Sigma_i,\Sigma_j,\Sigma_k) \rightarrow \beta (-1,1/2,1/2)$ as the 
singularity is approached,
and solutions with 
$(\Sigma_i,\Sigma_j,\Sigma_k) \rightarrow \beta (2,-1,-1)$, where the
latter concerns the case \Aplus\ since $\beta < 1$ is required.
Furthermore, in the cases \Bplus\ and \Cplus, there exist solutions
whose asymptotic behaviour is oscillatory: An orbit approaching 
the heteroclinic cycles/network 
($\partial\mathscr{D}_1$, $\partial\mathscr{D}_2$, $\partial\mathscr{D}_3$)
corresponds to a solution that is represented by 
a periodic sequence of ``Kasner epochs''
associated with the Kasner fixed points that lie on the cycles;
in the Kasner epochs the solution is characterized by 
$(\Sigma_i,\Sigma_j,\Sigma_k) = (-\beta,\beta/2 \pm \sqrt{3/4} \sqrt{4-\beta^2}, \beta/2 \mp \sqrt{3/4} \sqrt{4-\beta^2})$.
\end{Remark}

\begin{Remark}
The distinction into the different $\boldsymbol{+}$ scenarios is 
intimately connected with the energy conditions; see the list of cases in Section~\ref{boundaries}.
If the energy conditions hold (case \Aplus, $\beta < 1$), each Kasner solution represents a possible past asymptotic state for
anisotropic Bianchi type~I models. If the energy conditions are satisfied only marginally
(case \Bplus, $\beta =1$), i.e., at the onset of energy condition violation, the Taub solutions
are excluded as past asymptotic states.
Finally, in the case $\beta > 1$ (case \Cplus\ and case \Dplus) 
the energy conditions are violated;
in this context, the quantity $\beta - 1$ can be regarded as a measure for the magnitude of the violation.
Increasing the value of $\beta-1$ from zero to one leads to an ever increasing exclusion of
Kasner states from the past attractor; 
if $\beta-1$ reaches one ($\beta =2$), 
the set of Kasner states that represent past asymptotic states has shrunk to the empty set.
This is the onset of oscillatory behaviour of generic solutions.
\end{Remark}

\begin{proof}
In the proof we choose to restrict ourselves to the cases \Aplus, \Bplus, \Cplus, since these
are the difficult cases; the proof in the case \Dplus\ 
is similar (but simpler).
The first step to prove Theorem~\ref{pastthm1} is to show that
$\alpha(\gamma)$ must be contained in $\partial\mathcal{X}$.
Using Lemma~\ref{limitonboundary} this amounts to proving
that $\alpha(\gamma) = \mathrm{F}$ is impossible.
Therefore, assume that there exists $\gamma \subset \mathcal{X}\backslash\mathrm{F}$
such that $\alpha(\gamma) = \mathrm{F}$.
By Theorem~\ref{futurethm} we know that $\omega(\gamma) = \mathrm{F}$, hence
$\gamma$ must be a homoclinic orbit. However, this a contradiction to
the fact that the function $M$, which is well-defined in $\mathcal{X}$ and in
particular at $\mathrm{F}$, cf.~\eqref{Mdef}, is 
strictly monotonically decreasing along $\gamma$. 
Consequently, $\alpha(\gamma) = \mathrm{F}$ is impossible, and we conclude
$\alpha(\gamma) \subseteq \partial\mathcal{X}$.

In the next step of the proof we show that $\alpha(\gamma)$ is contained
in a subset of the boundary $\partial\mathcal{X}$: 
$\alpha(\gamma) \subseteq \big( \partial\mathscr{C}_1 \cup \partial\mathscr{C}_2 \cup \partial\mathscr{C}_3 \big) \cup 
\big( \mathrm{D}_1 \cup \mathrm{D}_2 \cup \mathrm{D}_3 \big)$.
By definition, $\partial \mathcal{X}$ is represented by the disjoint union
\begin{equation*}
\partial \mathcal{X} = 
\underbrace{%
\big( \partial\mathscr{C}_1 \cup \partial\mathscr{C}_2 \cup \partial\mathscr{C}_3 \big)}_{[\partial\mathcal{X}]_1}
\cup 
\underbrace{%
\big( \mathrm{D}_1 \cup \mathrm{D}_2 \cup \mathrm{D}_3 \big)}_{[\partial\mathcal{X}]_2}
\cup
\Big( \underbrace{%
\big(\mathscr{C}_1 \cup \mathscr{C}_2 \cup \mathscr{C}_3\big) \backslash
\big( \mathrm{D}_1 \cup \mathrm{D}_2 \cup \mathrm{D}_3 \big)}_{[\partial\mathcal{X}]_3} \Big) 
\cup
\underbrace{\big(\partial\mathscr{K} \times \mathscr{T} \big)}_{[\partial\mathcal{X}]_4}\:,
\end{equation*}
see Section~\ref{boundaries}.
Suppose that there exists an orbit $\gamma$ such that $\alpha(\gamma) \ni \mathrm{P} \in [\partial\mathcal{X}]_3$;
w.l.o.g.\ $\mathrm{P} \in \mathscr{C}_i$ ($\mathrm{P} \not\in\mathrm{D}_i$) for some $i$.
Since the $\alpha$-limit set $\alpha(\gamma)$ is invariant under the flow of the dynamical system, 
the entire orbit through $\mathrm{P}$, $\gamma_{\mathrm{P}}$, and its $\omega$-limit  
must be contained in $\alpha(\gamma)$ as well. By Lemma~\ref{DLemma}, $\omega(\gamma_{\mathrm{P}})$
is a point $\mathrm{P}_{\mathrm{D}_i}$ of the set $\mathrm{D}_i$.
However, $\mathrm{P}_{\mathrm{D}_i}$ is a saddle in $\mathcal{X}$, see Table~\ref{tab1}; 
therefore, since $\mathrm{P}_{\mathrm{D}_i}\in \alpha(\gamma)$, 
either $\gamma$ is contained in the unstable manifold of $\mathrm{P}_{\mathrm{D}_i}$, 
or the intersection of $\gamma$ with the unstable manifold is empty.
The first alternative contradicts the assumption $\mathrm{P} \in \alpha(\gamma)$.
In the second case we appeal to the Hartman-Grobman theorem 
to infer that $\alpha(\gamma)$ contains not only $\mathrm{P}_{\mathrm{D}_i}$ 
but also points of the unstable manifold of $\mathrm{P}_{\mathrm{D}_i}$.
(Consider a sequence $\tau_n$ such that $\tau_n \rightarrow -\infty$
and $\gamma(\tau_n) \rightarrow \mathrm{P}_{\mathrm{D}_i}$ as $n\rightarrow \infty$.
By the Hartman-Grobman theorem there exists 
a sequence $\bar{\tau}_n = \tau_n + \delta\tau_n$, with $\delta\tau_n > 0$, $\bar{\tau}_n  < \tau_{n-1}$,
such that the sequence $\gamma(\bar{\tau}_n)$ possesses an accumulation point on the
unstable manifold of $\mathrm{P}_{\mathrm{D}_i}$.)
The local dynamical systems analysis, cf.~Table~\ref{tab1},
thus implies that  $\alpha(\gamma)$ contains
points that do not lie on $\mathscr{C}_i$ but in the interior of $\mathcal{X}$.
This, however, 
is a contradiction to Lemma~\ref{limitonboundary}.
The assumption has led to a contradiction; therefore, the intersection
of $\alpha(\gamma)$ and $[\partial\mathcal{X}]_3$ must be empty.
The procedure to show that $\alpha(\gamma) \cap [\partial\mathcal{X}]_4 = \emptyset$ 
is identical, where we use the flow on $\partial\mathscr{K} \times \mathscr{T}$,
see Figure~\ref{vacuumflow}, and the stability analysis of the fixed points
as summarized in Table~\ref{tab1}.
We have therefore shown that
$\alpha(\gamma) \subseteq \big( \partial\mathscr{C}_1 \cup \partial\mathscr{C}_2 \cup \partial\mathscr{C}_3 \big) \cup 
\big( \mathrm{D}_1 \cup \mathrm{D}_2 \cup \mathrm{D}_3 \big)$.

In the last step of the proof we investigate which structures on 
the set 
$\partial\mathscr{C}_1 \cup \partial\mathscr{C}_2 \cup \partial\mathscr{C}_3$ are 
potential $\alpha$-limit sets. 
Since this set is two-dimensional, possible $\alpha$-limit sets are fixed points,
periodic orbits, homoclinic orbits, or heteroclinic cycles/networks.
The analysis of Section~\ref{boundaries}, see Figures~\ref{cylinder},~\ref{base}, and Figure~\ref{cylinder2},
shows that, in case \Aplus, there do not exist any other potential $\alpha$-limit sets on
$\partial\mathscr{C}_1 \cup \partial\mathscr{C}_2 \cup \partial\mathscr{C}_3$
than fixed points. (This concludes the proof of the theorem for case \Aplus.)
In case \Bplus, there exists an entangled network of heteroclinic cycles, a 
`heteroclinic network'; in case \Cplus, there are three 
independent heteroclinic cycles:
$\partial\mathscr{D}_1$, $\partial\mathscr{D}_2$, $\partial\mathscr{D}_3$ 
To complete the proof of the theorem 
we have to investigate whether $\alpha(\gamma)$ can coincide
with $\partial\mathscr{D}_i$ for some $i$ in case \Cplus\ (where $1<\beta<2$).

Consider the heteroclinic cycle $\partial\mathscr{D}_i$ for some $i$.
At the point $\mathrm{P}$ on $\partial\mathscr{D}_i$ given by
$(s_i, s_j, s_k) = (0,0,1)$, $\Sigma_i + \beta =0$, $\Sigma_j = \Sigma_k$, 
we consider the three-dimensional 
hyperplane $\mathcal{H}_{\mathrm{P}}: \Sigma_j = \Sigma_k$ orthogonal to $\partial\mathscr{D}_i$;
the natural axes on $\mathcal{H}_{\mathrm{P}}$ are $s_i$, $s_j$, and $\Sigma_i + \beta$.
We study the Poincar\'e map induced on $\mathcal{H}_{\mathrm{P}}$ by the dynamical system.
The subspace $s_i = 0$ is an invariant subspace of $\mathcal{H}_{\mathrm{P}}$
(since $\overline{\mathscr{C}}_i$ is an invariant subspace of $\overline{\mathcal{X}}$).
Equation~\eqref{monMi} and Figure~\ref{Dc1} reveal that $\mathrm{P}$ is a saddle point
for the dynamical system on $\mathcal{H}_{\mathrm{P}}$, the unstable manifold being
the $s_j$-axis, the stable manifold being the $(\Sigma_i + \beta)$-axis.

A straightforward calculation shows that
\begin{equation*}
s_i^{-1} s_i^\prime = 3 \beta - \sqrt{3} \sqrt{4 -\beta^2}\, \lambda\:,\qquad (\lambda\in [-1,1])
\end{equation*}
along $\partial\mathscr{D}_i$. Under the assumptions of case \Cplus\ the r.\ h.\ side is strictly 
positive for all $\lambda$, hence, for some $\epsilon >0$,
$s_i^{-1} s_i^\prime \geq  2 \epsilon$ along $\partial\mathscr{D}_i$
and $s_i^{-1} s_i^\prime \geq  \epsilon$ in a sufficiently small neighborhood $\mathscr{U}_i$ of $\partial\mathscr{D}_i$
in $\mathcal{X}$. 
Therefore, as long as a solution is contained in this neighborhood $\mathscr{U}_i$ of $\partial\mathscr{D}_i$,
we find that $s_i(\tau)$ is bounded by some constant times $e^{\epsilon \tau}$ (for decreasing $\tau$).
We infer that a solution either remains within $\mathscr{U}_i$ for all sufficiently small $\tau$  
($\tau\rightarrow -\infty$) or $|\Sigma_i + \beta|$ becomes large (and the solution leaves
$\mathscr{U}_i$ in this way).
Letting $d\varsigma = -(3/2) (1-w) \Omega d\tau$ (so that decreasing $\tau$-time corresponds to increasing
$\varsigma$-time) we find that
\begin{equation}\label{varsigmaODE}
\frac{d}{d\varsigma} \big(\Sigma_i + \beta \big) =  \big(\Sigma_i + \beta\big)  - \frac{2}{1-w} (w_i - v_-) \:,
\end{equation}
where $w_i = w_i(s_1,s_2,s_3)$ satisfies $w_i \rightarrow v_-$ as $s_i \rightarrow 0$.

Let us consider a simplified problem. A differential equation of the type
$f'(\varsigma) =  f(\varsigma) + g(\varsigma)$ with a function $g(\varsigma)$
that converges to zero as $\varsigma\rightarrow \infty$ possesses a unique solution such that
$f(\varsigma) \rightarrow 0$ as $\varsigma \rightarrow \infty$;
if, initially, $f$ is larger than this special solution, 
then $f(\varsigma) \rightarrow \infty$ as $\varsigma\rightarrow \infty$; if,
initially, $f$ is smaller, then $f(\varsigma) \rightarrow -\infty$ as $\varsigma\rightarrow \infty$.

Likewise, we obtain from~\eqref{varsigmaODE} that the quantity $\Sigma_i + \beta$ is
increasing if it is positive and sufficiently large initially,
while it is decreasing if it is negative and sufficiently small initially.
In the former cases, the solution eventually leaves $\mathscr{U}_i$, since
$\Sigma_i + \beta$ has become too large;
in the latter case, the solution leaves $\mathscr{U}_i$, because
$\Sigma_i + \beta$ has become too small.
Consider a one-parameter set of initial data, e.g., $s_i = \mathrm{const}$, $s_j = \mathrm{const}^{\prime}$
in $\mathcal{H}_{\mathrm{P}}$.
Let $(\Sigma_i + \beta)_0$ be an initial value of $\Sigma_i + \beta$ such that 
the associated solution satisfies 
$(\Sigma_i + \beta)(\varsigma) > \delta$ for sufficiently large $\varsigma$ (for some appropriately chosen
$\delta$). By continuous dependence on initial data, there exists an open interval
of initial data containing $(\Sigma_i + \beta)_0$ such that the analog holds for
each initial datum of this interval. Let $(\Sigma_i + \beta)_{\mathrm{pos}}$ denote the infimum
of $(\Sigma_i + \beta)$ of the maximally extended open interval. (This infimum exists because
there exists an analogous maximally extended open interval comprising the
small initial values of $\Sigma_i + \beta$ which lead to $(\Sigma_i + \beta)(\varsigma) < -\delta$.)
By construction, the solution with initial datum $(\Sigma_i + \beta)_{\mathrm{pos}}$ must
remain in $\mathscr{U}_i$ for all times, which implies that the $\alpha$-limit of this solution is $\partial\mathscr{D}_i$.
Accordingly, $\partial\mathscr{D}_i$ is a possible $\alpha$-limit set of orbits
in $\mathcal{X}$

To see that convergence to $\partial\mathscr{D}_i$ is non-generic (in the sense stated in the theorem)
we can invoke the Hartman-Grobman theorem for discrete flows (for the flow on
$\mathcal{H}_{\mathrm{P}}$); we obtain that
$\partial\mathscr{D}_i$ has the character of a saddle:
There exists solutions that converge to $\partial\mathscr{D}_i$, but
generically, solutions are driven away from this cycle.
Alternatively, we note that in order to converge to $\partial\mathscr{D}_i$
a solution must satisfy
\begin{equation*}
(\Sigma_i + \beta)(0) = (\Sigma_i + \beta)_0 = \frac{2}{1-w} \int_0^\infty e^{-\varsigma}\, 
\big[w_i(s_i(\varsigma),s_j(\varsigma),s_k(\varsigma)) - v_-\big] \:d\varsigma\:,
\end{equation*}
which is a direct consequence of~\eqref{varsigmaODE}.
Since $(\Sigma_i + \beta)_0 = o(1)$ as $s_i \rightarrow 0$ (and uniformly in $s_j$), 
there cannot exist an open neighborhood of $\mathrm{P}$ in $\mathcal{H}_{\mathrm{P}}$
such that solutions of the Poincar\'e map with initial data of that neighborhood possess $\mathrm{P}$
(i.e., $\partial\mathscr{D}_i$) as an $\alpha$-limit set.
This completes the proof of the theorem.
\end{proof}

\subsection*{The $\boldsymbol{-}$ cases}

To simplify the analysis of the $\boldsymbol{-}$ cases, we require the matter model to satisfy 
a typical `genericity' assumption which concerns the neighborhood of the isotropic state of the matter:
We assume definiteness of the Hessian $\mathfrak{h}$ of $\psi$ at the Friedmann point (i.e., we exclude the cases (iii)-(iv) of Lemma~\ref{stabilityF}).
The matter models we discuss in Section~\ref{mattermodels} satisfy the required assumption automatically.

\begin{Lemma}\label{FLemma}
In the $\boldsymbol{-}$ cases\/ \textnormal{\Aminus, \Bminus, \Cminus}, and \textnormal{\Dminus} 
the fixed
point\/ $\mathrm{F}$ is a hyperbolic saddle with a two-dimensional stable and a two-dimensional unstable manifold. 
\end{Lemma}

\begin{proof}
In terms of the coordinates~\eqref{stcoords} on $\mathscr{T}$,
the fixed point $\mathrm{F}$ is given by 
$(\Sigma_1,\Sigma_2,\Sigma_3) = (0,0,0)$ and $(t_i,t_j) = (\bar{t}_i,\bar{t}_j)$,
where $(\bar{t}_i,\bar{t}_j)$ is the (single) critical point of
the (positive) function $\psi$ on $\mathscr{T}$, see~\eqref{Fint}.
Using that $\psi = 0$ on $\partial\mathscr{T}$ (which we prove below)
it follows that $(\bar{t}_i,\bar{t}_j)$ is a global maximum of $\psi$.
Therefore, at least for a matter model with a generic characteristic function $\psi$, 
the Hessian $\mathfrak{h}$ of $\psi$ at $(\bar{t}_i,\bar{t}_j)$
is negative definite. Part~(ii) of Lemma~\ref{stabilityF} then implies 
the statement we have wished to prove.
It remains to show that $\psi\rightarrow 0$ as $\partial\mathscr{T}$ is approached.
To this end 
we use the representation~\eqref{wiinpsi2} of the rescaled matter quantities
and the fact that $v_- > w > v_+$, which holds by assumption, cf.~\eqref{vwv}.
Then there exists some $\epsilon > 0$ such that 
for all $t_j$ there exists $T>0$ and
\[
\frac{\partial}{\partial t_i} \log \psi = \frac{w_i -w}{2} \geq \epsilon\,,\qquad
\frac{\partial}{\partial t_i} \log \psi = \frac{w_i -w}{2} \leq -\epsilon
\]
for all $t_i < -|T|$ and $t_i > T$ respectively. 
Consequently, $\log\psi \rightarrow -\infty$ and thus $\psi\rightarrow 0$
as $t_i \rightarrow \pm\infty$. Hence $\psi = 0$ on $\partial\mathscr{T}$.
\end{proof}

\begin{Theorem}[\textbf{Future and past asymptotics}]\label{pastthm2}
Let $\gamma\subset \mathcal{X} \backslash \mathrm{F}$ 
be an orbit of the dynamical system~\eqref{dynamicalsystem}.
In the $\boldsymbol{-}$ cases\/ \textnormal{\Aminus, \Bminus, \Cminus},
and \textnormal{\Dminus}, 
one of the following possibilities occurs:
\vspace{-2ex}
\begin{itemize}
\item $\alpha(\gamma)$ is a fixed point on $\partial\mathcal{X}$ and
$\omega(\gamma)$ is a fixed point on $\partial\mathcal{X}$;
this is the generic case;
\item $\alpha(\gamma)$ is a fixed point on $\partial\mathcal{X}$ and
$\omega(\gamma) = \mathrm{F}$;
\item $\alpha(\gamma) = \mathrm{F}$ and $\omega(\gamma)$ is a fixed point
on $\partial\mathcal{X}$.
\end{itemize}
Table~\ref{tab2} gives a complete list of the fixed points on $\partial\mathcal{X}$
together with the number of orbits that converge to the particular points. 
\end{Theorem}

\begin{proof}
The theorem is essentially proved by repeating the first two steps
of the proof of Theorem~\ref{pastthm1}.
The main difference is the character of the fixed point $\mathrm{F}$,
which is described in Lemma~\ref{FLemma}.
Note that on $\partial\mathscr{C}_1 \cup \partial\mathscr{C}_2 \cup \partial\mathscr{C}_3$
(which is the relevant part of $\partial\mathcal{X}$ where $\alpha(\gamma)$/$\omega(\gamma)$ must
reside) there
do not exist any structures (like periodic orbits or
heteroclinic cycles) that qualify as possible
$\alpha$-/$\omega$-limits except the fixed points of Table~\ref{tab2}.
\end{proof}

\textit{Interpretation of Theorem~\ref{pastthm2} and conclusions}.
Theorem~\ref{pastthm2} (in conjunction with the results of
Table~\ref{tab2}) entails that
in each of the $\boldsymbol{-}$ cases, \textit{generic} solutions converge to fixed points
on the Kasner circle(s) as $\tau\rightarrow -\infty$, 
which implies that the past attractor
of the dynamical system~\eqref{dynamicalsystem} is located on the Kasner circle(s).
Therefore, each generic Bianchi type~I model with
anisotropic matter, where the matter is supposed to satisfy the requirements of the $\boldsymbol{-}$ 
cases, approaches a Kasner solution toward the singularity.
In contrast, toward the future, the asymptotic behaviour of generic solutions is quite diverse.
In particular, the asymptotic properties of the matter model, in the shape of
the function $v(s)$, play an important role, because these determine 
the number of fixed points in $\mathrm{D}_i$. Assuming the simplest case,
$\# \mathrm{D}_i = 1$, which is in accord with the examples of anisotropic
matter models discussed in Section~\ref{mattermodels}, we obtain
the following results:
In the cases \Aminus\ and \Bminus\ the behaviour of generic solutions 
towards the future is
$(\Sigma_i,\Sigma_j,\Sigma_k) \rightarrow \beta (2,-1,-1)$.
In case \Cminus, generic solutions converge to a Kasner solution towards
the future. However, there is only a subset of Kasner solutions that
qualify as possible future asymptotic states; it is only 
those Kasner solutions with $\max_l \Sigma_l > 2/\beta$ (which corresponds to 
$\max_l p_l > (2+\beta)/(3 \beta)$
when written in terms of the Kasner exponents)
that come into question.
This set of Kasner solutions that qualify as future asymptotic states becomes
larger with decreasing $\beta$.
Finally, in the case \Dminus ($\beta \leq -2$), generic solutions
converge to Kasner solutions toward the future, and conversely, 
each Kasner solution occurs as a future asymptotic state.

\begin{Remark}
Theorem~\ref{pastthm1} (in conjunction with the results of
Table~\ref{tab2}) also implies that there exist (non-generic) Bianchi type~I models
whose asymptotic behaviour is quite different.
In particular, in cases \Aminus, \Bminus, \Cminus, the orbits converging
to $\mathrm{D}_i$ as $\tau\rightarrow \infty$ give rise to solutions 
whose asymptotic behaviour toward the future 
is characterized by $(\Sigma_i,\Sigma_j,\Sigma_k) \rightarrow \beta (-1,1/2,1/2)$.
If $\#\mathrm{D}_i = 1$, this behaviour is non-generic; however, for 
matter models whose properties are such that $\#\mathrm{D}_i > 1$,
this asymptotic behaviour is shared by a generic set of solutions.
The most interesting non-generic behaviour concerns isotropization: 
There exist non-generic solutions (a one-parameter set)
that isotropize toward the past, and non-generic solutions (a one-parameter set)
that isotropize toward the future.
\end{Remark}

\begin{Remark}
In Section~\ref{boundaries} we have seen that the $\boldsymbol{-}$ cases \Aminus, \Bminus, \Cminus\
(and the special case $\beta = -2$ of case \Dminus) are compatible with the energy conditions.
However, despite this fact the assumptions of the $\boldsymbol{-}$ cases are representative of 
a matter model that is rather unconventional. 
This is due to the fact that the isotropic state of the matter is not energetically favorable
and thus unstable (cf.\ the proof of Lemma~\ref{FLemma}, where we showed that
the Hessian $\mathfrak{h}$ of $\psi$ is negative definite at the coordinates of $\mathrm{F}$).
\end{Remark}

\subsection*{The case \Azero}

The case \Azero\ represents a bifurcation between the $\boldsymbol{+}$ cases
and the $\boldsymbol{-}$ cases. 
The analysis of this case is difficult for several reasons, the most
obvious being the fact that center manifold analysis becomes ubiquitous, 
since several fixed points possess large center subspaces.
Another problem is the fact that the asymptotic behaviour of the 
function $\psi(s_1,s_2,s_3)$ as $(s_1,s_2,s_3)\rightarrow \partial\mathscr{T}$
is undetermined by the assumptions of case \Azero;
there are several alternatives: $\psi\rightarrow 0$, $\psi\rightarrow \infty$, 
$\psi$ might converge to some positive number, or $\psi$ might not
be convergent at all as $(s_1,s_2,s_3)\rightarrow \partial\mathscr{T}$.

Our expectation is that---under consistent assumptions on the behaviour of the function
$\psi$ on $\overline{\mathscr{T}}$---the subcase \Azeroplus\ resembles
the case \Aplus, and the subcase \Azerominus\ resembles the case \Aminus.
However, a careful analysis is required to establish these vague expectations
as facts; this goes beyond the scope of the present paper.

\section{Anisotropic matter models}
\label{mattermodels}

In this section we present in detail 
three important examples of matter models to which one can apply the main results of this paper: 
Collisionless matter, described by the Vlasov equation, elastic matter, and magnetic fields.
For a general introduction to collisionless matter and the Vlasov-Einstein system 
we refer to~\cite{A,R3}; the Bianchi type~I case is discussed in detail in~\cite{HU}.
For a thorough discussion on the general relativistic theory of elasticity we refer to~\cite{BS, CQ, KM, KS, T};
we will confine ourselves to deriving the energy-momentum tensor of elastic bodies.

\subsection{Collisionless matter}\label{vlasov}

Consider an ensemble of particles with mass $m$ that move along 
the geodesics of a spacetime $(M,\bar{g})$.
(The geodesic motion of the particles reflects the condition of absence 
of any interactions other than gravity; in particular, collisions are excluded.)
The ensemble of particles is represented by a distribution function (`phase space density') 
$f\geq 0$, which is defined on the mass shell, i.e., 
on the subset of the tangent bundle given by $\bar{g}(v,v) = -m^2$,
where $v$ denotes the (future directed) four momentum.
If $(t,x^i)$ is a system of coordinates on $M$ such that $\partial_t$ is timelike and $\partial_{x^i}$ is spacelike, 
then the spatial coordinates
$v^i$ of the four momentum are coordinates on the mass shell, and
we can regard $f$ as a function $f = f(t,x^i, v^j)$, $i,j =1,2,3$.
The distribution function $f$ 
satisfies the Vlasov equation
\begin{equation}\label{vlasoveq}
\partial_t f +\frac{v^j}{v^0}\partial_{x^j}f-\frac{1}{v^0}\Gamma^j_{\mu\nu}v^\mu v^\nu\partial_{v^j}f=0\:,
\end{equation}
where $v^0>0$ is determined in terms of the metric $\bar{g}_{\mu\nu}$ and the spatial coordinates $v^i$ 
of the momentum via the mass shell relation $\bar{g}_{\mu\nu} v^\mu v^\nu = -m^2$. 
The energy-momentum tensor is defined as
\begin{equation}\label{Tvlasov}
T^{\mu\nu}=\int f v^\mu v^\nu \sqrt{|\det\bar{g}|}\:|v_0|^{-1} \: dv^1 dv^2 dv^3\:.
\end{equation}
In the definition of a Bianchi type~I solution of the Einstein-Vlasov system,
where we use the notation and conventions of Section~\ref{setup},
$f$ is assumed to be independent of $x^i$ and so~\eqref{vlasoveq} takes the
form
\begin{equation}\label{vlasovbianchi}
\partial_t f+2 k^j_{\weg l} v^l \partial_{v^j} f=0\:.
\end{equation}
It is well known, see, e.g.,~\cite{MM, R1,R,R4}, that the general
solution of~\eqref{vlasovbianchi} is given by
\begin{equation}\label{solvlasov}
f = f(t,v^j)=f_0(v_j)\:,
\end{equation}
where $f_0$ is an arbitrary function which corresponds to the initial data;
usually, $f_0$ is assumed to be compactly supported, so that the integral~\eqref{Tvlasov} is well defined.
Using the spatial metric $g_{ij}$, cf.~\eqref{metric}, we may replace 
$\det \bar{g}$ by $\det g$ in~\eqref{Tvlasov} and we can write $|v_0| = \sqrt{m^2 + g_{ij} v^i v^j}$.

Accordingly, equation~\eqref{Tvlasov} means that the energy-momentum tensor is given as 
a function of the spatial metric $g_{ij}$ (which depends on the initial data $f_0$);
we have thus verified \textit{Assumption~\ref{assumptionT}\/} for collisionless matter
(in Bianchi type~I).
If $f_0$ satisfies the symmetry requirement
\begin{equation*}
f_0(v_1,v_2,v_3)=f_0(-v_1,-v_2,v_3)=f_0(-v_1,v_2,-v_3)=f_0(v_1,-v_2,-v_3)\:,
\end{equation*}
and the metric is diagonal, then the energy-momentum tensor $T^\mu_{\weg \nu}$ is diagonal as well, 
which proves that {\it Assumption} \ref{assumptiondiagonal} 
is satisfied.
Therefore, the Bianchi type~I Einstein-Vlasov system admits the class of diagonal models
as solutions, cf.~the remarks at the end of Section~\ref{setup}.

The rescaled principal pressures are given by
\begin{equation}\label{wivlasov}
w_i=\frac{s_i{\displaystyle\int} f_0 \,v_i^2 \Big[\frac{m^2}{\mathrm{x}} +\sum_k s_k v_k^2\Big]^{-1/2} dv_1 dv_2 dv_3}%
{{\displaystyle\int} f_0 \Big[\frac{m^2}{\mathrm{x}} + \sum_k s_k v_k^2\Big]^{1/2} dv_1 dv_2 dv_3}
=
\frac{s_i{\displaystyle\int} f_0 \,v_i^2 \Big[\sum_k s_k v_k^2\Big]^{-1/2} dv_1 dv_2 dv_3}%
{{\displaystyle\int} f_0 \Big[\sum_k s_k v_k^2\Big]^{1/2} dv_1 dv_2 dv_3}\:,
\end{equation}
when we assume that the particles have zero mass (see, however, the remark at the end of this section).
As a consequence, \textit{Assumption~\ref{assumptionwi}} holds with
\begin{equation}
w = \frac{1}{3}\:,
\end{equation}
and~\eqref{rhopi} takes the form
\begin{equation}
\rho = \rho(n,s_1,s_2,s_3) = n^{4/3} (s_1 s_2 s_3)^{-1/6} \int f_0 \Big[\sum_k s_k v_k^2\Big]^{1/2} dv_1 dv_2 dv_3\:.
\end{equation}
Taking the limit $s_i\to 0$ in~\eqref{wivlasov} we find that
\begin{equation}\label{wijkvlasov}
w_i = v_- = 0\:,\qquad 
w_j=\frac{s_j\int f_0 v_j^2 [s_j v_j^2+ s_k v_k^2]^{-1/2} d v_1 d v_2 d v_3}{\int f_0[s_jv_j^2+s_kv_k^2]^{1/2} d v_1 d v_2 d v_3}\:,
\qquad w_k = 1 - w_j
\end{equation}
on $\overline{\mathscr{C}}_i$, i.e., \textit{Assumptions~\ref{assumptiondominant}--\ref{a9}\/} are satisfied (with $v_- = 0$).
In order to make contact with~\eqref{wiandvI} we identify $I$ with $f_0$ and define 
\begin{equation}
u[I](z_1,z_2,z_3) :=  \frac{z_1{\displaystyle\int} f_0(v_1,v_2,v_3) \,v_1^2 \Big[\sum_k z_k v_k^2\Big]^{-1/2} dv_1 dv_2 dv_3}%
{{\displaystyle\int} f_0(v_1,v_2,v_3) \Big[\sum_k z_k v_k^2\Big]^{1/2} dv_1 dv_2 dv_3}
\end{equation}
for all $f_0$ and $(z_1,z_2,z_3)\in\overline{\mathscr{T}}$.
Using that $I_{(\sigma)} = f_0 \circ \sigma$ it is straightforward to show that~\eqref{wiandvI} holds,
i.e., $w_i = u[I_{(\sigma_i^{-1})}]\circ\sigma_i$ leads to~\eqref{wivlasov}.
The function $u_i(s)$ of Definition~\ref{v(s)def}${}^\prime$ can be read off easily; 
equation~\eqref{wijkvlasov} thus corresponds to~\eqref{wijku} with
$v_- = 0$, $v_+ =1$; cf.~also~\eqref{vpmvspez}.
We conclude that $\bm{\beta = 1}$, i.e., an ensemble of massless collisionless particles
constitutes an anisotropic matter model that is of class \Bplus.

Finally we note that \textit{Assumption~\ref{vassum}\/} is satisfied as well.
One can show that $u_i(s)$ (which replaces $v(s)$ in~\eqref{vassumeq})
is strictly monotonically increasing, hence there exists only one solution of~\eqref{diex}
and thus only one fixed point on $\mathscr{D}_i$, i.e., $\#\mathrm{D}_i = 1$.
Analogously, one can convince oneself that \textit{Assumption~\ref{F}\/} holds.

\begin{Remark}
If we consider ensembles of collisionless particles with positive mass, $m >0$,
$w_i$ and $w$ are functions of $(s_1,s_2,s_3)$ and an additional scale, which
can be taken to be $\mathrm{x}$ or $n = (\det g)^{-1/2}$ (where we
recall that $\mathrm{x} = n^{2/3} (s_1s_2s_3)^{-1/3}$).
(In~\cite{HU}, $\mathrm{x}$ is replaced by $z = m^2/(m^2 + \mathrm{x})$.)
In Section~\ref{conclusions} we will show that the analysis of this paper
carries over straightforwardly to this more general situation
(the main reason being that the length scale is a monotone function).
\end{Remark}

\subsection{Magnetic fields}\label{magnetic}

For an electromagnetic field represented by the antisymmetric electromagnetic
field tensor $F_{\mu\nu}$
the energy-momentum tensor is given by
\[
T^\mu_{\ \nu}=-\frac{1}{4\pi}\left(F^\mu_{\ \alpha}F^\alpha_{\ \nu}-\frac{1}{4}\delta^\mu_{\ \nu}F^\beta_{\ \alpha}F^\alpha_{\ \beta}\right)\:.
\]
The equations for the field are the Maxwell equations
\[
\nabla_\mu F^{\mu\nu}=0\:,\qquad \nabla_{[\sigma}F_{\mu\nu]}=0\:.
\] 
Consider specifically a purely magnetic field in a Bianchi type~I spacetime with metric~\eqref{metric} 
that is aligned along, say, the third axis. 
In this case the electromagnetic field tensor takes the form
\[
F_{\mu\nu}=\left(\begin{matrix}0 & 0 & 0 & 0\\ 0 & 0 & K & 0\\ 0 & -K & 0 & 0\\ 0 & 0 & 0 & 0 \end{matrix}\right),
\] 
where $K$ determines the magnetic field: $B^1 = 0$, $B^2 = 0$, $B^3= K (g^{11}g^{22}g^{33})^{1/2}$.
For a diagonal metric, the Maxwell equations imply that $K$ is a constant; hence the energy 
density $\rho$,
\[
\rho = \frac{1}{8 \pi} \: g^{11} g^{22}\: K^2\;,
\]
is a function of the metric (which depends on the initial data for the magnetic field).
Furthermore, $T^\mu_{\ \nu}$ is diagonal and
\begin{equation}\label{Tmagnetic}
T^1_{\ 1}=T^2_{\ 2}=\rho\:,\ T^3_{\ 3}=-\rho\:.
\end{equation}
Accordingly, \textit{Assumptions~\ref{assumptionT}} and~\textit{\ref{assumptiondiagonal}}
are satisfied and diagonal models exist.
It follows from~\eqref{Tmagnetic} that $w_1 =1$, $w_2 = 1$, $w_3 = -1$, and $w = 1/3$.
For these normalized anisotropic pressures all the remaining assumptions are satisfied straightforwardly 
and we have $v_- = 1$, $v_+ = 1$ and thus
\[
\beta=-2\:.
\] 
Therefore, in our classification, the asymptotic behaviour is of type \Dminus.
 
The conclusions are identical if we add an electric field parallel to the magnetic 
field. The Maxwell equations show that $E^1=E^2=0$ and $E^3 = L (g^{11}g^{22}g^{33})^{1/2}$, where
$L = \mathrm{const}$.
The energy density becomes $8 \pi \rho = g^{11} g^{22} (K^2 + L^2)$,
the energy-momentum tensor remains a functional of
the metric, and~\eqref{Tmagnetic} and its consequences remain valid.

\begin{Remark}
The magnetic field considered in~\cite{L} is not aligned with one axis and thus not included in our analysis. 
In fact, when not aligned, the magnetic field has to rotate and its dynamics becomes non-trivial
and has to be added to the system of equations; in addition, 
the energy-momentum tensor is no longer diagonal.  
\end{Remark}

\subsection{Elastic matter}\label{elasticity}

In elasticity theory, an elastic material in a completely relaxed state is 
represented by a three-dimensional Riemannian manifold $(N,\gamma)$, the \textit{material space}, 
where the points of $N$ identify the particles of the material (in the continuum limit) and 
$\gamma$ measures the distance between the particles in the completely relaxed state; 
let us denote by $X^A$, $A=1,2,3$, a system of local coordinates on $N$.
The coordinates on the spacetime $(M,\bar{g})$, on the other hand,  are denoted by $x^\mu$, 
$\mu=0,\dots,3$. 
The state of the elastic material is described by the \textit{configuration function} 
$\psi$, which is a (smooth) map
\[
\psi:M\to N\,,\qquad\quad x^\mu\mapsto X^A = \psi^A(x^\mu)\:,
\]
such that the kernel of the \textit{deformation gradient} 
$T\psi: TM \rightarrow TN$ is generated by a (future-directed unit) timelike 
vector field $u$, i.e., $\ker T\psi = \langle u \rangle$ or $u^\mu \partial_\mu \psi^A = 0$.
The vector field $u$ is the matter four-velocity; by construction, $\psi^{-1}(p)$ 
(i.e., the world-line of the particle $p \in N$) is an integral curve.

The pull-back of the material metric by the map $\psi$
is the \textit{relativistic strain tensor}
\[
h_{\mu\nu}=\partial_\mu\psi^A\partial_\nu\psi^B\,\gamma_{AB}\:;
\]
since $h_{\mu\nu}u^\mu=0$, $h$ represents a tensor in the orthogonal complement $\langle u \rangle^\perp$ of $u$ in $TM$.
$h_{\mu\nu}$ is a Riemannian metric on $\langle u \rangle^\perp$, hence
$h^\mu_{\ \nu}$ has three positive eigenvalues $h_1$, $h_2$, $h_3$. Note also that $\mathcal{L}_u h_{\mu\nu}=0$,
i.e., $h$ is constant along the matter flow.

The material is unstrained at the point $x$ iff $h_{\mu\nu} =g_{\mu\nu}$ holds $x$, where
\[
g_{\mu\nu}=\bar{g}_{\mu\nu}+u_\mu u_\nu
\]
is the Riemannian metric induced by the spacetime metric $\bar{g}$ on $\langle u \rangle^\perp$. 
The scalar quantity
\begin{equation}\label{numberdensity}
n=\sqrt{{\rm det}_gh}=\sqrt{h_1h_2h_3}
\end{equation}
is the \textit{particle density} of the material. 
This interpretation is justified by virtue of the continuity equation 
\[
\nabla_\mu\left(nu^\mu\right)=0\:.
\]

\begin{Definition}\label{elasticsymm}
An elastic material in a Bianchi type~I spacetime is said to be Bianchi type~I symmetric
iff $\mathcal{L}_{\xi_i} h_{\mu\nu}=0$, where $\xi_i$, $i=1,2,3$, are the Killing vectors of the spacetime. 
Furthermore, if the matter four velocity $u$ is orthogonal to the surfaces of homogeneity, 
the elastic material is said to be \textnormal{non-tilted}. 
\end{Definition}

According to Definition~\ref{elasticsymm} (and $\mathcal{L}_u h_{\mu\nu} = 0$),
the strain tensor of a Bianchi type~I symmetric and non-tilted elastic material satisfies
\begin{equation}\label{strainBianchiI}
h_{00}=h_{0k}=0\:,\quad h_{ij}= \mathrm{const}\:.
\end{equation}
in the coordinates of~\eqref{metric}.
Consequently, $h^i_{\weg j} = g^{ik} h_{jk}$ and thus $h_1$, $h_2$, $h_3$ depend only on $t$.

A specific choice of elastic material is made 
by postulating a \textit{constitutive equation}, i.e., the 
functional dependence of the (rest frame) energy density $\rho$ of 
the material on the configuration map, the deformation gradient 
and the spacetime metric. An important class of materials is the 
one for which this functional dependence enters only through 
the principal invariants of the strain 
tensor. In this case we have
\begin{equation}\label{lagrangian}
\rho=\rho(\mathfrak{q}_1,\mathfrak{q}_2,\mathfrak{q}_3)\:,
\end{equation}
where 
\[
\mathfrak{q}_1=\tr h\:,\qquad \mathfrak{q}_2=\tr h^2,\qquad \mathfrak{q}_3=\tr h^3\:;
\] 
since $n^2=(\mathfrak{q}_1^3-3\mathfrak{q}_1\mathfrak{q}_2+2\mathfrak{q}_3)/6$,
one of the principal invariants can be replaced by the particle density $n$.
The materials described by~\eqref{lagrangian} 
generalize the class of isotropic, homogeneous, 
hyperelastic materials from the classical theory of elasticity. 
In the following we shall refer to these material simply as elastic materials.
The stress-energy tensor associated with these 
materials is obtained as the variation with respect to the 
spacetime metric of the matter action $S_M=-\int\sqrt{|g|}\,\rho$. 
The general expression which results for the stress-energy tensor is 
\begin{equation}\label{generalT}
T_{\mu\nu}=2\frac{\partial\rho}{\partial \bar{g}^{\mu\nu}}-\rho\,\bar{g}_{\mu\nu}\:,
\end{equation}
which results in~\eqref{Tij} for the metric~\eqref{metric}.

In Bianchi type~I symmetry, and assuming that the elastic material is non-tilted, 
it follows from~\eqref{strainBianchiI} that the principal invariants of the strain are functions 
of the spatial metric $g$ only, hence non-tilted elastic materials satisfy \textit{Assumption~\ref{assumptionT}}.
In the following we introduce a natural class of constitutive equations that are compatible
with Assumptions~\ref{assumptionwi}--\ref{F} as well. 

\begin{Remark}
There exist elastic materials which do not satisfy these assumptions. 
For instance, for the constitutive equation considered in~\cite{CH}, which was taken from~\cite{KS}, 
the rescaled matter quantities (written in terms of the variables $s_1, s_2, s_3$)
do not have a continuous extension to the boundary $\partial\mathscr{T}$ of the space $\mathscr{T}$.
Therefore, strictly speaking, the analysis of this paper does not apply to this class of materials.
However, the reason for this problem is that the variables $(s_1, s_2, s_3)$
are ill-adapted to this particular constitutive equation; replacing
the `triangle variables' $(s_1, s_2, s_3)$ by `hexagon variables', cf.~\cite{CH}, remedies
this deficiency
and Assumptions~\ref{assumptionwi}--\ref{F} hold w.r.t.\ this 
alternative formulation of the problem.
The dynamics of the elastic matter models investigated in~\cite{CH} seems considerably 
more complicated than the dynamics of the matter models considered in this paper;
however, modulo the necessary reformulation of the problem, it appears that
the models of~\cite{CH} are of type $\beta =2$.
\end{Remark}

We consider elastic materials whose constitutive equation depends only on the number density $n$ 
and the dimensionless shear scalar $\mathfrak{s}$ defined by
\begin{equation}\label{shearscalar}
\mathfrak{s}=\frac{\mathfrak{q}_1}{n^{2/3}}-3=3\left(\frac{\mathscr{A}(h_1,h_2,h_3)}{\mathscr{G}(h_1,h_2,h_3)}-1\right),
\end{equation}
where $\mathscr{A}$ and $\mathscr{G}$ denote the arithmetic and geometric mean functions, respectively. 
The fundamental inequality $\mathscr{A}(a_1,a_2,a_3)\geq\mathscr{G}(a_1,a_2,a_3)$, which holds for all 
real non-negative numbers $a_1,a_2,a_3$, with equality iff $a_1=a_2=a_3$, see~\cite{HLP}, implies
that $\mathfrak{s}\geq 0$ and $\mathfrak{s}=0$ iff $h_1=h_2=h_3$. Thus the shear scalar $\mathfrak{s}$ measures deviations from isotropy. 
%
Using the requirement $\rho=\rho(n,\mathfrak{s})$ in~\eqref{generalT} leads to
\[
T_{\mu\nu}=\rho\,u_\mu u_\nu +\left(n\frac{\partial\rho}{\partial n}-\rho\right)g_{\mu\nu}+
\frac{2}{n^{2/3}}\frac{\partial\rho}{\partial\mathfrak{s}}\left(h_{\mu\nu}-\frac{1}{3}\mathfrak{q}_1g_{\mu\nu}\right).
\]

We restrict ourselves to constitutive equations of the \textit{quasi Hookean} form
\begin{equation}\label{constitutiveeq}
\rho= \check{\rho}(n)+\check{\mu}(n)\, f(\mathfrak{s})\:,
\end{equation}
where $\check{\rho}(n)$ 
is the \textit{unsheared energy density} and $\check{\mu}(n)$ the \textit{modulus of rigidity}. 
For such a constitutive equation we obtain
\begin{equation}\label{stress}
T_{\mu\nu}=\rho\, u_\mu u_\nu + p\,g_{\mu\nu}+
\frac{2\check{\mu}(n)}{n^{2/3}}\,f'(\mathfrak{s})\,\left(h_{\mu\nu}-\frac{1}{3} \mathfrak{q}_1 g_{\mu\nu}\right).
\end{equation}
Here, $p$ denotes the isotropic pressure $p$, which is given by
\begin{equation}\label{isopressure}
p=\check{p}(n)+\check\nu(n)\,f(\mathfrak{s})\:,\qquad \check{p}=n^2\frac{d}{dn}\left(\frac{\check\rho}{n}\right),\quad 
\check{\nu}=n^2\frac{d}{dn}\left(\frac{\check\mu}{n}\right).
\end{equation}
The quantity $\check{p}(n)$ is the \textit{unsheared pressure}. 
Note that if the constitutive equation~\eqref{constitutiveeq} 
depends only on $n$, i.e., if $f(\mathfrak{s})\equiv 0$, then the elastic material reduces to a 
perfect fluid with energy density $\check{\rho}$ and pressure $\check{p}$. 
  
To specify the functions $\check{\rho}$ and $\check{\mu}$ 
in the constitutive equation~\eqref{constitutiveeq} 
we postulate a linear equation of state between the unsheared pressure and the unsheared energy density, i.e., 
$\check{p} = a\check{\rho}\,$, and a linear equation 
of state between the modulus of rigidity and the unsheared pressure, i.e., $\check{\mu} =b\,\check{p}\,$. 
By~\eqref{isopressure} this is equivalent to setting
\[
\check{\rho} = \rho_0 n^{a+1}\,,  
\qquad \check{\mu}= \rho_0\, a b\, n^{a+1} 
\]
for some constant $\rho_0>0$.
Accordingly, the density $\rho$ and the pressure $p$ take the form
\begin{equation}\label{rho}
\rho=\rho_0 n^{a+1}\left(1+ab\, f(\mathfrak{s})\right)\:,\quad\text{and}\quad p=a\rho\,
\end{equation}
and the stress-energy tensor~\eqref{stress} becomes
\begin{equation}\label{stress2}
T_{\mu\nu}=\rho\left[u_\mu u_\nu+ag_{\mu\nu}+2(3+\mathfrak{s})\frac{ab\,f'(\mathfrak{s})}{1+ab\,f(\mathfrak{s})}
\left(\frac{h_{\mu\nu}}{\mathfrak{q}_1}-\frac{1}{3}g_{\mu\nu}\right)\right].
\end{equation}

We make a number of assumptions on the parameters $a,b$ and the function $f$:
\begin{itemize}
\item[(A1)] $a\in [-1,1)$, $a\,b\geq 0$.
\item[(A2)] $f(\mathfrak{s})\geq 0$ and $f(\mathfrak{s})=0$ if and only if $\mathfrak{s}=0$.
\item[(A3)] $f'(\mathfrak{s})>0$. 
\end{itemize} 
Conditions (A1) and (A2) imply that the energy density is non-negative and has a global minimum at zero shear. 
Condition (A3) then corresponds to the (physically reasonable) assumption that the rest energy of the body 
is an increasing function of the shear.

\begin{Remark}
If $b = 0$, the modulus of rigidity $\check{\mu}$ vanishes and the elastic matter
reduces to a perfect fluid with linear equation of state $p = a \rho$. 
If $a=0$ (so that $p = 0$), the choice of $b$ is irrelevant, since $a b = 0$; 
this is clear because shear cannot occur for dust. 
\end{Remark}

Denoting by $w_i$ the rescaled principal pressures, equation~\eqref{stress2} leads to
\begin{equation}\label{normalizedpressure}
w_i = \frac{p_i}{\rho}=
a+2 \,\underbrace{(3+\mathfrak{s})\frac{ab\,f'(\mathfrak{s})}{1+ab\,f(\mathfrak{s})}}_{\text{\normalsize $=:Q(\mathfrak{s})$}}
\left(\frac{h_i}{\mathfrak{q}_1}-\frac{1}{3}\right)\:.
\end{equation}
\begin{itemize}
\item[(A4)] We assume that $\sup_{z>0}Q(z)<\infty$ and that the limit $Q_\infty:=\lim_{z\to \infty}Q(z)$ exists.
\end{itemize}

For elastic matter that is Bianchi type~I symmetric and non-tilted the relativistic strain
is given by~\eqref{strainBianchiI}. 
We assume that $h_{ij}$ (which plays the role of the initial data for the matter) 
is diagonal; by scaling the spatial coordinates we can achieve $h_{ij}=\delta_{ij}$.
Substituting into~\eqref{stress2} it follows immediately that $T^i_{\weg j}(t_0)$ is diagonal
(since $g_{ij}(t_0)$ is diagonal), hence $T^i_{\weg j}$ remains diagonal for all times
by the evolution equations~\eqref{evolution}. 
Therefore, the class of diagonal Bianchi type~I models with elastic matter is well-defined,
cf.~the remarks at the end of Section~\ref{setup}.

Using $h_{ij}=\delta_{ij}$ we further obtain
\begin{subequations}\label{hsands}
\begin{equation}
h_1=g^{11}\,,\quad h_2=g^{22}\,,\quad h_3=g^{33}\,,\quad
\mathfrak{q}_1 = \mathrm{x}\,,\quad
n = (\det g)^{-1/2}\:,
\end{equation}
and therefore
\begin{equation}
\mathfrak{s}= \frac{\mathrm{x}}{n^{2/3}} -3  =  (s_1 s_2 s_3)^{-1/3}-3\:.
\end{equation}
\end{subequations}
Inserting~\eqref{hsands} into~\eqref{normalizedpressure} we find 
\begin{equation}\label{wielastic}
w_i(s_1,s_2,s_3)=a+2\left(s_i-\frac{1}{3}\right)Q(\mathfrak{s})
\end{equation}
and $w = a$.
This shows that the elastic materials under consideration satisfy \textit{Assumption~\ref{assumptionwi}}. 
Furthermore, by~(A4), the functions $w_i$ can be extended to $\partial\mathscr{T}$;
on $\overline{\mathscr{C}}_i$ we obtain
\[
w_i=a-\frac{2}{3}Q_\infty\:,\quad w_j=a+2\left(s_j-\frac{1}{3}\right)Q_\infty\:,\quad w_k=a+2\left(s_k-\frac{1}{3}\right)Q_\infty\:;
\]
hence \textit{Assumptions~\ref{assumptiondominant}--\ref{a9}} hold as well, where $v_- = a - 2 Q_\infty/3$.
Evidently, the rescaled pressures are of the form~\eqref{wiandv} with
\[
v(z_1,z_2,z_3)=a+2\left(z_1-\frac{1}{3}\right)Q_\infty\:;
\]
the function $v(s)$ is given by $v(s) = a + 2 (s-1/3) Q_\infty$, see~\eqref{vdef}; cf.~also~\eqref{wijkv}.
We find that
\begin{equation}\label{betaQ}
v_-=a-\frac{2}{3}Q_\infty\:,\quad v_+=a+\frac{4}{3}Q_\infty\:,\quad \text{and} \quad \beta=\frac{4 Q_\infty}{3(1-a)}\:.
\end{equation}
(A1)--(A4) imply that $\beta\geq 0$, i.e., elastic models are anisotropic models
of the $\boldsymbol{+}$ type (\Aplus, \Bplus, \Cplus, or \Dplus) 
or of the \Azero\ type. However, the latter case occurs only if $Q_\infty = 0$,
which implies that $v(s) \equiv w$ ($=a$), which is the special subcase of \Azero\
introduced in Assumption~\ref{vassum}.
It is easy to see from the linearity of the function $v(s)$ that \textit{Assumption~\ref{vassum}}
is satisfied for all values of $Q_\infty$.
In particular, there exists only one solution of~\eqref{diex}
and thus only one fixed point on $\mathscr{D}_i$, i.e., $\#\mathrm{D}_i = 1$.

\begin{Remark}
Assumption (A3) excludes the possibility $Q_\infty < 0$, hence
neither of the $\boldsymbol{-}$ cases can occur for the elastic materials. 
A formal way to obtain an elastic material that is of type~$\boldsymbol{-}$ 
would be to allow $f'<0$; this is unphysical, however, since it implies that 
the rest energy of the elastic body has a maximum at zero shear (instead of a minimum).  
\end{Remark}

It remains to check the validity of Assumption~\ref{F}. By~\eqref{wielastic}, the equations $w_i=a$ $\forall i$ are 
satisfied only if $s_1=s_2=s_3=1/3$ (i.e., at the center of $\mathscr{T}$) or
at points $(s_1,s_2,s_3)$ where $Q(\mathfrak{s})=0$. 
By~(A3), the latter equation has no solutions for finite values of $\mathfrak{s}$
(i.e., in $\mathscr{T}$), hence \textit{Assumption~\ref{F}} holds.

\begin{Remark}
It is immediate from~\eqref{wielastic} that the
dominant energy condition is satisfied if
\begin{itemize}
\item[(A5)] $\sup_{z>0}Q(z)\leq \mathcal{A}(a) := \min\Big\{\frac{3}{4}(1-a),\frac{3}{2}(1+a)\Big\}$.
\end{itemize}
Clearly, $0 \leq \mathcal{A}(a) \leq 1$; $\mathcal{A}(a) = 0$ if and only if $a = \pm 1$,
while $\mathcal{A}(a)=1$ if and only if $a=-1/3$.
If in addition $a\geq -1/3$ holds, then the strong energy condition is also satisfied; 
note in particular that the validity of the strong energy condition does not impose 
any condition on the constant $b$. 
\end{Remark}

\begin{Example}[John materials]
$f(\mathfrak{s})=\mathfrak{s}$.
This class of elastic materials was introduced in~\cite{T}.
In this case $Q_\infty = 1$, hence $\beta = 4/[3(1-a)]$ by~\eqref{betaQ}.
Consequently, this class of materials is of type \Aplus\ if $a\in[-1,-1/3)$,
of type \Bplus\ if $a =-1/3$, of type \Cplus\ if $a \in (-1/3,1/3)$, and of type \Dplus\  
if $a \geq 1/3$. 
Condition~(A5) is satisfied if and only if
\begin{equation}\label{john}
a=-\textfrac{1}{3}\:,\quad -1\leq b<0\:,
\end{equation}
i.e., in case \Bplus.
\end{Example}

\begin{Example}[Logarithmic materials]
$f(\mathfrak{s})=\log(\varepsilon+\mathfrak{s})-\log \varepsilon$ with $\varepsilon>0$;
for simplicity, $\varepsilon=3$.
Since $Q_\infty = 0$, these materials are of type \Azero\ where $v(s) \equiv w$ ($=a$).
Condition (A5) reads 
\begin{equation}\label{log}
a b\leq \mathcal{A}(a)\:,
\end{equation}
hence these materials satisfy the dominant energy condition if $b$ is sufficiently small ($a \neq \pm 1$).
\end{Example}

\begin{Example}[Power-Law materials] 
$f(\mathfrak{s})=(\varepsilon+\mathfrak{s})^\lambda-\varepsilon^\lambda$ with $\varepsilon >0$ and $\lambda>0$;
for simplicity, $\varepsilon=3$. 
In this case $Q_\infty = \lambda$, hence $\beta = 4 \lambda/[3(1-a)]$ by~\eqref{betaQ}.
Consequently, this class of materials is of one of the types \Aplus, \Bplus, \Cplus, or \Dplus,  
depending on the values of $a$ and $\lambda$.
Since $\sup_{z>0}Q(z) = \lambda \min\{1, a b \,3^\lambda\}$,
condition (A5) is satisfied if
\begin{equation}\label{power}
3^{-\lambda}\leq a b \leq\frac{3^{-\lambda}}{\lambda}\mathcal{A}(a)
\quad\text{ or if }\quad \big( a b\leq 3^{-\lambda}\big)\: \wedge\: \big(\lambda \leq \mathcal{A}(a)\big)\,.
\end{equation}
This condition reduces to~\eqref{john} if $\lambda=1$;
while~\eqref{power} is never satisfied if $\lambda>1$,
it holds for a certain range of $a$ and $b$, if $\lambda < 1$.
For instance, with $a=1/3$ we have $\mathcal{A}(a) = 1/2$ and~\eqref{power}
reads $b \leq \sqrt{3}$, $\lambda \leq 1/2$.
\end{Example}

\section{Concluding remarks}
\label{conclusions}

In this paper we have discussed the dynamics of Bianchi type~I solutions of the Einstein equations 
with anisotropic matter. 
The focus of our analysis has been the asymptotic behaviour of solutions 
in the limit of late times and toward the initial singularity. 
The matter model was not specified explicitly, but only through a set of mild 
assumptions that are motivated by basic physical considerations.
In this way, our analysis applies to a wide variety of anisotropic matter models
including matter models as different from each other as 
collisionless matter, 
elastic materials and magnetic fields.

A basic assumption on the matter model was to require 
that the isotropic pressure $p$
and the density $\rho$ obey a linear equation of state $p = w \rho$, where $w =\mathrm{const}\in (-1,1)$, 
see Assumption~\ref{assumptionwi}. 
It is important to note, however, that this assumption is not necessary.
On the contrary, based on the results derived in this paper
it is relatively simple to treat nonlinear equations of state.
Let us elaborate. Assume that $\rho$ is given by a more general
function instead of~\eqref{rhopi}, i.e., $\rho = \rho(n,s_1,s_2,s_3)$.
In this case we obtain from~\eqref{wwiintermsof} that $w = w(n,s_1,s_2,s_3)$ and $w_i = w_i(n, s_1,s_2,s_3)$, $i=1,2,3$;
an interesting subcase is $\rho(n,s_1,s_2,s_3) = \varphi(n) \psi(s_1,s_2,s_3)$; 
here, $w = w(n) = (\partial \log \varphi/\partial n) -1$
and $w_i = w_i(n,s_1,s_2,s_3)$ is given by~\eqref{wiinpsi}, where $w$ is replaced by $w(n)$.
The Einstein evolution equations, written in terms of the dynamical systems variables of Section~\ref{dynamicalsection},
decouple into an equation for $H$ and a reduced system of equations for the remaining variables,
where we can replace $\mathrm{x}$ by $n$, since $\mathrm{x} = n^{2/3} (s_1 s_2 s_3)^{-1/3}$. 
Accordingly, the dynamical system that encodes the dynamics is given by~\eqref{dynamicalsystem}, which 
is supplemented by the equation $n^\prime = -3 n$.
When we choose to compactify the variable $n$, i.e., when we replace $n$ by $N = n/(1+n)$,
the state space of this dynamical system is $\mathscr{K} \times \mathscr{T} \times (0,1)$.
If we assume an equation of state such
that $w(N,s_1,s_2,s_3)$ and $w_i(N,s_1,s_2,s_3)$ possess well-defined limits as $N\rightarrow 0$
and $N\rightarrow 1$, we can extend the dynamical system to the boundaries 
$\mathcal{X}_0 = \mathscr{K}\times\mathscr{T} \times \{0\}$ and
$\mathcal{X}_1 = \mathscr{K}\times\mathscr{T} \times \{1\}$ of the state space.
The dynamical system on each of these boundary subsets coincides with the system~\eqref{dynamicalsystem}
that we have discussed so extensively in this paper.
Since the variable $N$ is strictly monotone, the asymptotic dynamics of solutions of
the dynamical system is associated
with the limits $N\rightarrow 0$ and $N\rightarrow 1$.
Accordingly, asymptotically, the flow of the boundary subsets 
$\mathcal{X}_0$ and $\mathcal{X}_1$ (and thus the results of the present paper)
constitute the key to an understanding of 
the dynamics of the more general problem with nonlinear equations of state. 

It is conceivable that several of our assumptions can be relaxed (however, physics might disapprove),
which could lead to interesting extensions of the results of this paper. 
For instance, if Assumption~\ref{F} is removed, then there exist
several isotropic states of the matter and thus there might exist
Bianchi type~I solutions that isotropize both toward the past and
toward the future.
Assumption~\ref{assumptionT}, on the other hand, 
could be replaced by the condition that the 
energy-momentum tensor depends not only on the spatial metric but also on the 
second fundamental form. In this way it is possible
to extend the analysis to even larger classes of matter fields.
This might lead to valuable generalizations of the results presented
in this paper.


\begin{center}
{\it Acknowledgement}
\end{center} 
We gratefully acknowledge the hospitality of the Institut Mittag-Leffler.
Also, we would like to thank an anonymous referee for useful suggestions.
S.\ C.\ is supported by  Ministerio Ciencia e Innovaci\'on, Spain (Project MTM2008-05271).

\end{document}